\documentclass[11pt]{article}
\pdfoutput=1
\usepackage{jcapmod}

\usepackage{amsmath, amsfonts}
\usepackage{booktabs}
\usepackage[english]{babel}
\usepackage{graphicx}
\usepackage{multirow}
\usepackage{slashed}
\usepackage{subfig}
\usepackage{hyperref}
\usepackage[load-configurations=astronomy, range-units=brackets, range-phrase=-, per-mode=reciprocal, mode=math]{siunitx}

\numberwithin{equation}{section}

\def\beq{\begin{equation}}
\def\eeq{\end{equation}}

\def\d{\mathrm{d}}
\def\ell{l}

\def\As{A_\mathrm{s}}
\def\ns{n_\mathrm{s}}
\def\Neff{{N_\mathrm{eff}}}
\def\NeffLSS{\tilde{N}_\mathrm{eff}}
\def\thetasLSS{\tilde{\theta}_s}
\def\aeq{a_\mathrm{eq}}

\def\Pnw{P^\mathrm{nw}}
\def\Pw{P^\mathrm{w}}
\def\Pobs{P_g}
\def\Oobs{O_g}

\def\fsky{{f_\mathrm{sky}}}
\def\lmin{\ell_\mathrm{min}}
\def\lmax{\ell_\mathrm{max}}
\def\kmin{k_\mathrm{min}}
\def\kmax{k_\mathrm{max}}
\def\zmax{z_\mathrm{max}}

\DeclareSIUnit{\parsec}{pc}
\DeclareSIUnit{\Mpc}{\mega\parsec}
\DeclareSIUnit{\Gpc}{\giga\parsec}
\DeclareSIUnit{\h}{\mathit{h}}
\DeclareSIUnit{\hPerMpc}{\h\per\Mpc}
\DeclareSIUnit{\MpcPerh}{\per\h\Mpc}
\DeclareSIUnit{\muKelvin}{\mu\kelvin}

\definecolor{Blue}{rgb}{0.25, 0.41, 0.88}
\definecolor{DarkOrange}{rgb}{1.0, 0.549, 0.}
\definecolor{Purple}{rgb}{0.5, 0., 0.5}
\definecolor{Red}{rgb}{0.92, 0., 0.}
\definecolor{pyBlue}{RGB}{31, 119, 180}
\definecolor{pyRed}{RGB}{214, 39, 40}
\definecolor{pyGray}{rgb}{0.5, 0.5, 0.5}
\definecolor{pyLightBlue}{RGB}{30, 144, 255}
\definecolor{pyDarkBlue}{RGB}{0, 0, 128}
\definecolor{pyBlue2}{RGB}{0, 111, 237}
\definecolor{pyRed2}{RGB}{224, 52, 36}

\DeclareRobustCommand{\SkipTocEntry}[4]{}

\begin{document}
	
\pagenumbering{roman}
\begin{titlepage}
	\baselineskip=15.5pt \thispagestyle{empty}
	
	\bigskip\
	
	\vspace{1cm}
	\begin{center}
		{\fontsize{20.74}{24}\selectfont \sffamily \bfseries Searching for Light Relics\\[8pt] with Large-Scale Structure} 
	\end{center}
	
	\vspace{0.2cm}
	\begin{center}
		{\fontsize{12}{30}\selectfont Daniel Baumann,$^{\bigstar}$ Daniel Green$^{\spadesuit,\blacklozenge}$ and Benjamin Wallisch$^{\clubsuit,\bigstar}$} 
	\end{center}

	\begin{center}
		\vskip8pt
		\textsl{$^{\bigstar}$ Institute of Theoretical Physics, University of Amsterdam,\\Science Park 904, Amsterdam, 1098 XH, The Netherlands}
		
		\vskip8pt
		\textsl{$^\spadesuit$ Department of Physics, University of California, San Diego,\\9500 Gilman Drive, La Jolla, CA 92093, US}
		
		\vskip8pt
		\textsl{$^\blacklozenge$ Department of Physics, University of California, Berkeley,\\366 LeConte Hall, Berkeley, CA 94720, US}
		
		\vskip8pt
		\textsl{$^\clubsuit$ Department of Applied Mathematics and Theoretical Physics,\\University of Cambridge, Cambridge, CB3 0WA, UK}
	\end{center}

	\vspace{1.2cm}
	\hrule \vspace{0.3cm}
	\noindent {\sffamily \bfseries Abstract}\\[0.1cm]
	Light thermal relics of the hot big bang, often quantified by the parameter $\Neff$, are one of the primary targets of cosmological measurements. At present, the energy density in such relics is constrained to be less than ten percent of the total energy density in radiation. Upcoming cosmic microwave background (CMB) experiments, however, have the potential to measure the radiation density at the one-percent level, which is close to well-motivated theoretical targets. In this paper, we explore to what degree the CMB observations can be enhanced by future large-scale structure surveys. We carefully isolate the information encoded in the shape of the galaxy power spectrum and in the spectrum of baryon acoustic oscillations (BAO). We find that measurements of the shape of the power spectrum can significantly improve on current and near-term CMB experiments. We also show that the phase shift of the BAO spectrum induced by relic neutrinos can be detected at high significance in future experiments.
	\vskip10pt
	\hrule
	\vskip10pt
\end{titlepage}

\thispagestyle{empty}
\setcounter{page}{2}
\tableofcontents

\clearpage
\pagenumbering{arabic}
\setcounter{page}{1}

\clearpage
\section{Introduction}
\label{sec:introduction}

Future cosmological observations have the potential to measure the radiation density of the early universe at the subpercent level. This order of magnitude improvement over current constraints would provide a new window into the very early universe and allow us to search for extra light particles with very weak couplings to the Standard Model. Small changes to the radiation density of the early universe lead to well-understood changes in the anisotropy spectrum of the cosmic microwave background (CMB)~\cite{Bashinsky:2003tk, Hou:2011ec, Baumann:2015rya, Pan:2016zla}. The same effects also create imprints in the initial conditions for the clustering of matter and, hence, may be observable in the late universe. It is therefore natural to ask how much the constraints on extra relativistic species can be improved by including future observations of the large-scale structure (LSS) of the universe. 

\vskip4pt
Within the Standard Model (SM) of particle physics, neutrinos make a significant contribution to the radiation density of the early universe. The cosmic neutrino background (C$\nu$B) was created about one second after the Big Bang, when the expansion rate of the universe dropped below the weak interaction scale. Shortly after neutrino decoupling, electrons and positrons annihilated, transferring their entropy to photons, but not to the neutrinos. This slightly reduced the energy density of the neutrinos relative to that of the photons. Nevertheless, \SI{41}{\percent} of the total radiation density of the universe is still expected to be in the form of cosmic neutrinos. The gravitational effect of the C$\nu$B has recently been observed in the damping~\cite{Hou:2011ec} and the phase shift~\cite{Follin:2015hya, Baumann:2015rya} of the CMB anisotropy spectrum. 

\vskip4pt
An interesting consequence of many proposals for physics beyond the Standard Model (BSM) are extra light particles~\cite{Essig:2013lka}, such as axions~\cite{Peccei:1977hh, Weinberg:1977ma, Wilczek:1977pj}, axion-like particles (ALPs)~\cite{Arvanitaki:2009fg}, dark photons~\cite{Holdom:1985ag, Galison:1983pa} and light sterile neutrinos~\cite{Abazajian:2012ys}. These particles are often so weakly coupled to the SM that they escape detection in terrestrial experiments. However, in astrophysics and cosmology, we have access to high-density environments which can overcome the small cross sections and allow a significant production of the extra species. For example, new light particles can be produced in the interior of stars~\cite{Raffelt:1996wa}. The absence of an anomalous extra cooling over the lifetime of stars puts some of the best current constraints on weakly coupled species. A similar argument can be applied to cosmology~\cite{Brust:2013xpv, Chacko:2015noa, Baumann:2016wac}. The high densities of the early universe allow these particles to have been in thermal equilibrium with the SM and can therefore make a significant contribution to the total radiation density of the universe. New particles that are more weakly coupled than neutrinos would have decoupled before the QCD phase transition. Their contribution to the final radiation density is then suppressed, explaining why these particles have not been detected yet. In this paper, we will explore the sensitivity of future cosmological observations to this type of BSM physics.

\vskip4pt
The search for light thermal relics has been adopted as one of the main science targets of the next generation of CMB experiments, such as the CMB-S4 mission~\cite{Abazajian:2016yjj}. Through improved measurements of small-scale anisotropies and polarization, future CMB observations will be extremely sensitive to the damping and the phase shift of the anisotropy spectrum. In this work, we explore the additional constraining power provided by current and future LSS experiments, such as {(e)BOSS}~\cite{Dawson:2012va, Dawson:2015wdb}, DES~\cite{Abbott:2005bi}, DESI~\cite{Aghamousa:2016zmz}, LSST~\cite{Ivezic:2008fe} and Euclid~\cite{Laureijs:2011gra}. It was established in~\cite{Font-Ribera:2013rwa, Dodelson:2016wal, Obuljen:2017jiy} that these surveys carry information about relativistic species. We will examine how this information is encoded in both the shape of the matter power spectrum and the spectrum of baryon acoustic oscillations~(BAO). We find that measurements of the shape of the power spectrum can significantly improve on the current CMB constraints, although the largest improvements are subject to the usual challenge of modeling the power spectrum. The peak locations of the BAO spectrum carry additional information about light relics that is robust to corrections to the overall shape of the power spectrum~\cite{Baumann:2017lmt}, such as those arising from nonlinear gravitational evolution~\cite{Eisenstein:2006nj, Crocce:2007dt, Sugiyama:2013gza}. We will explore in detail how this information can be isolated in the BAO~spectrum. This protected information may play a useful role in elucidating apparent discrepancies between CMB and low-redshift measurements, and be a valuable tool in the search for exotic physics in the dark sector.

\vskip10pt
The outline of the paper is as follows. In Section~\ref{sec:motivation}, we present the theoretical motivation for a precise measurement of the radiation density in the early universe, focusing on the effects of extra light species on the spectrum of acoustic oscillations. We highlight that these effects are imprinted in both the CMB and BAO spectra. In Section~\ref{sec:forecast}, we forecast CMB and LSS~constraints on the number of relativistic species, $\Neff$, for a number of future observations. In Section~\ref{sec:phaseShift}, we isolate the information encoded in the phase shift of the BAO spectrum and study the prospects for extracting this information in upcoming surveys. Our conclusions are presented in Section~\ref{sec:conclusions}.

\vskip4pt
A series of appendices contain technical details of our analysis: In Appendix~\ref{app:CMB}, we describe our CMB forecasts and present results for a range of experimental configurations. In Appendix~\ref{app:specs}, we provide details of our LSS forecasts. We define the specifications for the galaxy surveys used in this work and present results for a range of data combinations and cosmologies. In Appendix~\ref{app:broadband+phaseShiftExtraction}, we outline our method for extracting the broadband spectrum and the phase shift. Finally, in Appendix~\ref{app:convergence}, we show a few of the convergence tests that we performed to establish the stability of our numerical analysis.

\section{Cosmological Signatures of Light Relics}
\label{sec:motivation}

It is rather remarkable that all current cosmological data (e.g.~\cite{Hinshaw:2012aka, Ade:2015xua, Alam:2016hwk, Abbott:2017wau}) is fit by a simple six-parameter model---the $\Lambda\mathrm{CDM}$ model. In this section, we introduce the standard cosmological model and its extension to include extra relativistic species. We review the imprints that light particles leave on the cosmic microwave background and the large-scale structure of the universe. We will pay particular attention to the unique signature that these particles leave on the spectrum of acoustic oscillations. In the next section, we will quantify the level of constraints on extra light species to be expected from future cosmological observations.

\subsection{The Standard Model}

The $\Lambda\mathrm{CDM}$ model includes two parameters characterizing the initial conditions, namely the amplitude $\As$ and the tilt $\ns$ of the spectrum of primordial curvature perturbations. The remaining four parameters are associated with the geometry and composition of the universe: The matter content of the universe is described by the physical baryon and dark matter densities, $\omega_b \equiv \Omega_b h^2$ and $\omega_c \equiv \Omega_c h^2$, where $h$ is the reduced Hubble constant $h \equiv H_0/\!\left(\SI{100}{\kilo\meter\per\second\per\Mpc}\right)$. Instead of the Hubble constant $H_0$, we use the angular size of the sound horizon at decoupling, $\theta_s \equiv r_s(z_*)/D_A(z_*)$, where $r_s$ is the physical sound horizon and $D_A$ is the angular diameter distance, both evaluated at the redshift of decoupling, $z_*$. The parameter $\theta_s$ receives a contribution from the dark energy density $\Omega_\Lambda$. The standard six-parameter model is completed by the optical depth~$\tau$. In Table~\ref{tab:cosmologicalParameters}, we list the fiducial values of the $\Lambda\mathrm{CDM}$ parameters, based on the Planck best-fit cosmology~\cite{Ade:2015xua}.
\begin{table}[h!]
	\centering
	\sisetup{group-digits=false}
	\begin{tabular}{c S[table-format=1.5] l} 
			\toprule
		Parameter 				& {Fiducial Value}	& Description															\\
			\midrule[0.065em]
		$\omega_b$ 				& 0.02230 			& Physical baryon density $\omega_b \equiv \Omega_b h^2$				\\
		$\omega_c$				& 0.1188			& Physical dark matter density $\omega_c \equiv \Omega_c h^2$			\\
		$100\,\theta_s$ 		& 1.04112 			& $100\,\times\,$angular size of the sound horizon at decoupling 		\\ 
		$\tau$ 					& 0.066 			& Optical depth due to reionization										\\
		$\ln(\num{e10}\As)$		& 3.064 			& Log of scalar amplitude (at pivot scale $k_0 = \SI{0.05}{\per\Mpc}$)	\\
		$\ns$					& 0.9667 			& Scalar spectral index (at pivot scale $k_0 = \SI{0.05}{\per\Mpc}$)	\\
			\midrule[0.065em]
		$\Neff$					& 3.046 			& Effective number of (free-streaming) relativistic species				\\
		$Y_p$					& 0.2478			& Primordial helium fraction 											\\ 
			\bottomrule 
	\end{tabular}
	\caption{Parameters of the reference cosmological model and their fiducial values based on~\cite{Ade:2015xua}.}
	\label{tab:cosmologicalParameters}
\end{table}

\vskip4pt
In this work, we are interested in future measurements of the radiation density of the universe. The contribution from photons, $\rho_\gamma$, is fixed by the measured value of the CMB temperature. In addition, the Standard Model of particle physics predicts a contribution from neutrinos. The expected radiation density from each neutrino species is
\beq
\rho_{\nu_i} = \frac{7}{8} \left(\frac{4}{11}\right)^{\!4/3} \rho_\gamma \equiv a_\nu^{-1} \rho_\gamma\, .	\label{eq:Nnu}
\eeq
The three neutrino species of the Standard Model (and their antiparticles) therefore contribute a significant amount to the total radiation density in the early universe: $\rho_\nu/\rho_r = \sum_i \rho_{\nu_i}/\rho_r \approx \SI{41}{\percent}$. Although neutrinos decoupled at early times, their gravitational effects are still relevant and have recently been observed in the CMB~\cite{Follin:2015hya, Baumann:2015rya}.

\subsection{Extra Light Relics}

Physics beyond the Standard Model may add an extra radiation density $\rho_X$ to the early universe.\footnote{This energy density may even be negative if it is not associated with a new particle species, but rather with non-standard properties of neutrinos or changes to the conventional thermal history.} It is conventional to measure this radiation density relative to the density $\rho_{\nu_i}$ of a single SM neutrino species:
\beq
\Delta\Neff \equiv \frac{\rho_X}{\rho_{\nu_i}} = a_\nu \frac{\rho_X}{\rho_\gamma}\, ,
\eeq
and define $\Neff = 3.046 + \Delta\Neff$ as the \textit{effective number of neutrinos}, although $\rho_X$ may have nothing to do with neutrinos. Current measurements of the CMB anisotropies and the light element abundances find~\cite{Ade:2015xua, Cyburt:2015mya}
\begin{align}
\Neff &= 3.04 \pm 0.18 \quad \text{(CMB)}\, ,	\label{eq:NCMB}\\[4pt]
\Neff &= 2.85 \pm 0.28 \quad \text{(BBN)}\, ,
\end{align}
which is consistent with the SM prediction of $\Neff = 3.046$. We expect that future cosmological observations will improve these constraints by up to an order of magnitude. Any non-zero~value for~$\Delta\Neff$ would indicate physics beyond the standard models of particle physics and/or cosmology.

\begin{figure}[t!]
	\begin{center}
		\includegraphics{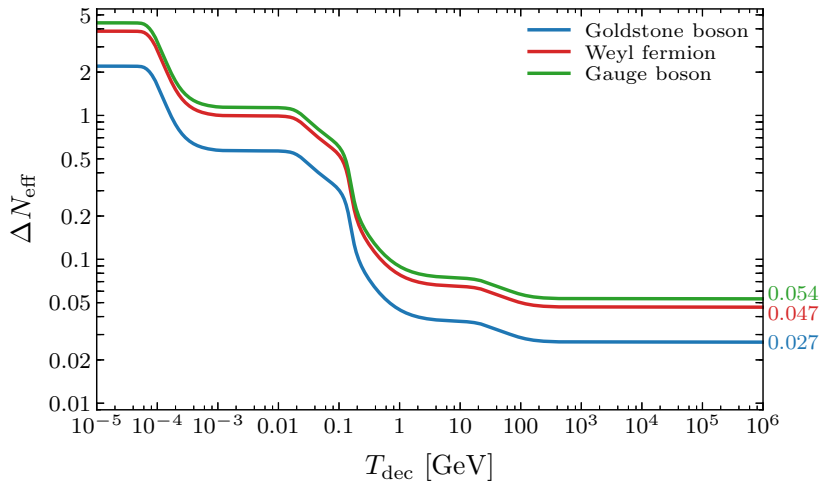}
		\caption{Contributions of a single thermally-decoupled Goldstone boson, Weyl fermion or massless gauge boson to the effective number of neutrinos, $\Delta\Neff$, as a function of its decoupling temperature $T_\mathrm{dec}$. The drop in $\Delta\Neff$ around $\SI{150}{\mega\electronvolt}$ is due to the QCD phase transition, where we employed the lattice QCD calculation of~\cite{Borsanyi:2016ksw}.}
		\label{fig:deltaNeff}
	\end{center}
\end{figure}

\vskip4pt
A natural source for $\Delta\Neff \neq 0$ are extra relativistic particles. Figure~\ref{fig:deltaNeff} %
shows the contribution to~$\Delta\Neff$ from a single thermally-decoupled species as a function of the decoupling temperature~$T_\mathrm{dec}$ and the spin of the particle. The plot assumes that the extra species was in thermal equilibrium at some point in the history of the universe and that the number of relativistic degrees of freedom at decoupling was not significantly larger than the SM value. We also assumed no significant entropy production after decoupling. We see that decoupling after the QCD phase transition produces a contribution to $\Neff$ that is comparable to that of a single neutrino species, which is in tension with current observations. Decoupling before the QCD phase transition, however, creates an abundance that is smaller by an order of magnitude and hence still consistent with current limits. Future observations will therefore give us access to particles that are more weakly coupled than neutrinos. The exclusion of the minimal thermal abundance $\Delta\Neff = 0.027$ would have important consequences for BSM physics~\cite{Brust:2013xpv, Chacko:2015noa, Baumann:2016wac}. We find it intriguing that this threshold seems to be within reach of future CMB and LSS observations. In this paper, we will quantify this expectation.
\begin{figure}[t!]
	\begin{center}
		\includegraphics{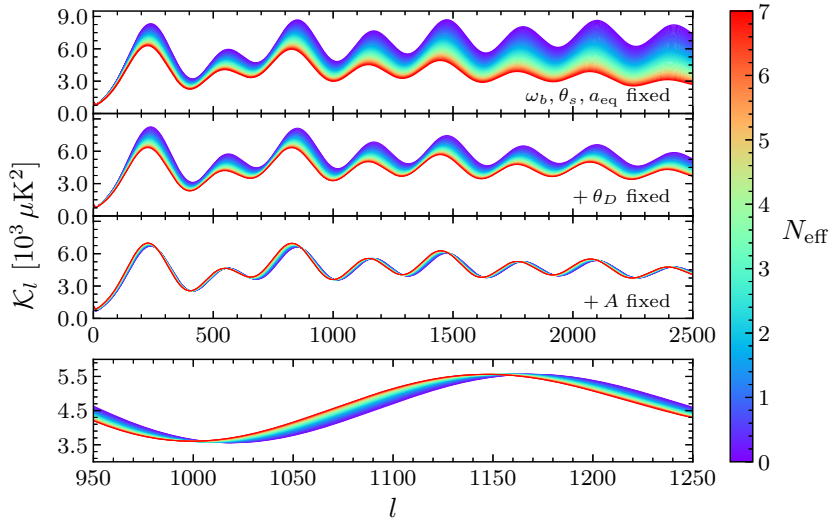}
		\caption{Variation of the CMB power spectrum as a function of $\Neff$. The spectra have been rescaled, so that the fiducial spectrum for $\Neff=3.046$ is undamped, i.e.~the exponential Silk damping was removed. Following~\cite{Follin:2015hya}, the physical baryon density~$\omega_b$, the scale factor at matter-radiation equality $\aeq \equiv \omega_m/\omega_r$ and the angular size of the sound horizon~$\theta_s$ are held fixed in all panels. The dominant effect in the first panel is the variation of the damping scale~$\theta_D$. In the second panel, we fixed $\theta_D$ by adjusting the primordial helium fraction $Y_p$. The dominant variation is now the amplitude perturbation $\delta A$. In the third panel, the spectra are normalized at the fourth peak. The remaining variation is the phase shift $\phi$ (see the zoom-in in the fourth panel).}
		\label{fig:CMBPhaseShift}
	\end{center}
\end{figure}

\subsection{Phases of New Physics}
\label{sec:phases}

Keeping the acoustic scale $\theta_s$ fixed (e.g.~by adjusting the Hubble constant $H_0$), an increase in the radiation density of the early universe reduces the mean free path of fluctuations in the photon-baryon fluid and increases the \textit{damping} of small-scale fluctuations~\cite{Hou:2011ec} (see Fig.~\ref{fig:CMBPhaseShift}). The constraint in~\eqref{eq:NCMB} is mostly derived from measurements of the CMB damping tail~\cite{Ade:2015xua, Keisler:2011aw}. However, the damping tail is also affected by changes to the primordial helium fraction, $Y_p$, which induces a variation in the free electron fraction and hence the mean free path of photons. In our forecasts, we will both fix $Y_p$ to the value demanded by BBN consistency ($\Lambda\mathrm{CDM}$+$\Neff$) and vary it ($\Lambda\mathrm{CDM}$+$\Neff$+$Y_p$) to explore the degeneracy with $\Neff$. The part of $\Neff$ that is associated with \textit{free-streaming} relativistic particles leads to a characteristic \textit{phase shift} in the CMB spectrum~\cite{Bashinsky:2003tk, Baumann:2015rya} (see Fig.~\ref{fig:CMBPhaseShift}), which helps to break the degeneracy between~$\Neff$ and~$Y_p$. The phase shift associated with SM neutrinos has recently been measured in the Planck spectrum~\cite{Follin:2015hya, Baumann:2015rya}. 

\begin{figure}[t!]
	\begin{center}
		\includegraphics{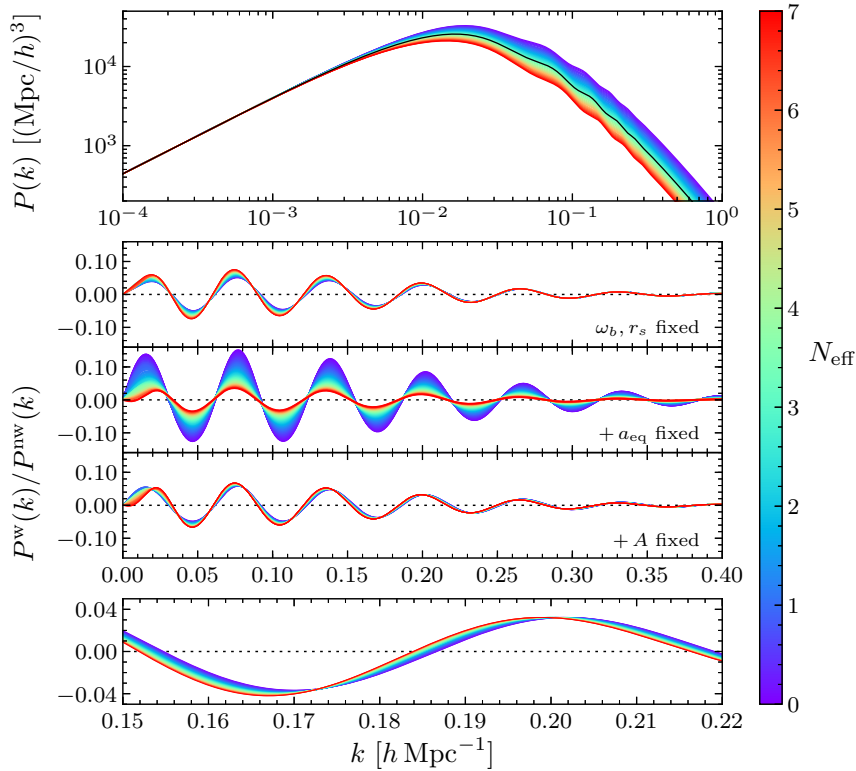}
		\caption{Variation of the matter power spectrum $P(k)$ (\textit{top}) and the BAO spectrum $\Pw(k)/\Pnw(k)$~(\textit{bottom}) as a function of $\Neff$. The physical baryon density~$\omega_b$ and the physical sound horizon at the drag epoch, $r_s$, are held fixed in all panels of the BAO spectrum. In the second BAO panel, we fixed the scale factor at matter-radiation equality, $\aeq \equiv \omega_m/\omega_r$. The variation in the BAO amplitude $\delta A$ is then the dominant contribution. In the third BAO panel, the spectra are normalized at the fourth peak and the bottom panel shows a zoom-in illustrating the remaining phase shift.\vspace{-2pt}}
		\label{fig:BAOPhaseShift}
	\end{center}
\end{figure}

\vskip4pt
In this work, we pay particular attention to the information about $\Neff$ contained in the BAO~spectrum. To isolate the BAO signal, we split the power spectrum into a smooth (`no-wiggle') part and an oscillatory (`wiggle') part,
\beq
P(k) \equiv \Pnw(k) + \Pw(k)\, .
\eeq 
Our method for performing this separation is described in Appendix~\ref{app:broadband+phaseShiftExtraction}. We will demonstrate that the most robust information about $\Neff$ lives in $\Pw(k)$. In particular, it was shown in~\cite{Baumann:2017lmt} that the phase of the BAO spectrum is immune to the effects of nonlinear gravitational evolution. In Figure~\ref{fig:BAOPhaseShift}, we show the dependence of the phase of the BAO spectrum on the number of relativistic species $\Neff$. We claim that this information is preserved after nonlinear corrections are taken into account. 

\section{Future Constraints on Light Species}
\label{sec:forecast}

We have argued that measuring the radiation density at the percent level provides an interesting window into early universe cosmology and beyond the Standard Model particle physics. In this section, we will further quantify the constraining power of future cosmological observations. We will consider two types of forecasts based on $P(k)$ and $\Pw(k)$. We will refer to these as `$P(k)$-forecasts' and `BAO-forecasts', respectively.

\subsection{Fisher Methodology}

We will use standard Fisher information theory to forecast the constraints of future observations. While Fisher forecasts have to be used with care, they provide useful guidance for the sensitivities and design of future experiments. In this section, we recall the basic elements of the Fisher methodology and its application to galaxy surveys~\cite{Bassett:2009mm, Verde:2009tu}. The relatively standard Fisher forecasting of CMB observations is summarized in Appendix~\ref{app:CMB}. Further details on the LSS forecasting can be found in Appendix~\ref{app:specs}.

\vskip4pt
Given a likelihood function $\mathcal{L}(\vec{\theta}\hskip1pt)$ for the model parameters $\vec{\theta} \equiv \{\omega_b, \omega_c, \theta_s, \tau, \As, \ns, \Neff, Y_p\}$, we define the Fisher matrix as the average curvature of the log-likelihood around the fiducial point in parameter space,
\beq
F_{ij} = - \left\langle \frac{\partial^2 \ln\mathcal{L}}{\partial \theta_i\, \partial \theta_j} \right\rangle ,
\eeq
where the expectation value denotes an average over all possible realizations of the data. If the likelihood is Gaussian, then the inverse Fisher matrix gives the covariance matrix. This means that $F_{ii}^{-1/2}$ is the error on the parameter $\theta_i$, when all other parameters $\theta_{j\ne i}$ are known, while $\sigma(\theta_i) = (F^{-1})_{ii}^{1/2}$ is the error on $\theta_i$ after marginalizing over the other parameters. More generally, the Cram\'er-Rao bound,
\beq
\sigma(\theta_i) \ge \sqrt{(F^{-1})_{ii}} \, ,	\label{eq:CramerRaoBound}
\eeq
gives a lower limit on the marginalized constraints.

\vskip4pt
The Fisher matrix for a galaxy survey is~\cite{Tegmark:1997rp} 
\beq
F_{ij} = \int_{-1}^1 \frac{\d\mu}{2}\, \int_{\kmin}^{\kmax} \frac{\d k\, k^2}{(2\pi)^2}\, \frac{\partial \ln \Pobs(k,\mu)}{\partial \theta_i} \frac{\partial \ln \Pobs(k,\mu)}{\partial \theta_j} \,V_\mathrm{eff}(k,\mu)\, , 	\label{eq:galaxyFisherMatrix}
\eeq
where $\Pobs(k,\mu)$ is the anisotropic galaxy power spectrum, $\mu$ is the cosine between the wavevector~$\vec{k}$ and the line-of-sight, and $V_\mathrm{eff}$ is the effective survey volume,
\beq
V_\mathrm{eff}(k,\mu) \equiv \int\!\d^3 r \left[\frac{n_g(\vec{r}\hskip1pt) \Pobs(k,\mu)}{n_g(\vec{r}\hskip1pt) P_g(k,\mu)+1} \right]^2 \approx \left[\frac{\bar{n}_g \Pobs(k,\mu)}{\bar{n}_g \Pobs(k,\mu)+1} \right]^2 V \, .	\label{eq:V_eff}
\eeq
In the second equality, we have assumed that the comoving number density of galaxies is independent of position, $n_g(\vec{r}\hskip1pt) \approx \bar{n}_g = const$, and introduced the actual survey volume $V$. To derive the constraints from independent redshift bins, we take $V$ to be the volume within each bin and add the corresponding Fisher matrices. The minimum wavenumber accessible in a survey is given by the volume of the survey\hskip1pt\footnote{We assume that the survey volume has a spherical geometry. The geometry of a given redshift bin (or the full survey volume) is neither spherical nor cubic, but we have checked that all of our results are essentially unaffected by this choice.} as $\kmin = 2\pi\,\left[3V/(4\pi)\right]^{-1/3}$.

\subsubsection{Modeling the Power Spectrum}
\label{sec:baoModeling}

In~\S\ref{sec:phases}, we introduced the linear matter power spectrum $P_\mathrm{lin}(k)$, and separated it into its smooth and oscillatory parts. In order to obtain semi-realistic constraints on most parameters of the cosmological model, it is often sufficient to model the observed galaxy power spectrum as $\Pobs(k) \approx b^2 P_\mathrm{lin}(k)$, where $b$ is the linear biasing parameter. However, the constraints on extra relativistic species are particularly sensitive to the way degeneracies are broken and to the nonlinear damping of the oscillatory feature, so we need to be more careful in the modeling of the signal~\cite{Blas:2016sfa, Hand:2017ilm, Ding:2017gad}. Moreover, since observations only determine the angular positions and redshifts of objects, we need to take into account the corresponding redshift space distortions~(RSD) and geometric projection effects.

\vskip4pt
Our model for the observed galaxy power spectrum is the following remapping of the linear matter power spectrum:
\beq
\Pobs(k,\mu) = {\color{Blue}b^2 F^2(k,\mu)} {\color{darkgreen}\Pnw(k,\mu)} \Big[ 1+ {\color{Red}O(k,\mu)}\,{\color{DarkOrange}D(k,\mu)}\Big] {\color{Purple}Z(k,\mu)}\, .	\label{eq:Pobs}
\eeq
All functions in this expression have an implicit redshift dependence. We now define the different elements of~\eqref{eq:Pobs}:
\vskip4pt
\begin{itemize}
\item $\color{Red}{O(k,\mu)}$: This function encodes the BAO signal and can be written as
\beq
O(k,\mu) \equiv B(k)\,O_\mathrm{lin}(k'(k,\mu)) + A(k) \, ,	\label{eq:O}
\eeq
where $O_\mathrm{lin}(k') \equiv \Pw_\mathrm{lin}(k')/\Pnw_\mathrm{lin}(k')$ is the normalized wiggle spectrum evaluated at the rescaled wavenumbers~\cite{Ballinger:1996cd}
\beq
k' = k\, \sqrt{(1-\mu^2)/q_\perp^2 + \mu^2/q_\parallel^2}\, , \quad \text{with}\quad q_\perp \equiv \frac{D_A(z)}{D_A^\mathrm{fid}(z)}\, , \quad q_\parallel \equiv \frac{H^\mathrm{fid}(z)}{H(z)}\, .	\label{eq:rescaledWavenumbers}
\eeq
This rescaling reflects the fact that the wavenumbers $k$ cannot be measured directly, but instead have to be derived from the measured angles and redshifts using the angular diameter distance $D_A^\mathrm{fid}(z)$ and Hubble rate $H^\mathrm{fid}(z)$ of a fiducial cosmology. This is often referred to as anisotropic geometric effects. In the limit of spherically-averaged clustering measurements, these become isotropic and $k' = k/q$, where $q = q_\perp^{2/3} q_\parallel^{1/3} = D_V(z)/D_V^\mathrm{fid}(z)$, with the radial BAO dilation given by $D_V \propto (D_A^2/H)^{1/3}$.

To model uncertainties in the BAO extraction, we have introduced two free functions $B(k)$ and $A(k)$ in~\eqref{eq:O}, which we take to be smooth polynomials in $k$ (see~\S\ref{sec:Broad}). Ultimately, we will marginalized over these polynomials to remove any information that is not robust to the BAO signal itself. 

\item $\color{Blue}{b(z)}$: The bias of the target galaxies (e.g.~luminous red galaxies, emission line galaxies or quasars) sets the overall amplitude of the signal in each redshift bin. We will make the common assumption that $b(z) \propto 1/D_1(z)$, where $D_1(z)$ is the linear growth function. This means that the bias is larger at high redshifts, which implies that the galaxy power spectrum may get significant corrections from nonlinear biasing even at high redshifts.

\item $\color{Blue}{F(k,\mu)}$: This function characterizes the effect of redshift space distortions. Following~\cite{Kaiser:1987qv}, we write
\beq
F(k,\mu) = \frac{1}{\left(q_\perp^2 q_\parallel\right)^{1/2}} \left[ 1+\beta\, \mu'(k,\mu)^2 R(k)\right] ,	\label{eq:linearRSD}
\eeq
where $\beta \equiv f/b$, with the linear growth rate $f \equiv \d\ln D_1/\,\d\ln a$. The factors of $q_i$ account for differences in the cosmic volume in different cosmologies. Projection effects on the angle to the line-of-sight are included as~\cite{Ballinger:1996cd}
\beq
\mu'(k,\mu) = \mu / \sqrt{ \mu^2 + (1-\mu^2) Q^2}\, ,
\eeq
where $Q \equiv q_\parallel/q_\perp$, which becomes unity in the isotropic case. BAO reconstruction removes redshift space distortions on large scales, which we have modeled by adding the factor $R(k) = 1-\exp[-(k\Sigma_s)^2/2]$ in~\eqref{eq:linearRSD}, where the value of $\Sigma_s$ depends on the experimental specifications, in particular the noise levels. In our baseline forecasts, we take $\Sigma_s \to \infty$, i.e.~$R \equiv 1$, but we comment on finite values of $\Sigma_s$ in~\S\ref{sec:constraints_planned}.

\item $\color{DarkOrange}{D(k,\mu)}$: This function models the nonlinear damping of the BAO signal~\cite{Eisenstein:2006nj, Seo:2007ns}
\beq
D(k,\mu) \equiv \exp\left[- \frac{1}{2}\Big(k^2\mu^2 \Sigma^2_\parallel +k^2(1-\mu^2) \Sigma_\perp^2\Big) \right] ,	\label{eq:Dk}
\eeq
where the damping scales perpendicular and parallel to the line-of-sight are given by
\begin{align}
\Sigma_\perp(z) 	&= 9.4\,(\sigma_8(z)/0.9)\, \si{\MpcPerh}\, , \\
\Sigma_\parallel(z) &= (1+f(z)) \,\Sigma_\perp(z)\, ,
\end{align}
with $\sigma_8$ being the amplitude of (linear) matter fluctuations at a scale of \SI{8}{\MpcPerh}. We account for BAO reconstruction by decreasing these damping scales by an appropriate factor, e.g.~$0.5$ for \SI{50}{\percent} reconstruction. Following~\cite{White:2010qd, Font-Ribera:2013rwa}, we include the degradation in the reconstruction due to shot noise using a reconstruction multiplier $r(x)$, i.e.~$\Sigma_i \to r(x) \Sigma_i$. We obtain $r(x)$ by interpolating over the table
\beq
\begin{aligned}
r &= (1.0,\, 0.9,\, 0.8,\, 0.7,\, 0.6,\, 0.55,\, 0.52,\, 0.5)\,, \\
x &= (0.2,\, 0.3,\, 0.5,\, 1.0,\, 2.0,\, 3.0,\, 6.0,\, 10.0)\,,
\end{aligned}
\eeq
with $r(x<0.2)=1.0$ and $r(x>10.0)=0.5$, which depends on the number density $\bar{n}_g$ via $x \equiv \bar{n}_g \Pobs(k_0,\mu_0)/0.1734$ evaluated at $k_0=\SI{0.14}{\hPerMpc}$ and $\mu_0 =0.6$. This means that we assume \SI{50}{\percent} reconstruction at high number densities and no reconstruction for low densities.

\item $\color{darkgreen}{\Pnw(k,\mu)}$: The linear no-wiggle spectrum $\Pnw_\mathrm{lin}(k,\mu)$ is determined from the linear power spectrum using the method described in Appendix~\ref{app:broadband+phaseShiftExtraction}. Nonlinear corrections to this spectrum can be parameterized as
\beq
\Pnw(k,\mu) = \tilde B(k)\Pnw_\mathrm{lin}(k'(k,\mu)) + \tilde A(k)\, ,	\label{eq:PnwX}
\eeq
where $\tilde B(k)$ and $\tilde A(k)$ are smooth functions (see~\S\ref{sec:Broad}). For the purpose of our BAO-forecasts, $\tilde A(k)$ and $\tilde B(k)$ are degenerate with $A(k)$ and $B(k)$ in~\eqref{eq:O} and it is therefore consistent to use the linear spectrum. 

\item $\color{Purple}{Z(k,\mu)}$: For photometric surveys, we take the uncertainty in the redshift determination of the targets into account through the following function:
\beq
Z(k,\mu) = \exp\left[-k^2 \mu^2 \Sigma_{z}^2 \right] ,
\eeq
where $\Sigma_z = c\,(1+z)\,\sigma_{z0} / H(z)$ is given in terms of the root-mean-square redshift error~$\sigma_{z0}$~\cite{Seo:2003pu, Zhan:2005ki}. The redshift error, which depends on the experimental specifications, reduces the effective resolution for modes along the line-of-sight. We neglect this effect for spectroscopic surveys.
\end{itemize}
When evaluating the derivatives in the Fisher matrix~\eqref{eq:galaxyFisherMatrix}, the parameters $b(z)$, $\beta$, $R(k)$, $D(k,\mu)$ and $Z(k,\mu)$ are always computed using the fiducial cosmology. We are assuming that, after accounting for modeling uncertainties, no relevant cosmological information can be recovered from these functions.

\subsubsection{Accounting for Broadband Effects}
\label{sec:Broad}

Nonlinear evolution and biasing can change the shape of the power spectrum at high wavenumbers in a way that cannot be modeled from first principles. We account for this uncertainty by marginalizing over polynomials in $k$ in both the $P(k)$- and BAO-forecasts. In particular, the functions introduced in~\eqref{eq:PnwX} are defined as
\beq
\tilde A(k,z_i) = \sum_{n=0}^{N_a} \tilde a_{n,i} \, k^{n} \, , \qquad
\tilde B(k,z_i) = \sum_{m=0}^{N_b} \tilde b_{m,i} \, k^{2m}\, .	\label{eq:AB}
\eeq
As indicated, we allow independent polynomials in each redshift bin centered around $z_i$. The coefficients $\tilde a_{n,i}$ and $\tilde b_{m,i}$ are included in the list of parameters $\theta_i$. Derivatives with respect to these parameters are determined analytically, using the fiducial values $\tilde b_{0,i} =1$ and $\tilde a_{n,i} =\tilde b_{m\neq0,i}= 0$. A more careful treatment would replace this polynomial model with a perturbative model for the dark matter and biasing, and would marginalize over the bias parameters. In practice, this has been shown to give qualitatively similar forecasts~\cite{Gleyzes:2016tdh}. Our marginalization procedure is therefore sufficient to illustrate the sensitivity of our forecasts to broadband information.

\vskip4pt
Our BAO-forecasts will marginalize over the `broadband corrections' in~\eqref{eq:O}, with $A(k)$ and~$B(k)$ defined as in~\eqref{eq:AB}.\footnote{To avoid a proliferation of parameters, we will use $a_n$ and $b_n$ for the parameters in both~\eqref{eq:O} and~\eqref{eq:PnwX}, i.e.~we will drop the tildes from now on. Which parameter set is meant will be clear from the context.} At the level of the Fisher matrix, marginalizing over a polynomial and an exponential are equivalent. As a result, the function $B(k)$ captures the uncertainty in the damping scales $\Sigma_\parallel$ and $\Sigma_\perp$ in~\eqref{eq:Dk}. This implies that our marginalization procedure will eliminate any cosmological information associated with the nonlinear damping of the power spectrum, leaving the distinct information contained in the oscillating part of the spectrum~$O_\mathrm{lin}(k'(k,\mu))$. This type of procedure is used in the analysis of BAO data to correct for errors made in the modeling of $\Pnw(k)$, see e.g.~\cite{Beutler:2016ixs}.

\vskip4pt
We will choose various levels of marginalization in our forecasts. This will help to distinguish the information encoded in the smooth shape of the spectrum, $\Pnw(k)$, from that contained in the frequency and phase of the BAO spectrum, $\Pw(k)$. In addition, these marginalizations also give a sense for the level of robustness of each type of information when accounting for the various uncertainties in modeling the data of a realistic galaxy survey.

\subsubsection{Extracting the BAO Signal}
In describing the power spectrum, we introduced the idea of marginalizing over polynomials to remove the information in $\Pobs(k)$ that is thought to be degenerate with nonlinear evolution and galaxy biasing. The BAO spectrum is known to be robust to these effects and should therefore survive any such treatment. In principle, the BAO signal could be isolated with sufficient marginalization. However, in practice, it is more useful to extract the information associated with the BAO signal before any marginalization. The robustness of the BAO spectrum to nonlinearities means we can be more aggressive with our choice of $\kmax$ and less cautious with our marginalization. Consequently, it is convenient to treat the BAO signal and the broadband information independently. 

\vskip4pt
The observed BAO spectrum is defined by 
\beq
\Oobs(k,\mu) \equiv \frac{\Pobs^\mathrm{w}(k,\mu)}{\Pobs^\mathrm{nw}(k,\mu)} = D(k,\mu) \, O(k,\mu) \, ,	\label{eq:baoSpectrum}
\eeq
where $D(k,\mu)$ and $O(k,\mu)$ were introduced in~\eqref{eq:Pobs}. To derive the new Fisher matrix for the BAO spectrum directly, we first write the derivatives of $\Pobs(k,\mu)$ as
\beq
\frac{\partial \ln \Pobs(k,\mu)}{\partial \theta_i} = \frac{1}{\Pobs^\mathrm{nw} + \Pobs^\mathrm{w}} \left( \frac{\partial \Pobs^\mathrm{nw}}{\partial \theta_i} + \frac{\partial \Pobs^\mathrm{w}}{\partial \theta_i}\right) .	\label{eq:dP}
\eeq
We then drop the term proportional to $\partial_{\theta_i} \Pobs^\mathrm{nw}$ since it is degenerate with the marginalization over the broadband corrections. For the same reason, we write $\partial_{\theta_i} \Pobs^\mathrm{w} \approx b^2 F^2 \Pnw D\, \partial_{\theta_i} O$, i.e.\ we do not act with the derivatives on the functions $D(k,\mu)$ and $b F(k,\mu)$. The derivative in~\eqref{eq:dP} therefore becomes
\beq
\frac{\partial \ln \Pobs(k,\mu)}{\partial \theta_i} \approx \frac{D(k,\mu)}{1+D(k,\mu) \, O(k,\mu)} \frac{\partial O(k,\mu)}{\partial \theta_i} \, .
\eeq
While the derivatives that we have dropped are non-zero, the marginalization procedure described above is designed to remove them and the forecasts for cosmic parameters should consequently be the same. Removing this information by hand (and marginalizing) ensures that our BAO-forecasts do not include these broadband effects, as we will show in Fig.~\ref{fig:marginalizationPlanned}. The resulting Fisher matrix is then given by
\beq
F_{ij} = \int_{-1}^1 \frac{\d\mu}{2}\, \int_{\kmin}^{\kmax} \frac{\d k\, k^2}{(2\pi)^2}\, \frac{D(k,\mu)^2 }{(1+D(k,\mu) \, O(k,\mu))^2} \frac{\partial O(k,\mu)}{\partial \theta_i} \frac{\partial O(k,\mu)}{\partial \theta_j} \,V_\mathrm{eff}(k,\mu)\, .	\label{eq:baoFisherMatrix}
\eeq
We note that this Fisher matrix depends on $\Pobs^\mathrm{nw}(k,\mu)$ only through $V_\mathrm{eff}(k,\mu)$, which determines the signal-to-noise. For photometric surveys, we replace $V_\mathrm{eff}(k,\mu) \to Z(k,\mu)^2 V_\mathrm{eff}(k,\mu)$ to account for the redshift error and the associated reduction of power along the line-of-sight. In principle, we should model $\Pobs^\mathrm{nw}(k,\mu)$ using the nonlinear (galaxy) power spectrum, given that we will work close to the nonlinear regime. However, nonlinear evolution also correlates the modes and produces a non-Gaussian covariance matrix. Since most of the surveys under consideration in this paper are limited by shot noise, using the nonlinear power spectrum without taking into account the associated mode coupling in the covariance would artificially increase the number of signal-dominated modes. To be consistent with the use of a Gaussian covariance, our forecasts will therefore use the linear broadband spectrum.

\subsection{Summary of Results}

We are now ready to forecast the constraints of current and future CMB and LSS observations~on the effective number of relativistic species $\Neff$. Unless stated otherwise, our baseline analysis assumes a $\Lambda\mathrm{CDM}$+$\Neff$ cosmology in which the primordial helium fraction $Y_p$ is fixed by consistency with BBN.  At the end of the section, we will also present results with $Y_p$ as a free parameter. We will further dissect the information content of the BAO spectrum in the next section.

\vskip4pt
In Appendix~\ref{app:CMB}, we present detailed forecasts for current and future CMB experiments. The expected $1\sigma$ constraints for representative versions of the Planck satellite, a near-term CMB-S3 experiment and a future CMB-S4 mission are $\sigma(\Neff) = 0.18$, $0.054$, $0.030$, respectively. In~\S\ref{app:futureCMB}, we show how these constraints depend on variations of the experimental configurations. We would like to know how much these CMB constraints would improve with the addition of LSS~data.

\vskip4pt
We will give the results of two types of forecasts based on $P(k)$ and $\Pw(k)$. Our $P(k)$-forecasts apply the Fisher matrix~\eqref{eq:galaxyFisherMatrix} with $\kmax = \SI{0.2}{\hPerMpc}$ and marginalize over $b_{m\leq1}$. To be conservative about nonlinear biasing, we do not increase $\kmax$ at large redshifts, despite the (near-)linearity of the matter power spectrum. Our BAO-forecasts use the Fisher matrix~\eqref{eq:baoFisherMatrix} with $\kmax = \SI{0.5}{\hPerMpc}$ and marginalize over $a_{n\leq4}$, $b_{m\leq3}$. We will also show how these forecasts depend on the choice of $\kmax$ and the level of marginalization.

\subsubsection{Constraints from Planned Surveys}
\label{sec:constraints_planned}

A number of galaxy surveys are expected to take place over the next decade. The power of these surveys to constrain $\Neff$ is most sensitive to the survey volume, the number densities of galaxies and the redshift errors (spectroscopic versus photometric). The precise specifications of the surveys used in our analysis are given in Appendix~\ref{app:specs}, where we also present more detailed forecasts for the full set of parameters.

\paragraph{Baseline results}
In Table~\ref{tab:CMB+Pk_Neff}, %
\begin{table}[b!]
	\centering
	\begin{tabular}{c S[table-format=2.4] S[table-format=2.4] S[table-format=2.4] S[table-format=2.4] S[table-format=2.4] S[table-format=2.4] S[table-format=2.4]}
			\toprule
					& 		& \multicolumn{4}{c}{spectroscopic}				& \multicolumn{2}{c}{photometric}	\\
			\cmidrule(lr){3-6} \cmidrule(lr){7-8}
					& CMB 	& {BOSS}	& {eBOSS}	& {DESI}	& {Euclid}	& {DES}		& {LSST}				\\
			\midrule[0.065em] 
		Planck		& 0.18	& 0.14		& 0.13		& 0.087		& 0.079		& 0.17		& 0.14					\\
		CMB-S3		& 0.054	& 0.052		& 0.051		& 0.045		& 0.043		& 0.054		& 0.052					\\
		CMB-S4		& 0.030	& 0.030		& 0.030		& 0.028		& 0.027		& 0.030		& 0.030					\\
			\bottomrule
	\end{tabular}
	\caption{Forecasted $1\sigma$ constraints on $\Neff$ for various combinations of current and future CMB and LSS experiments using $P(k)$-forecasts with $\kmax = \SI{0.2}{\hPerMpc}$.}
	\label{tab:CMB+Pk_Neff}
\end{table}
we present the $1\sigma$ constraints on $\Neff$ for various combinations of current and future CMB and LSS experiments using the full $P(k)$-forecast. In Table~\ref{tab:CMB+BAO_Neff}, we compare these results to the same experiments using our BAO-forecasts. At BOSS levels of sensitivity and number densities, the BAO feature makes the most significant impact on constraints, particularly when combined with a CMB experiment like Planck. In contrast, with the larger volume and redshift range of DESI, the broadband shape carries most of the information and can lead to a significant improvement in the constraint on $\Neff$ both for Planck and a typical CMB-S3 experiment. Finally, photometric redshift surveys like DES and LSST generally perform worse than spectroscopic surveys because they are effectively two-dimensional for the scales of interest. However, the employed redshift error is conservative and we do not take the full potential of these surveys into account as we are only considering observations of galaxy clustering and have not included weak gravitational lensing measurements, for instance. We expect the constraints to improve with these additional LSS observables, but quantifying this is beyond the scope of this work.
\begin{table}[t]
	\centering
	\begin{tabular}{c S[table-format=2.4] S[table-format=2.4] S[table-format=2.4] S[table-format=2.4] S[table-format=2.4] S[table-format=2.4] S[table-format=2.4]}	
			\toprule
					& 		& \multicolumn{4}{c}{spectroscopic}				& \multicolumn{2}{c}{photometric}	\\
			\cmidrule(lr){3-6} \cmidrule(lr){7-8}
					& {CMB} & {BOSS}	& {eBOSS}	& {DESI}	& {Euclid}	& {DES}		& {LSST}				\\
			\midrule[0.065em] 
		Planck		& 0.18	& 0.15		& 0.15		& 0.14		& 0.14		& 0.16		& 0.15					\\
		CMB-S3		& 0.054	& 0.052		& 0.052		& 0.050		& 0.050		& 0.054		& 0.052					\\
		CMB-S4		& 0.030	& 0.030		& 0.030		& 0.029		& 0.029		& 0.030		& 0.030					\\
			\bottomrule
	\end{tabular}
	\caption{Forecasted $1\sigma$ constraints on $\Neff$ for various combinations of current and future CMB and LSS experiments using BAO-forecasts with $\kmax = \SI{0.5}{\hPerMpc}$. }
	\label{tab:CMB+BAO_Neff}
\end{table}

\begin{figure}[t!]
	\begin{center}
		\includegraphics{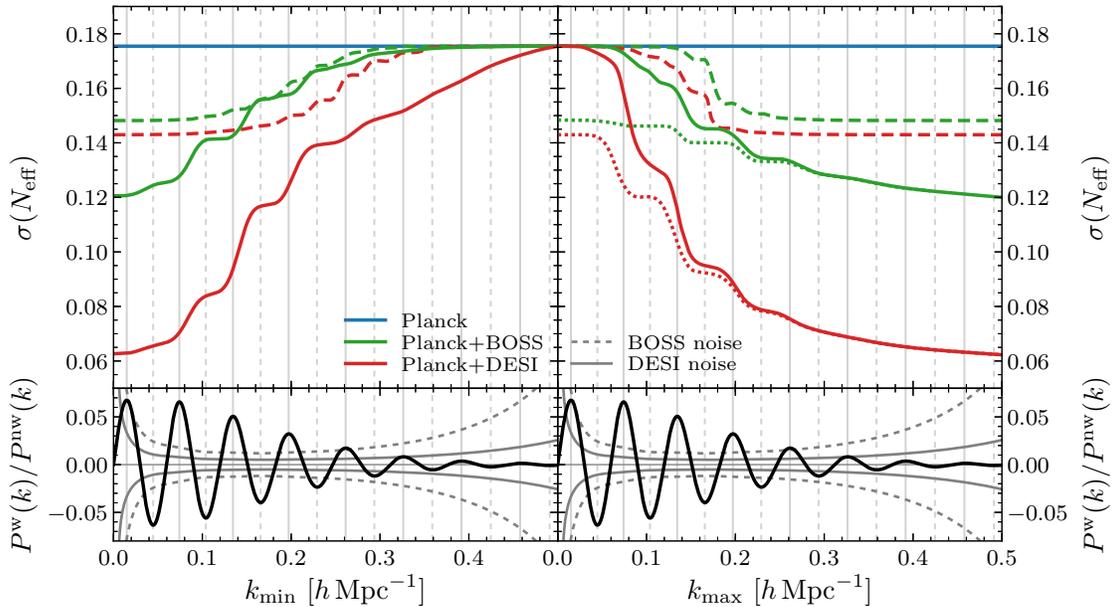}
		\caption{Forecasts for BOSS and DESI combined with Planck as a function of the smallest (\textit{left}) and largest (\textit{right}) Fourier modes used in the forecast, $\kmin$ and $\kmax$, with $\kmax=\SI{0.5}{\hPerMpc}$ in the left panel. The solid and dashed lines indicate the constraints from the $P(k)$- and BAO-forecasts, respectively. Shown as the dotted lines are the ``optimal constraints'' as described in the main text. The lower panel displays the linear BAO spectrum and an estimate of the noise levels.}
		\label{fig:kminkmax}
	\end{center}
\end{figure}

\paragraph{Sensitivity to $\boldsymbol{k_\mathrm{max}}$}
The broadband signal is sensitive to nonlinear effects and we should therefore understand how sensitive these results are to the choice of $\kmax$. In particular, we have chosen $\kmax = \SI{0.2}{\hPerMpc}$ in Table~\ref{tab:CMB+Pk_Neff}, but the usable range of scales is uncertain. Figure~\ref{fig:kminkmax} shows how the constraints vary as a function of the maximal wavenumbers included in the analysis,~$\kmax$, for both the $P(k)$- and BAO-forecasts. For the BAO-forecasts, we see a clear plateau for $\kmax > \SI{0.2}{\hPerMpc}$. This behavior is due to the damping of the oscillations at higher $k$ relative to the smooth power spectrum. Cosmic variance is ultimately determined by the amplitude of the smooth power spectrum and one cannot recover the high-$k$ oscillations even by lowering the shot noise. In contrast, the $P(k)$-forecasts show improvements out to $\kmax > \SI{0.3}{\hPerMpc}$.

\vskip2pt
Given that the BAO spectrum is robust to nonlinear evolution, it is natural to consider an optimal combination of the $P(k)$ and BAO spectra that uses all the available information. This means using $P(k)$ up to a certain $\kmax$ and adding BAO-only information for larger $k$. The~$\kmax$ of the $P(k)$ analysis then becomes the $\kmin$ of the BAO analysis to avoid double counting the information. Results for this optimal combination are shown as the dotted line in Fig.~\ref{fig:kminkmax}.
\begin{figure}[t!]
	\begin{center}
		\includegraphics{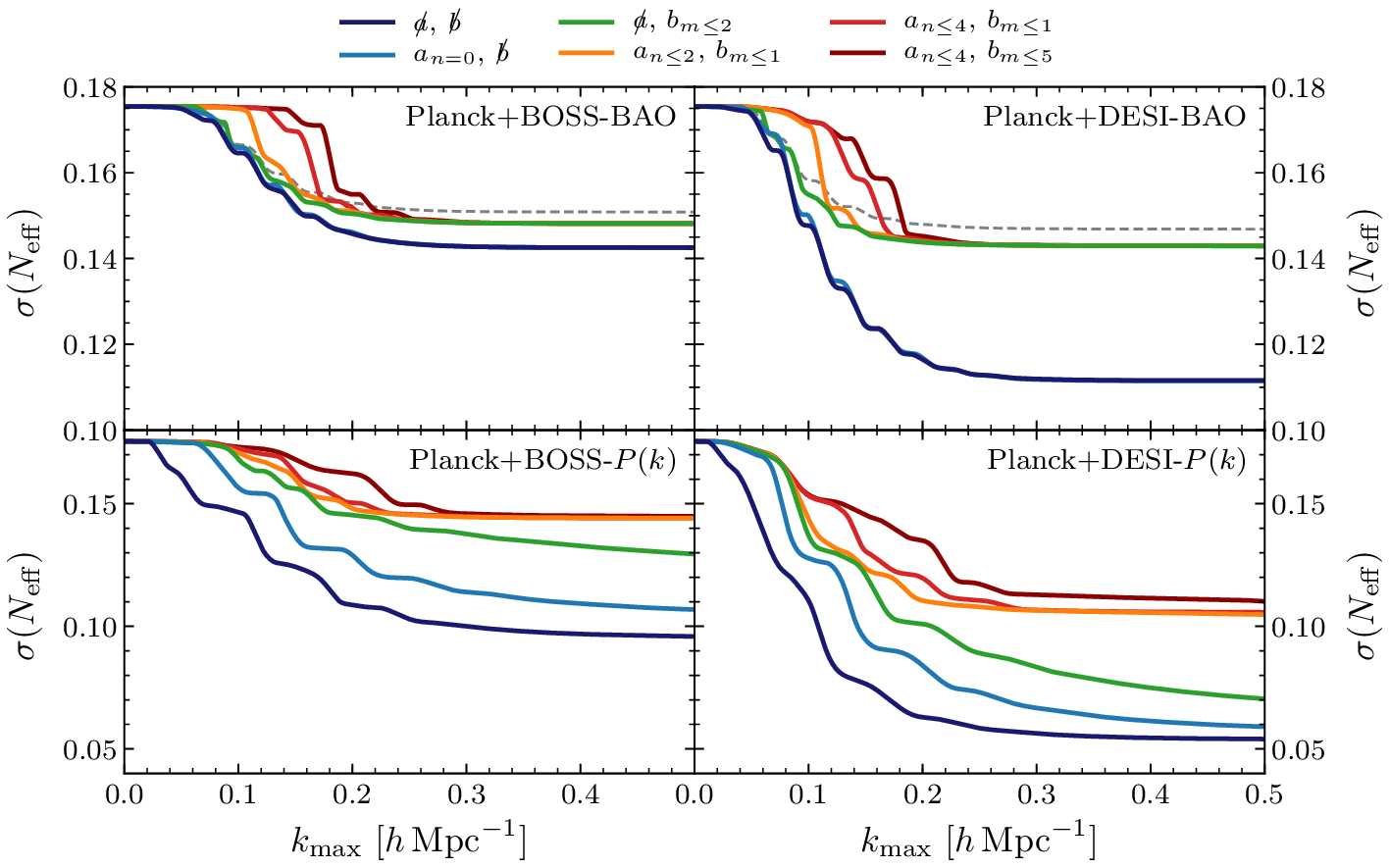}
		\caption{Forecasts for BOSS and DESI combined with Planck as a function of the largest Fourier modes used in the forecast, $\kmax$, using various levels of both additive and multiplicative marginalization, cf.~the $a_i$ and $b_i$-terms in~\eqref{eq:AB}. We have varied the number of parameters in the marginalization from none~($\slashed{a}$) to five~($a_{n\leq4}$) and none~($\slashed{b}$) to six~($b_{m\leq5}$), respectively. The dashed line shows the constraints from a standard isotropic BAO analysis for comparison.}
		\label{fig:marginalizationPlanned}
	\end{center}
\end{figure}

\paragraph{Sensitivity to marginalization}
High-redshift galaxy surveys benefit significantly from measuring highly biased objects. These large biases can offset the growth function, $b(z) D_1(z) = const$, and keep the amplitude of the galaxy power spectrum effectively fixed at high redshift. This boost is important for maintaining a signal above the shot noise, which we have assumed is redshift-independent. As a consequence, high-redshift and low-redshift galaxy power spectra are equally sensitive to uncertainties in the biasing coefficients. This is particularly significant when determining the largest wavenumbers that carry useful cosmological information. While taking $\kmax > \SI{0.2}{\hPerMpc}$ is appealing to maximize the constraints on $\Neff$, we must also marginalize over successively more bias parameters. Figure~\ref{fig:marginalizationPlanned} shows how the results depend on the marginalization scheme. While both the $P(k)$- and BAO-constraints degrade significantly when going from no marginalization to a few bias parameters, the BAO-forecasts quickly become robust to the marginalization. In contrast, the $P(k)$-forecasts weaken notably with additional biasing, but always lie below the BAO-only results, as one would expect. This confirms the intuition that the information that is primarily driving the constraints derived from $P(k)$ is present in the no-wiggle power spectrum, $\Pnw(k)$, instead of the BAO spectrum.

\vskip4pt
It is instructive to compare the results of our BAO-forecasts with those of a standard BAO analysis. Specifically, it is conventional to use the BAO signal to constrain only $q_i$, $i=\perp,\parallel$, defined in~\eqref{eq:rescaledWavenumbers} and derive parameter constraints from them.\footnote{We also compared the constraints coming from the full anisotropic treatment (cf.~\S\ref{sec:baoModeling}) with the isotropic approximation. The BAO-forecasts only weaken at small wavenumbers depending on the marginalization procedure, but reach the same plateau values at large wavenumbers as our baseline analysis. In contrast, the constraints on $\Neff$ are systematically weaker in the isotropic $P(k)$-forecasts at the level of \SI{15}{\percent} for $\kmax = \SI{0.2}{\hPerMpc}$.} These derived limits on $\Neff$ are shown as the dashed lines in Fig.~\ref{fig:marginalizationPlanned}. The fact that the standard BAO constraints are slightly weaker than those of our full BAO-forecasts, even after marginalization, suggests there is information in the BAO spectrum beyond the BAO scale. We will explore this further in Section~\ref{sec:phaseShift}.

\paragraph{Degeneracy with $\boldsymbol{Y_{p}}$}
To explore possible degeneracies between the effective number of relativistic species $\Neff$ and the primordial helium fraction $Y_p$, we now consider a $\Lambda\mathrm{CDM}$+$\Neff$+$Y_p$ cosmology. In Tables~\ref{tab:CMB+Pk_Neff+Yp} and~\ref{tab:CMB+BAO_Neff+Yp}, we present the $1\sigma$ constraints on $\Neff$ and $Y_p$ for various combinations of current and future CMB and LSS experiments using $P(k)$-forecasts and BAO-forecasts, respectively. As expected, the CMB-only constraint on $\Neff$ become worse due to the well-known degeneracy between $\Neff$ and $Y_p$ in the CMB damping tail. When broadband information is included, we find significant improvements in the constraints on both $\Neff$ and $Y_p$. However, this improvement cannot be attributed to the phase shift as we see only modest improvements in our BAO-forecasts. The broadband shape of the matter distribution is sensitive to the expansion history and to free-streaming neutrinos, but is not significantly affected by $Y_p$. As a result, the broadband information in $P(k)$ can break CMB degeneracies even without the phase shift information.
\begin{table}[t]
	\centering
	\begin{tabular}{c c S[table-format=2.4] S[table-format=2.4] S[table-format=2.4] S[table-format=2.4] S[table-format=2.4] S[table-format=2.4] S[table-format=2.4]}	
			\toprule
								&			& 			& \multicolumn{4}{c}{spectroscopic}				& \multicolumn{2}{c}{photometric}	\\
			\cmidrule(lr){4-7} \cmidrule(lr){8-9}
								& Parameter	& {CMB}		& {BOSS}	& {eBOSS}	& {DESI}	& {Euclid}	& {DES}		& {LSST}				\\
			\midrule[0.065em] 
		\multirow{2}{*}{Planck} & $\Neff$	& 0.32		& 0.25		& 0.22		& 0.14		& 0.13		& 0.29		& 0.23					\\
								& $Y_p$		& 0.018		& 0.016		& 0.016		& 0.013		& 0.012		& 0.017		& 0.015					\\
			\midrule[0.065em]
		\multirow{2}{*}{CMB-S3} 	& $\Neff$	& 0.12		& 0.12		& 0.11		& 0.094		& 0.088		& 0.12		& 0.11					\\
								& $Y_p$		& 0.0069	& 0.0068	& 0.0067	& 0.0060	& 0.0058	& 0.0069	& 0.0066				\\
			\midrule[0.065em]
		\multirow{2}{*}{CMB-S4}	& $\Neff$	& 0.081		& 0.079		& 0.078		& 0.070		& 0.067		& 0.081		& 0.078					\\
								& $Y_p$		& 0.0047	& 0.0046	& 0.0046	& 0.0043	& 0.0042	& 0.0047	& 0.0046				\\
			\bottomrule
	\end{tabular}
	\caption{Forecasted $1\sigma$ constraints on $\Neff$ and $Y_p$ for various combinations of current and future CMB and LSS experiments using $P(k)$-forecasts with $\kmax = \SI{0.2}{\hPerMpc}$.}
	\label{tab:CMB+Pk_Neff+Yp}
\end{table}
\medskip
\begin{table}
	\centering
	\begin{tabular}{c c S[table-format=2.4] S[table-format=2.4] S[table-format=2.4] S[table-format=2.4] S[table-format=2.4] S[table-format=2.4] S[table-format=2.4]}	
			\toprule
								& 			& 			& \multicolumn{4}{c}{spectroscopic}				& \multicolumn{2}{c}{photometric}	\\
			\cmidrule(lr){4-7} \cmidrule(lr){8-9}
								& Parameter	& {CMB}		& {BOSS}	& {eBOSS}	& {DESI}	& {Euclid}	& {DES}		& {LSST}				\\
			\midrule[0.065em] 
		\multirow{2}{*}{Planck} & $\Neff$	& 0.32		& 0.29		& 0.29		& 0.28		& 0.28		& 0.30		& 0.29					\\
								& $Y_p$		& 0.018		& 0.018		& 0.018		& 0.018		& 0.018		& 0.018		& 0.018					\\
			\midrule[0.065em]
		\multirow{2}{*}{CMB-S3} 	& $\Neff$	& 0.12		& 0.12		& 0.12		& 0.12		& 0.12		& 0.12		& 0.12					\\
								& $Y_p$		& 0.0069	& 0.0069	& 0.0069	& 0.0069	& 0.0069	& 0.0069	& 0.0069				\\
			\midrule[0.065em]
		\multirow{2}{*}{CMB-S4}	& $\Neff$	& 0.081		& 0.080		& 0.080		& 0.079		& 0.079		& 0.081		& 0.080					\\
								& $Y_p$		& 0.0047	& 0.0047	& 0.0047	& 0.0046	& 0.0046	& 0.0047	& 0.0047				\\
		\bottomrule
	\end{tabular}
	\caption{Forecasted $1\sigma$ constraints on $\Neff$ and $Y_p$ for various combinations of current and future CMB and LSS experiments using BAO-forecasts with $\kmax = \SI{0.5}{\hPerMpc}$.}
	\label{tab:CMB+BAO_Neff+Yp}
\end{table}

\paragraph{Comments on reconstruction}
In our baseline forecasts, we took $R \equiv 1$ in~\eqref{eq:linearRSD}, which is equivalent to taking $\Sigma_s \to \infty$. A few comments are in order regarding the effect of a finite $\Sigma_s$. As discussed in~\cite{Seo:2015eyw}, the optimal smoothing scale $\Sigma_s$ used in the BAO reconstruction depends on the noise levels of the experiment. Having said that, we have found only small changes in our results when going from $\Sigma_s=\infty$ to finite $\Sigma_s$. The constraints quoted in Tables~\ref{tab:CMB+Pk_Neff} to~\ref{tab:CMB+BAO_Neff+Yp} are basically unaffected, except for DESI and Euclid in the $P(k)$-forecasts, where the impact is also mild. Changing $\Sigma_s$ from \SI{30}{\MpcPerh} to \SI{15}{\MpcPerh} and \SI{10}{\MpcPerh}, the constraint on~$\Neff$ slightly weakens from \num{0.090} to \num{0.093} and \num{0.096} for Planck+DESI (\num{0.082}, \num{0.086} and \num{0.090} for Planck+Euclid) in $\Lambda\mathrm{CDM}$+$\Neff$ compared to the quoted \num{0.087} (\num{0.079}) in Table~\ref{tab:CMB+Pk_Neff}. In practice, this roughly \SI{10}{\percent} effect has to be compared to the impact on the reconstruction efficiency.

\subsubsection{Designer's Guide for Future Surveys}
\label{sec:constraints_future}

One of the main benefits of a Fisher forecast is that it can inform the design of future experiments. For spectroscopic surveys, the basic parameters are the total number of objects,~$N_g$, the maximal redshift, $\zmax$, and the sky area in square degrees, $\Omega$. From these, we derive the survey volume,~$V$, and the comoving number density, $\bar{n}_g$.\footnote{For simplicity, we will assume that the comoving number density can be approximated by a constant over the complete survey volume. However, very similar results are obtained for BOSS and DESI when using the specific redshift-dependent number densities.} In this section, we will explore how the constraints on~$\Neff$ depend on these parameters.

\vskip4pt
Most of the survey characteristics are encoded in the effective survey volume,\footnote{The effective survey volume also depends on the linear bias parameter $b$ through $\bar{n}_g \Pobs \propto \bar{n}_g b^2$. This dependence is degenerate with a rescaling of $\bar{n}_g$, so we will take $b(z=0) \equiv 1$ and vary $\bar{n}_g$. This ignores the impact that changes in $b$ may have on redshift space distortions.} $V_\mathrm{eff}$, cf.~\eqref{eq:V_eff} and~\eqref{eq:baoFisherMatrix}. The dependence of $V_\mathrm{eff}$ on the survey parameters is somewhat non-trivial. Increasing~$V$ (by increasing $\zmax$ and/or $\Omega$), at fixed $N_g$, will also reduce $\bar{n}_g$. For signal-dominated modes, $\bar{n}_g\Pobs \gg 1$, this effect is not important and the effective volume scales approximately as $V_\mathrm{eff} \propto V$. However, for $\bar{n}_g\Pobs \ll 1$, the shot noise is important and the reduction in the comoving density is more important than the increase in the volume, so that the effective volume scales as $V_\mathrm{eff} \propto V^{-1}$. This means that we will only benefit from an increase in the volume as long as the modes of interest, $k \in \SIrange[range-phrase={,}, range-units=brackets, open-bracket=[, close-bracket=]]{0.1}{0.3}{\hPerMpc}$, are signal dominated.

\vskip4pt
As mentioned before, the increased linearity of the matter distribution at high redshifts is undermined by the larger biasing. As a result, the main benefit of large $\zmax$ is the increased survey volume and hence the total number of modes. Unfortunately, the survey volume only grows slowly with redshift for $z>2$ and the resulting improvements in parameters is relatively modest for large increases in $\zmax$. The situation is slightly different for the BAO spectrum as the nonlinear damping factor $D(k,\mu)$ depends on the clustering of the matter directly and is therefore less important at high redshifts. However, the BAO signal alone has a relatively modest effect on $\Neff$ forecasts in general and the change to the damping factor consequently does not make a visible difference in our forecasts. 

\begin{figure}[t!]
	\begin{center}
		\includegraphics{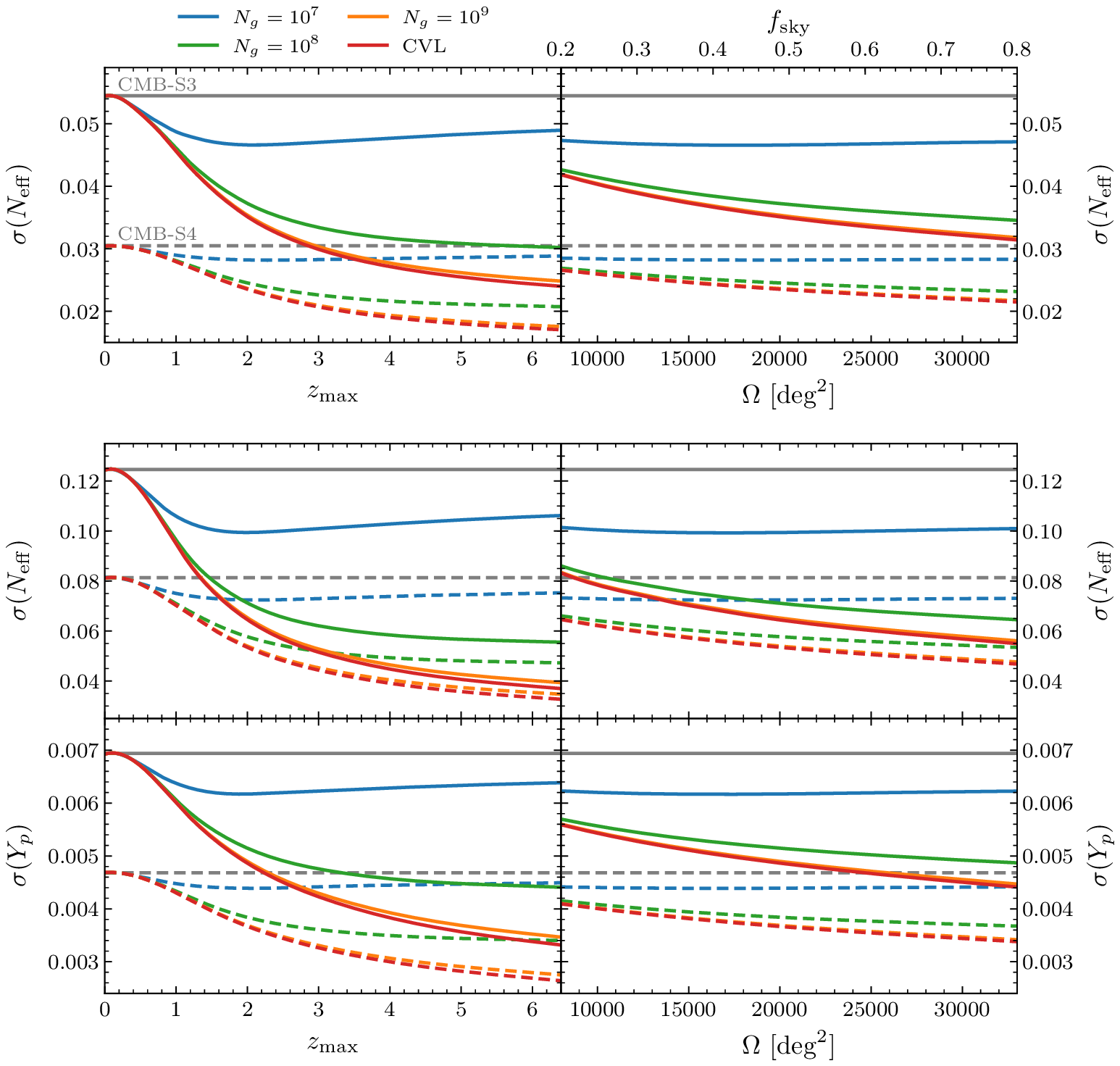}\vspace{-2pt}
		\caption{Future constraints for $\Lambda\mathrm{CDM}$+$\Neff$~(\textit{top}) and $\Lambda\mathrm{CDM}$+$\Neff$+$Y_p$~(\textit{bottom}) from the full galaxy power spectrum, $\Pobs(k)$, up to $\kmax=\SI{0.2}{\hPerMpc}$ as a function of the total number of objects, $N_g$, at fixed survey area $\Omega = \SI{20000}{deg^2}$~(\textit{left}) and as a function of the survey area $\Omega$ (or sky fraction $\fsky$) for fixed $\zmax=2$~(\textit{right}). The comoving number density is assumed to be constant and given by the total volume of the survey. For ``CVL'' (\textcolor{pyRed}{red}), all modes in the survey are assumed to be measured up to the limit set by cosmic variance. Solid and dashed lines correspond to combining the LSS data with CMB-S3 and CMB-S4 data, respectively. The \textcolor{pyGray}{gray} lines indicate the level of sensitivity of the respective CMB experiments alone.\vspace{-8pt}}
		\label{fig:futureConstraints}
	\end{center}
\end{figure}

\vskip4pt
In the top panel of Fig.~\ref{fig:futureConstraints}, we present $P(k)$-forecasts for $\Neff$ for a variety of survey configurations, assuming $Y_p$ is fixed by BBN consistency. We see that the largest improvement comes from increasing $N_g$ from \num{e7} to \num{e8}. As we increase the number of objects further, we reach the cosmic variance limit for all modes of interest. We see that an optimistic future survey combined with a near-term CMB experiment can provide constraints that are comparable to (or slightly stronger than) those projected for CMB-S4 alone. Having said that, it does not appear that one can push the measurement of $\Neff$ well beyond the CMB-S4 target. Moreover, as in the case of the planned experiments, the improvements from the BAO signal alone are rather small.

\vskip4pt
The value of LSS becomes more significant as we expand the space of parameters. The bottom panel of Fig.~\ref{fig:futureConstraints} shows $P(k)$-forecasts for $\Lambda\mathrm{CDM}$+$\Neff$+$Y_p$. We again see that the most significant jump in sensitivity arises when $N_g$ increases from \num{e7} to \num{e8}. We note that a factor of two improvement in $\sigma(\Neff)$ over CMB-S4 seems possible. We also see that the $P(k)$-forecasts for $\Neff$ marginalized over $Y_p$ are competitive with CMB-only forecasts with~$Y_p$ held fixed. In this sense, $P(k)$ adds robustness to the measurement of $\Neff$ under broader extensions of $\Lambda\mathrm{CDM}$. The improvement in $Y_p$ is slightly weaker, but shows the same general trend.

\begin{figure}[t!]
	\begin{center}
		\includegraphics{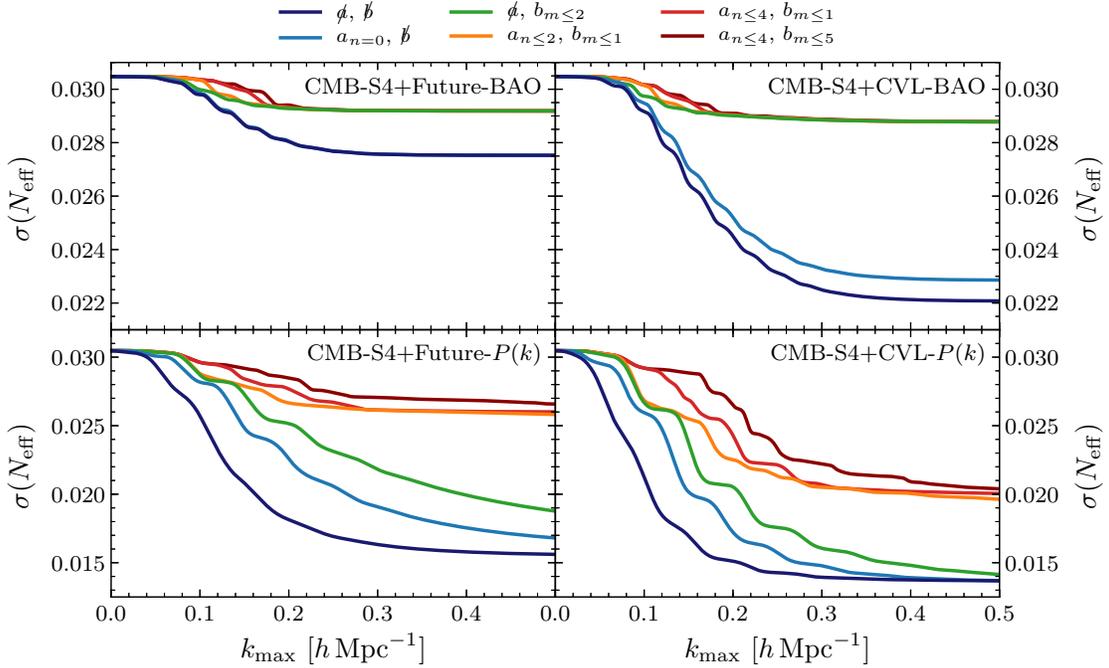}
		\caption{Forecasts for two future surveys combined with CMB-S4 as a function of the largest Fourier modes used in the forecast, $\kmax$, using various levels of both additive and multiplicative marginalization. We have varied the number of parameters in the marginalization from none~($\slashed{a}$) to five~($a_{n \leq 4}$) and none~($\slashed{b}$) to six~($b_{n \leq 5}$), respectively. The employed experimental specifications for the ``Future''-survey are $N_g = \num{e8}$, $\zmax = 3$ and $\fsky = 0.5$, whereas the ``CVL''-survey is cosmic variance limited for all $k$ up to $\kmax$ over $\fsky = 0.5$ and $\zmax=6$.}
		\label{fig:marginalizationFuture}
	\end{center}
\end{figure}

\vskip4pt
The range of accessible modes in near-term galaxy surveys is limited by their reliance on highly biased objects, but more futuristic surveys may not have the same limitations. Future surveys can also have high signal-to-noise beyond $k = \SI{0.2}{\hPerMpc}$, making it worth to consider the impact of increasing $\kmax$. In Figure~\ref{fig:marginalizationFuture}, we show the potential reach of two representative surveys. The first, denoted ``Future'', is characterized by $N_g = \num{e8}$, $\fsky = 0.5$ and $\zmax=3$, which is roughly the same as a spectroscopic follow-up to LSST. The second, denoted ``CVL'', is cosmic variance limited for all $k\leq \kmax$ over $\fsky = 0.5$ and $\zmax=6$. In principle, a \SI{21}{cm}~intensity mapping survey could achieve similar performance~\cite{Obuljen:2017jiy}. We see that $\sigma(\Neff) \sim 0.015$ is achievable through the measurement of $P(k)$ in either survey for $\kmax = \SI{0.5}{\hPerMpc}$, although the improvement with CVL is more robust to marginalization.

\section{Measurements of the Phase Shift}
\label{sec:phaseShift}

In the previous section, we showed how much the combination of future CMB and LSS measurements can improve the sensitivity to extra relativistic species. The dominant source of improvement came from the broadband shape of the power spectra, $\Pnw(k)$, rather than the BAO~spectrum,~$\Pw(k)$. Nevertheless, the shift of the acoustic peaks is a particularly robust signature of free-streaming, relativistic species~\cite{Baumann:2017lmt} and is therefore an interesting observable in its own right. In this section, we will isolate the signal coming from the phase shift and forecast our ability to measure it in future surveys. Measuring the BAO phase shift provides an independent test of pre-recombination physics in a low-redshift observable. This could be used to shed light on possible discrepancies between low- and high-redshift measurements or as a discovery channel for exotic new physics.

\subsection{Isolating the Phase Shift}
\label{sec:template}

The BAO feature in Fourier space can be written as 
\beq
O_\mathrm{lin}(k) = A(k) \sin\! \left[ \alpha^{-1} r_s k + \phi(k) \right] ,	\label{eq:phase_model}
\eeq 
where the parameter $\alpha$ represents changes in the BAO scale $r_s$, and the amplitude modulation~$A(k)$ and the phase shift $\phi(k)$ encode a number of physical effects that alter the time evolution of the baryons. While $\alpha$ and $A(k)$ are implicit functions of redshift, $\phi(k)$ is redshift independent. Relativistic species are the unique source of a constant shift in the locations of the BAO peaks in the limit of large wavenumbers, i.e.~$\phi(k \to \infty) = \phi_\infty$~\cite{Bashinsky:2003tk, Baumann:2015rya}. In practice, however, the measurement of the BAO spectrum occurs over a relatively small range of scales with a small number of (damped) acoustic oscillations. On these scales, the $k$-dependence of the shift can be relevant. Furthermore, additional $k$-dependent shifts from other cosmological parameters may also have to be taken into account~\cite{Pan:2016zla}.

\vskip4pt
To measure the phase shift $\phi(k)$, we will construct a template for the $k$-dependence as a function of the relevant parameters. For small variations around their fiducial values, it is a good approximation to treat the shifts arising from each cosmological parameter independently. By varying one parameter at a time and measuring the change in the peak locations, we can construct a template $\phi(k) = \sum_i \beta_i(\vec{\theta}\hskip1pt) f_i(k)$. For $\Lambda\mathrm{CDM}$+$\Neff$, the parameters $\As$, $\ns$, and $\tau$ do not affect the evolution of the baryons prior to recombination and, therefore, do not change the phase of the oscillations. The parameters $\omega_b$ and $\theta_s$ do alter the BAO spectrum, but are effectively negligible for any realistic parameter range. The shifts induced by $\omega_c$ and $\Neff$, on the other hand, can be significant.

\vskip4pt
The parameter that is most independent of $\Neff$ is not the dark matter density $\omega_c$, but the scale factor at the time of matter-radiation equality, $\aeq$. Since CMB data essentially fixes $\aeq$, our template model can be reduced to
\beq
\phi(k) = \beta(\Neff) f(k) \, ,	\label{eq:PhaseTemplate}
\eeq
namely the shift induced by changing $\Neff$ at fixed $\aeq$. This is the same choice made by Follin et al.~\cite{Follin:2015hya} in their CMB measurement of the phase shift. Fixing $\aeq$ also reproduces the expected constant phase shift at large wavenumbers. The template for the phase shift at fixed $\omega_c$, in contrast, does not approach a constant at large wavenumbers, which implies that the change of~$\aeq$ to maintain constant $\omega_c$ is introducing a phase shift of comparable size to the constant shift induced by varying $\Neff$. For our applications, this additional shift plays no role, but it could be useful in future investigations. 

\begin{figure}[t!]
	\begin{center}
		\includegraphics{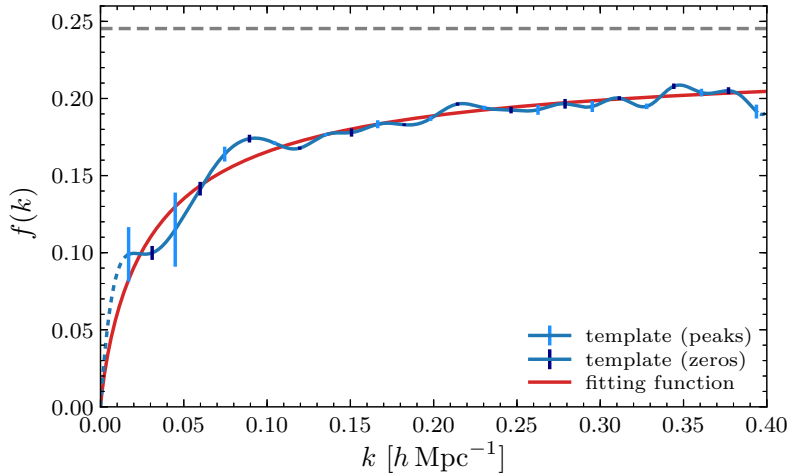}
		\caption{Template of the phase shift $f(k)$ as defined in~\eqref{eq:PhaseTemplate}. The numerical phase shifts~(\textcolor{pyBlue}{blue}) were obtained by sampling from 100 different cosmologies with varying $\Neff$ and rescaling by~$\beta(\Neff)$ as defined in~\eqref{eq:phi_norm}. The bars indicate the standard deviation in these measurements at the positions of the peaks~(\textcolor{pyLightBlue}{light blue}) and zeros~(\textcolor{pyDarkBlue}{dark blue}) compared to the fiducial BAO~spectrum. The \textcolor{pyRed}{red} line shows the fitting function defined in~\eqref{eq:phase_fit}. The dashed \textcolor{pyGray}{gray} line is an analytic approximation to the constant phase shift~\cite{Bashinsky:2003tk, Baumann:2015rya}.}
		\label{fig:phaseShiftTemplate}
	\end{center}
\end{figure}

\vskip4pt
We describe the measurement of the phase shift and the construction of the template in Appendix~\ref{app:broadband+phaseShiftExtraction}. In short, we determine the shift in the locations of the peaks/troughs and zeros of the BAO spectrum compared to the fiducial cosmology with $\Neff=3.046$ and sample 100~different cosmologies with varying $\Neff$ at fixed $\aeq$. It is convenient to normalize the template~$f(k)$ such that $\beta=0\text{ and }1$ for $\Neff = 0\text{ and }3.046$, respectively. In Figure~\ref{fig:phaseShiftTemplate}, we illustrate how the peaks/troughs and zeros of the BAO spectrum change in response to this variation in $\Neff$. We see that the phase shift created by $\Neff$ approaches a constant at large wavenumbers in line with physical expectations. 

\vskip4pt
The measurement of the phase shift is challenging because it requires a very accurate model of the no-wiggle spectrum $\Pnw(k)$ across a wide range of cosmological parameters. Errors in $\Pnw(k)$ effectively change the functions $A(k)$ and $B(k)$ in~\eqref{eq:O} and lead to errors in the measurement of the BAO peaks and zeros, respectively. The small size of the phase shift in Fig.~\ref{fig:phaseShiftTemplate} only exacerbates this problem. Fortunately, while the template is difficult to generate, our forecasts using the template are very stable. Furthermore, the template is well approximated by a simple fitting function,
\beq
f(k) = \frac{\phi_\infty}{1+(k_\star/k)^\xi} \, ,	\label{eq:phase_fit}
\eeq
where $\phi_\infty = 0.227$, $k_\star = \SI{0.0324}{\hPerMpc}$ and $\xi = 0.872$ were obtained by a weighted fitting procedure. From the analytic treatment at high wavenumbers $k$, we expect $\phi_\infty = 0.191\pi\,\epsilon_\mathrm{fid} + \mathcal{O}(\epsilon_\mathrm{fid}^2) \approx 0.245$ to linear order~\cite{Bashinsky:2003tk, Baumann:2015rya}, where $\epsilon(\Neff) = \Neff/(a_\nu +\Neff)$ is a measure of the excess radiation density, $(\rho_r-\rho_\gamma)/\rho_r$, with $a_\nu \approx 4.40$ as introduced in~\eqref{eq:Nnu}. This approximation overestimates the value obtained using the fitting formula by about \SI{8}{\percent}, which is consistent with the expected corrections from higher orders in $\epsilon_\mathrm{fid}\approx0.41$. Around $k\sim\SI{0.1}{\hPerMpc}$, where BOSS and DESI have the largest signal-to-noise ratio, the relative difference is almost~\SI{50}{\percent}, which makes it evident that the offset from the analytic approximation has to be taken into account in an analysis such as the one proposed below, whereas the precise shape of the template plays a sub-dominant role. We also note that this template is basically independent of changes to the BAO scale $r_s$, for example due to changes in the dark matter density.

\vskip4pt
We use the measured phase template to write the BAO~spectrum in terms of the spectrum in the fiducial cosmology:
\beq
O(k) = O_\mathrm{fid} \big(\alpha^{-1} k + (\beta-1)\, f(k)/r_s^\mathrm{fid} \big) \, ,	\label{eq:Oscale}
\eeq
where $\alpha\equiv\alpha(z_i)$ takes an independent value in each redshift bin centered around $z_i$ and $\beta$ is a single parameter for the entire survey. A measurement of $\alpha(z_i)$ and $\beta$ can then be translated into constraints on cosmological parameters using
\begin{align}
\alpha(\vec{\theta}; z) &\equiv q\, r_s^\mathrm{fid}/r_s = \left[D_V(z)/r_s\right] / \left[D_V(z)/r_s\right]_\mathrm{fid} \, ,\\
\beta(\Neff) 			&\equiv {\epsilon/\epsilon_\mathrm{fid}} \, ,	\label{eq:phi_norm}
\end{align}
where the parameters $q$ and $D_V$ were introduced in~\S\ref{sec:baoModeling}. With this normalization, the largest possible phase shift due to $\Neff$ is given by $\beta(\Neff\to\infty) = 2.45$.

\vskip4pt
In~\S\ref{sec:phase_param}, we will show that the forecasts produced using only these templates are in agreement with the forecasts using the full BAO spectrum. From a measurement of $\beta>0$, one gets a constraint on $\Neff$ that is only associated to the size of the phase shift. This approach is analogous to the template-based measurement of the phase shift in the CMB by Follin et al.~\cite{Follin:2015hya}. The measurement of $\Neff$ from the phase alone ignores the effects of $\Neff$ on $\alpha$, but has the advantage that any detection is unambiguously\hskip1pt\footnote{We have explicitly checked that our template gives an unbiased measurement of $\beta$. In particular, we have verified that we reproduce $\beta\approx0$ for a cosmology with $\Neff=0$.} a measurement of free-streaming relativistic particles.

\vskip4pt
We will also be interested in the measurement of $\beta$ when a prior on $\alpha$ is included, e.g.\ from the~CMB.\footnote{We also indirectly use the CMB data to constrain other cosmological parameters, in particular the scale factor at matter-radiation equality $\aeq$, so that we can ignore any additional phase shifts not associated with $\Neff$.} In a given cosmological model, the parameter $\alpha$ is fully determined by the set of cosmological parameters, $\alpha = \alpha(\vec\theta\hskip1pt)$. Since the $\alpha(z_i)$ inferred from the CMB are correlated between the $n$ redshift bins of a galaxy survey and $n$ is in general larger than the number of cosmological parameters, we compute the $n$-dimensional inverse covariance matrix according to $C_\alpha^{-1} = A^T F A$, where~$F$ is the Fisher matrix and $A$ is the pseudo-inverse of~$\nabla_{\!\vec\theta}\,\vec\alpha$. We use the CMB Fisher matrices for the $\Lambda\mathrm{CDM}$+$\Neff$ cosmology as in Section~\ref{sec:forecast}. We can then impose the $\alpha(z_i)$-prior on the redshift-binned likelihood function $\mathcal{L}(\alpha,\beta;z_i)$ according to $\mathcal{L}(\beta) \propto \int\prod_{z_i}\!\d\alpha_i\,\prod_{z_i}\!\mathcal{L}(\alpha_i,\beta;z_i)\,\pi(\alpha_1,\ldots,\alpha_n)$, where $\alpha_i \equiv \alpha(z_i)$ and $\pi$ is the $n$-dimensional Gaussian prior with covariance matrix $C_\alpha$. The observed posterior distribution of $\alpha(z_i)$ could also be constructed by evaluating $\alpha(z_i)$ for each point in a given CMB Markov chain.

\subsection{Constraints from Planned and Future Surveys}
\label{sec:phase_planned}

We will now show how well the phase shift can be measured in planned galaxy surveys. It is useful to first understand the parameter space $\alpha$-$\beta$ without imposing a prior on $\alpha$. Both parameters affect the locations of the acoustic peaks and are therefore quite degenerate. We will use likelihood-based forecasts to ensure accuracy. We will confirm that the posterior \mbox{distributions\hskip1pt\footnotemark}\footnotetext{Since we assume flat priors for the parameters, we can identify the posteriors with the likelihoods.} of~$\alpha$ and~$\beta$ are Gaussian, while the constraints on $\Neff$ derived from this parameterization are significantly non-Gaussian. This suggests that a Fisher matrix forecast in terms of $\alpha$ and $\beta$ would be more reliable than one that starts directly from $\Neff$.

\vskip4pt
We define the phase shift relative to the fiducial model with $\Neff = 3.046$. The broadband spectrum for the fiducial model can be isolated by using the method in Appendix~\ref{app:broadband+phaseShiftExtraction} or through the use of a fitting function along the lines of~\cite{Beutler:2016ixs}. These methods generate the BAO spectrum~$O_\mathrm{fid}(k)$ and hence~$O(k)$ via~\eqref{eq:Oscale}. We compute the log-likelihood using the same noise and modeling as in the Fisher matrix~\eqref{eq:baoFisherMatrix}.

\subsubsection*{Planned surveys}
\begin{figure}[b!]
	\begin{center}
		\includegraphics{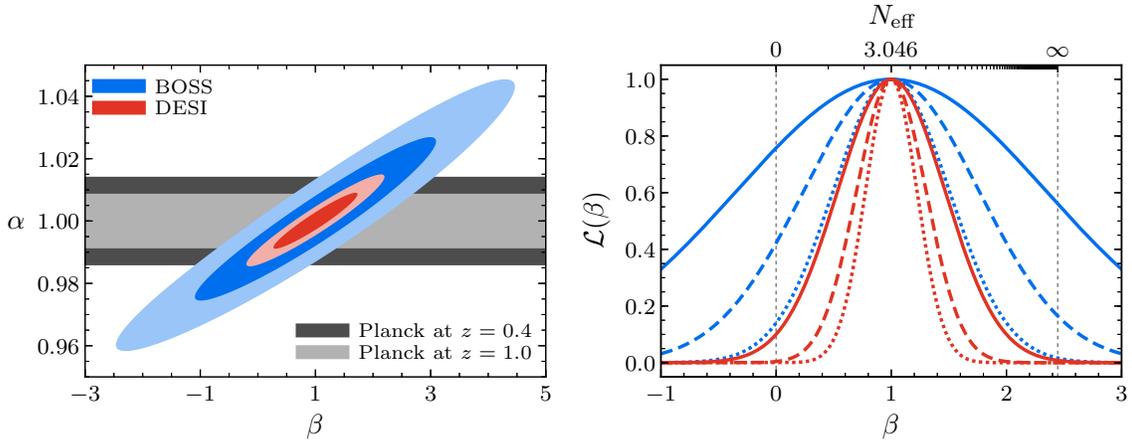}
		\caption{\textit{Left:} Contours showing $1\sigma$ and $2\sigma$ exclusions in the $\alpha$-$\beta$ plane for BOSS and DESI. For purpose of illustration, we have reduced these surveys to a single redshift bin (and therefore a single $\alpha$-parameter). The gray bands indicate Planck priors for $\alpha$ assuming the median redshift is $z=0.4\text{ and }1.0$ for BOSS and DESI, respectively. \textit{Right:} One-dimensional posterior distributions of $\beta$ for BOSS and DESI. The dashed and dotted lines indicate the use of a redshift-dependent CMB prior on $\alpha$ from Planck and CMB-S4, respectively.}
		\label{fig:likelihoodAlphaBeta}
	\end{center}
\end{figure}

Forecasts for the one- and two-dimensional posteriors are shown in Fig.~\ref{fig:likelihoodAlphaBeta} for both BOSS and DESI. %
We see that for both surveys the posterior distributions are Gaussian. The best-fit Gaussian for BOSS and DESI has $\sigma(\beta) = 1.3\text{ and }0.47$, respectively, which corresponds to a rejection of $\beta=0$ at \SI{77}{\percent} and \SI{98}{\percent} confidence. Clearly, BOSS cannot exclude $\beta=0$ (and hence $\Neff = 0$) without any prior information from the CMB. Since the weakness of the constraint on~$\beta$ is driven by the degeneracy with~$\alpha$ (see the left panel in Fig.~\ref{fig:likelihoodAlphaBeta}), we expect to get significant improvements in the constraints on $\beta$ after imposing a CMB prior on $\alpha$. Inspection of the two-dimensional contours already shows that we will sizeably limit the range of~$\beta$. The posterior distribution with the prior from Planck (CMB-S4) is shown in the right panel of Fig.~\ref{fig:likelihoodAlphaBeta}. For BOSS, we find $\sigma(\beta) = 0.76$~$(0.50)$ which implies that $\beta > 0$ at \SI{81}{\percent}~(\SI{95}{\percent}) confidence. Evidence for this signature of free-streaming neutrinos has been seen in existing data~\cite{Baumann:2018qnt}. For DESI, we should find strong evidence for a phase shift with $\sigma(\beta) = 0.30$ $(0.22)$ which excludes $\beta = 0$ at $3.5\hskip1pt\sigma$~($4.6\hskip1pt\sigma$).

\begin{figure}[t!]
	\begin{center}
		\includegraphics{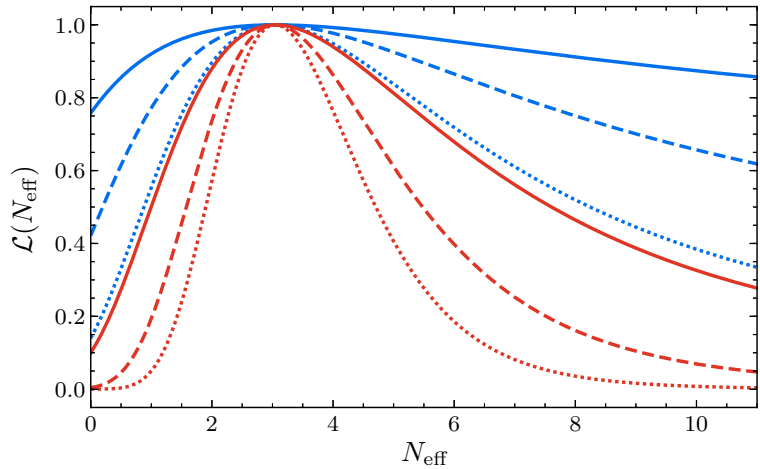}
		\caption{Posterior distributions of $\Neff$ for BOSS (\textcolor{pyBlue2}{blue}) and DESI (\textcolor{pyRed2}{red}) derived from the phase shift in the BAO spectrum, i.e.\ via the measurement of $\beta$. The dashed and dotted lines indicate that a redshift-dependent CMB prior on $\alpha$ has been imposed using Planck and CMB-S4, respectively.}
		\label{fig:likelihoodNeff}
	\end{center}
\end{figure}

\vskip4pt
To translate these results into constraints on $\Neff$, we use the relationship between~$\beta$ and~$\Neff$ given in~\eqref{eq:phi_norm}. This map is nonlinear over the measured range of $\beta$ and we therefore anticipate the posterior to be non-Gaussian. The derived $\Neff$-posteriors in Fig.~\ref{fig:likelihoodNeff} indeed show a highly non-Gaussian distribution. As anticipated from the $\beta$-posterior for BOSS, the constraints on~$\Neff$ are relatively weak without imposing a Planck prior on $\alpha$.

We also see that the constraining power is significantly weaker at bounding large values of~$\Neff$ than small ones. This asymmetry is simply a reflection of the fact that increasing~$\Neff$ does not produce proportionally larger phases shifts. This asymmetry was also seen in the CMB constraints of Follin et al.~\cite{Follin:2015hya}, likely for the same reason. Recall that we have an upper limit on the phase shift of $\beta < 2.45$, which is saturated for $\Neff \to \infty$. In practice, this means that for $\Neff \gg a_\nu \approx 4.40$, we will have an equal likelihood\hskip1pt\footnote{Realistic values of $\Neff$ are not quite in the asymptotic regime, but still show the weakened distinguishing power for larger $\Neff$.} for every value of $\Neff$ because they produce identical spectra. As a result, a flat prior on $\Neff$ (rather than $\beta$) will lead to ill-defined results because the integral $\int^{\infty}\!\d\Neff\,\mathcal{L}(\Neff)$ will diverge. On the other hand, for highly-significant detections of $\beta>0$, a flat prior over any reasonable range of $\Neff$ will produce stable results. We are not quite in this regime with BOSS, which is why we will only quote constraints on $\beta$.

\vskip4pt
Table~\ref{tab:betaConstraints} shows the projected constraints on $\beta$ for a variety of planned surveys with and without priors from the CMB. We see that roughly a factor of three improvement can be achieved in spectroscopic surveys going from BOSS to Euclid. Both DESI and Euclid should have sufficient sensitivity to reach a more than $5\sigma$ exclusion of $\beta=0$ when imposing a Planck prior. As before, galaxy clustering measurements in photometric surveys do not lead to competitive constraints as they are effectively two-dimensional on the relevant scales.
\begin{table}
	\centering
	\begin{tabular}{l S[table-format=2.3] S[table-format=2.3] S[table-format=2.3] S[table-format=2.3] S[table-format=2.3] S[table-format=2.3]}
			\toprule
							& \multicolumn{4}{c}{spectroscopic}				& \multicolumn{2}{c}{photometric}	\\
			\cmidrule(lr){2-5} \cmidrule(lr){6-7}
							& {BOSS}	& {eBOSS}	& {DESI}	& {Euclid}	& {DES}		& {LSST}				\\
			\midrule[0.065em]
		BAO					& 1.3		& 1.0		& 0.47		& 0.40		& 2.6		& 1.0					\\
			\midrule[0.065em]
		+ Planck prior		& 0.76		& 0.70		& 0.30		& 0.26		& 1.1		& 0.50					\\
			\midrule[0.065em]
		+ CMB-S4 prior		& 0.50		& 0.48		& 0.22		& 0.19		& 1.0		& 0.42					\\
			\bottomrule
	\end{tabular}
	\caption{Forecasted $1\sigma$ constraints on the amplitude of the phase shift~$\beta$  for current and future LSS experiments. We also show the constraints on $\beta$ after imposing a redshift-dependent prior on the BAO parameter $\alpha$ from Planck and CMB-S4.}
	\label{tab:betaConstraints}
\end{table}

\subsubsection*{Future surveys}
Given the robustness of the phase shift as a probe of light relics, a high-significance detection of the phase shift in LSS would be a valuable piece of cosmological information. We have seen that current and planned surveys can detect the phase shift, but are not expected to produce constraints on $\Neff$ that are competitive with those from the CMB. It is natural to ask if future surveys can reach this level of sensitivity.

\begin{figure}[b!]
	\begin{center}
		\includegraphics{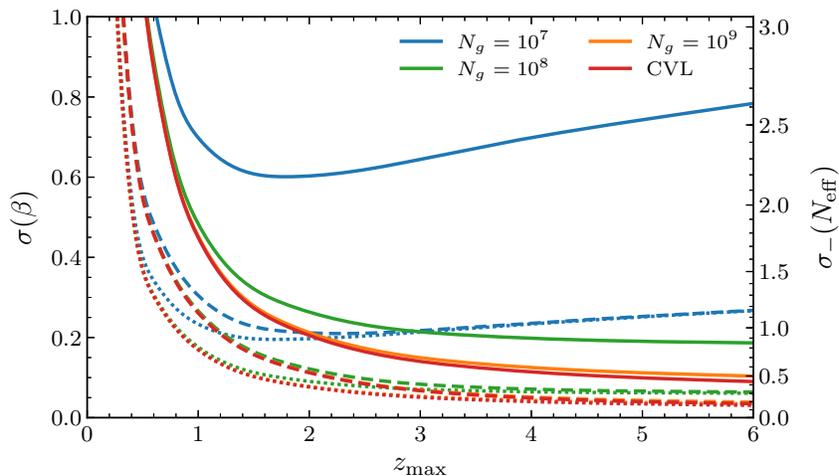}
		\caption{Future constraints on the amplitude of the phase shift $\beta$ as a function of~$\zmax$ and~$N_g$, assuming $\fsky=0.5$. The dashed and dotted lines indicate that a CMB prior on~$\alpha$ has been imposed using Planck and CMB-S4, respectively. The corresponding $1\sigma$ lower limit on~$\Neff$, which is $\Neff = 3.046 - \sigma_-(\Neff)$, is indicated by the right axis.}
		\label{fig:likelihoodBetaFuture}
	\end{center}
\end{figure}

\vskip4pt
Like the measurement of the BAO scale, the measurement of the phase requires large signal-to-noise for $\SI{0.1}{\hPerMpc} \lesssim k \lesssim \SI{0.3}{\hPerMpc}$. As long as the number density is sufficiently large to keep the shot noise below cosmic variance, we gain primarily by increasing $\zmax$ to achieve larger survey volumes. At larger levels of the shot noise, we only measure a few peak locations well which increases the degeneracy between $\alpha$ and $\beta$. Figure~\ref{fig:likelihoodBetaFuture}%
shows results for a variety of possible survey configurations. As before, the constraints on $\beta$ can be mapped into constraints on $\Neff$ using~\eqref{eq:phi_norm}. We see that with \num{e8} objects and $\zmax > 3$, we consistently obtain $\sigma(\Neff) < 0.5$~$(1.0)$ with (without) a prior on $\alpha$ included.

\vskip4pt
To put these results into context, the measurement of Follin et al.~\cite{Follin:2015hya} of $N_\mathrm{eff}^\phi = 2.3_{-0.4}^{+1.1}$ from the Planck TT spectrum is comparable to a survey with $N_g=\num{e9}$ objects out to redshift $\zmax=3$. Follin et al.~also forecasted $\sigma(N_\mathrm{eff}^\phi)=0.41$ for Planck TT+TE+EE which is near the sensitivity of future LSS surveys when increasing the redshift range to $\zmax=6$. Reaching this level of sensitivity will be extremely challenging with an optical survey, but could potentially be achieved with 21cm intensity mapping~\cite{Obuljen:2017jiy}.

\subsection{Comparison to Parameter-Based Approach}
\label{sec:phase_param}
	
It is instructive to compare the results of our template-based forecasts to a more direct parameter-based approach to isolating the phase shift. In the parameter-based approach, we define two new parameters $\thetasLSS$ and $\NeffLSS$ that play the role of $\theta_s$ and $\Neff$ in the BAO signal, but are taken to be independent of the same parameters in the CMB. We will then fix all remaining cosmological parameters in the BAO spectrum using the CMB, except $\omega_c$ which we traded for $\aeq$. As with our template extraction, holding $\aeq$ fixed ensures that the phase shift approaches a constant at large wavenumbers, whose value is determined by $\NeffLSS$. Beside measuring the phase of the BAO signal, the parameter $\NeffLSS$ also contributes to the scale parameter~$\alpha$ and could therefore be constrained by the standard BAO-scale measurement if all the other cosmological parameters are fixed to their Planck best-fit values. Introducing the additional parameter $\thetasLSS$ gives enough freedom to remove this effect and any constraint on $\NeffLSS$ must be coming from the phase shift alone. This is analogous to isolating the phase shift in the CMB by marginalizing over $Y_p$ or any other parameters that are degenerate with the $\Neff$-induced change to the damping tail~\cite{Baumann:2015rya}. We will confirm this expectation in our forecasts. 

\vskip4pt
Typically, the advantage of the parameter-based approach is that it is easy to implement. However, in this case, we found it more difficult to set up reliably. The phase shift ultimately controls the breaking of the degeneracy between $\thetasLSS$ and $\NeffLSS$ and, as we discussed in~\S\ref{sec:template}, $\Pnw(k)$ must therefore be determined sufficiently accurately to not produce errors in this shift. To compute the likelihood directly, we must re-compute $\Pnw(k)$ for every value of the cosmological parameters. Producing stable results for the BAO spectrum across a wide range of parameters can be very computationally expensive and technically challenging. Simpler and faster methods can work well near the fiducial cosmology (such as the use of a fitting function), but often produce noisy results as the parameters vary significantly and typically underestimate the likelihood as we depart from the fiducial cosmology (and, hence, overestimate the constraining power).

\vskip4pt
Despite the challenge presented by a parameter-based approach, it has the advantage that it should capture all of the cosmological information available. It is therefore useful to compare the results of the parameter-based and template-based approaches to see if the template is missing information. Fortunately, we will see that the posterior distributions of $\NeffLSS$ and $\thetasLSS$ can be largely reproduced as a derived consequence of the template-based forecasts. From the previous subsections, we should anticipate that the posteriors for $\NeffLSS$ and $\thetasLSS$ will be non-Gaussian, and will therefore require the calculation of the likelihood for $\NeffLSS$ and $\thetasLSS$ directly and not only the Fisher matrix. We will follow the same approach as described in~\S\ref{sec:phase_planned}. Computing the full likelihood is quite involved, which is the reason why we will assume that the CMB data fixes the other cosmological parameters to their fiducial values, except for $\NeffLSS$ and $\thetasLSS$.
\begin{figure}[t!]
	\begin{center}
		\includegraphics{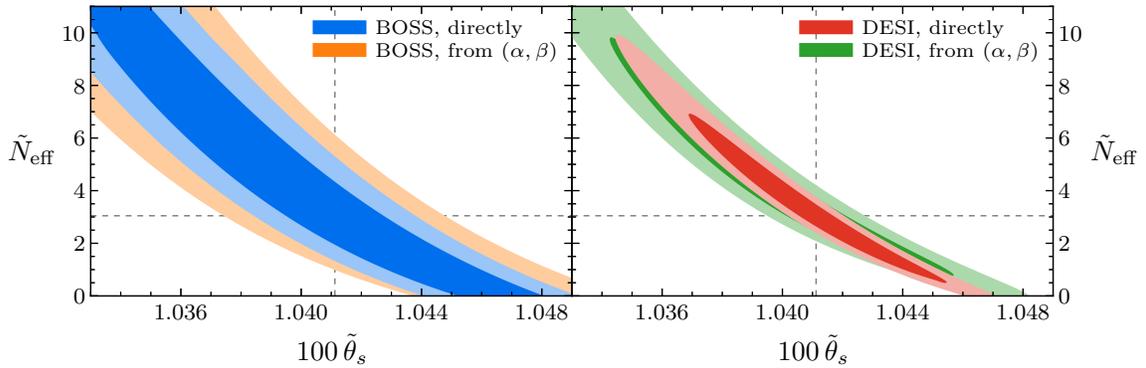}
		\caption{BOSS (\textit{left}) and DESI (\textit{right}) two-dimensional $1\sigma$ and $2\sigma$ contours for $\NeffLSS$ and~$\thetasLSS$, determined (`directly') from the likelihood for the BAO spectrum for each value of the parameters and derived (`from $(\alpha,\beta)$') from the redshift-binned likelihood for $\alpha$ and $\beta$. We find good agreement between both methods, suggesting that the two-dimensional parameterization is capturing most of the relevant information. The dashed lines indicate the fiducial values.}
		\label{fig:likelihoodNeffThetas}
	\end{center}
\end{figure}

\vskip4pt
Results of the likelihood analysis in terms of these parameters for both BOSS and DESI are shown in Fig.~\ref{fig:likelihoodNeffThetas}. We see that the results are similar, which establishes that our templates are capturing most of the information available in the BAO spectrum. This is an important observation because it allows us to simplify the analysis to a two-parameter template without much loss of information. In fact, the conclusion that these likelihoods are the same is not easily reproduced with any method of BAO extraction, but requires the robustness and stability of a method such as the one we use. Given instead our phase shift template, one can reliably compute Fisher matrices or likelihoods for $\alpha$ and $\beta$, and derive the implications for cosmological parameters from them. Future surveys such as DESI show somewhat larger differences between the two methods, which suggests that more information could potentially be extracted by using additional and/or alternative templates.

\vskip4pt
The doubling of cosmological parameters to treat the CMB and LSS independently, like in the case of $\NeffLSS$ and $\thetasLSS$, has useful conceptual advantages even if we derive constraints on these parameters from the posterior of $\alpha$ and $\beta$. Growing tensions between the CMB and certain \mbox{low-$z$} measurements have garnered much attention, but lack a compelling explanation. Measuring~$\thetasLSS$ and $\NeffLSS$ in the BAO spectrum may provide a new perspective on this issue without the need for a CMB anchor.

\section{Conclusions}
\label{sec:conclusions}

Large-scale structure surveys are an untapped resource in the search for light relics of the hot big bang. The growing statistical power of these surveys will make them competitive with the CMB in terms of the constraints they will provide on a broad range of cosmological parameters. Moreover, the combination of CMB and LSS observations will allow powerful and robust tests of the physical laws that determined the structure and evolution of the early universe.

\begin{figure}[t!]
	\begin{center}
		\includegraphics{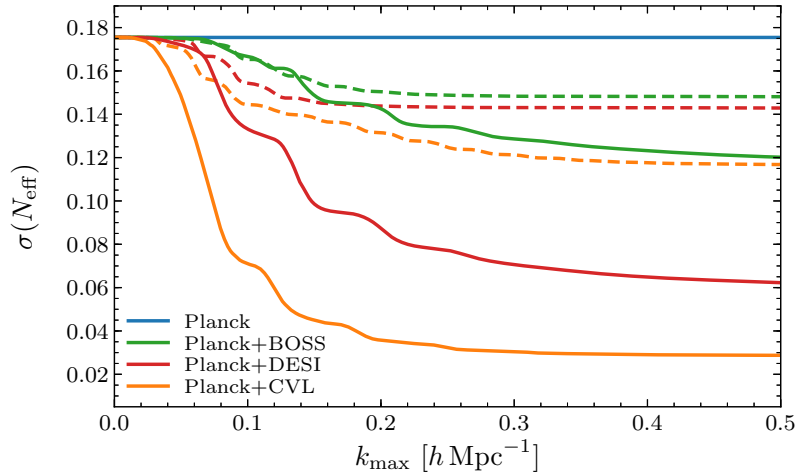}
		\caption{Sensitivity of planned and future LSS surveys to $\Neff$ using the galaxy power spectrum (solid) and the BAO spectrum (dashed) marginalized over two bias parameters, $b_{m\leq1}$. }
		\label{fig:neffPkBAO}
	\end{center}
\end{figure}

\vskip4pt
In this paper, we have explored the potential impact of LSS surveys on measurements of the parameter~$\Neff$. We have found that the dominant statistical impact of future surveys lies in the shape of the galaxy power spectrum. The distribution of dark matter in the universe is altered through the gravitational influence of the free-streaming radiation, leading to changes in the shape of the power spectrum that can be detected at high significance. A summary of the reach of selected planned and future surveys is given in Fig.~\ref{fig:neffPkBAO}. %
We see that BOSS and DESI can extend results significantly beyond the current CMB constraints. Futuristic surveys combined with a future CMB-S4 mission could achieve $\sigma(\Neff) \sim 0.015$, which is close to reaching the target of $\Delta\Neff = 0.027$ at a significance of $2\sigma$. 

\vskip4pt
Future LSS surveys will also be able to detect the coherent shift in the peak locations of the BAO spectrum. This would be an intriguing measurement as this phase shift is a highly robust and unambiguous probe of light relics and the cosmic neutrino background~\cite{Baumann:2017lmt}. Moreover, it is sensitive to extensions of $\Lambda\mathrm{CDM}$ without requiring the CMB as an anchor. Improved measurements of the phase shift may therefore play a useful role in elucidating apparent discrepancies between CMB and low-redshift measurements of the Hubble parameter~$H_0$~\cite{Freedman:2017yms}.

\vskip4pt
Future CMB and LSS observations could have a significant impact on our understanding of fundamental physics. In this paper, we have argued that, in the case of $\Neff$, these observations can play complimentary roles by both increasing the raw sensitivity and adding to the robustness of the measurement. We have also shown that the BAO spectrum carries more accessible cosmological information than only the acoustic scale. A broader exploration will likely reveal more targets that benefit from this complementarity.

\vskip20pt
\paragraph{Acknowledgements}
We thank Raphael Flauger and Joel Meyers for initial collaboration and helpful discussions. We are grateful to Matteo Biagetti, Diego Blas, Shirley Ho, Patrick \mbox{McDonald}, Uro\v{s} Seljak and Matias Zaldarriaga for helpful discussions. We thank Florian Beutler, Raphael Flauger, Mariana Vargas-Maga\~{n}a, An\v{z}e Slosar and Christophe Y\`{e}che for collaboration on related work, and Florian Beutler for numerous cross-checks and validations of our results. B.\,W.~is grateful to the Institute of Physics at the University of Amsterdam and the CERN theory group for their hospitality. D.\,B.~thanks the Aspen Center for Physics, which is supported by National Science Foundation Grant~PHYS-1066293, for hospitality when part of this work was being performed. D.\,B.~and B.\,W.~were supported by a Starting Grant of the European Research Council (ERC STG Grant~279617). D.\,B.~is supported by a Vidi grant of the Netherlands Organisation for Scientific Research~(NWO) that is funded by the Dutch Ministry of Education, Culture and Science~(OCW). B.\,W.~also acknowledges support by a Cambridge European Scholarship of the Cambridge Trust, an STFC Studentship and a Visiting PhD Fellowship of the Delta-ITP consortium, a program of NWO. Parts of this work were undertaken on the COSMOS Shared Memory System at DAMTP (University of Cambridge), operated on behalf of the STFC DiRAC HPC Facility. This equipment is funded by BIS National E-Infrastructure Capital Grant ST/J005673/1 and STFC Grants~ST/H008586/1, ST/K00333X/1. We acknowledge the use of \texttt{CLASS}~\cite{Blas:2011rf}, \texttt{IPython}~\cite{Perez:2007ipy}, and the Python packages \texttt{Matplotlib}~\cite{Hunter:2007mat} and \texttt{NumPy}/\texttt{SciPy}~\cite{Walt:2011num}.

\clearpage
\appendix
\section{Forecasting CMB Constraints} 
\label{app:CMB}

Forecasting the sensitivities of future CMB observations is by now a standard exercise; see e.g.~\cite{Wu:2014hta, Galli:2014kla, Allison:2015qca, Abazajian:2016yjj}. For completeness, this appendix collects the basic elements of our CMB Fisher analysis, as well as the specifications of the CMB experiments that were used in our analysis.	

\subsection{Fisher Matrix}

The Fisher matrix for CMB experiments can be written as
\beq
F_{ij} = \sum_{X,Y}\, \sum_{\ell=\lmin}^{\lmax} \frac{\partial C_\ell^{X}}{\partial \theta_i} \left[\mathbf{C}_\ell^{XY}\right]^{\!-1}\frac{\partial C_\ell^{Y}}{\partial \theta_j} \, .
\eeq
The covariance matrix $\mathbf{C}_\ell^{XY}$ for each multipole $\ell$ and $X=ab$, $Y=cd$, with $a,b,c,d=T,E,B$, is defined by
\beq
\mathbf{C}_\ell^{ab\hskip1pt cd} = \frac{1}{(2\ell+1) \fsky} \left[ (C_\ell^{ac} + N_\ell^{ac})(C_\ell^{bd} + N_\ell^{bd}) + (C_\ell^{ad} + N_\ell^{ad})(C_\ell^{bc} + N_\ell^{bc})\right] ,	\label{eq:cmbCovarianceMatrix}
\eeq
where $C_\ell^X$ are the theoretical CMB power spectra, $N_\ell^X$ are the (Gaussian) noise spectra of a given experiment and $\fsky$ is the effective sky fraction that is used in the cosmological analysis. We employ perfectly delensed power spectra and omit the lensing convergence for simplicity as it is sufficient for our purposes. We however comment on the effects of these assumptions below. The noise power spectra are
\beq
N_\ell^X = (\Delta X)^2 \exp\left\{ \frac{\ell (\ell+1)\, \theta_b^2}{8\ln2} \right\} ,	\label{eq:noiseSpectrum}
\eeq
where $\Delta X = \Delta T,\Delta P$ are the map sensitivities for temperature and polarization spectra, respectively, and $\theta_b$ is the beam width (taken to be the full width at half maximum). Note that we set $N^{TE}_\ell\equiv0$ as we assume the noise in temperature and polarization to be uncorrelated. For a multi-frequency experiment, the noise spectrum~\eqref{eq:noiseSpectrum} applies for each frequency channel separately. The effective noise after combining all channels is
\beq
N_\ell^X = \left[\sum_\nu \left(N_\ell^{X,\nu}\right)^{\!-1}\right]^{\!-1}\, ,
\eeq
where $N_\ell^{X,\nu}$ are the noise power spectra for the separate frequency channels $\nu$.

\subsection{Experimental Specifications}

Our specifications for the Planck satellite are collected in Table~\ref{tab:planckSpecs}. The adopted configuration is the same as that used in the CMB-S4 Science Book~\cite{Abazajian:2016yjj}. For the low-$\ell$ data, we use the unlensed TT~spectrum with $\lmin = 2$, $\lmax=29$ and $\fsky = 0.8$. We do not include low-$\ell$ polarization data, but instead impose a Gaussian prior on the optical depth, with $\sigma(\tau) = 0.01$. For the high-$\ell$ data, we use the unlensed TT, TE, EE spectra with $\lmin=30$, $\lmax = 2500$ and $\fsky = 0.44$. Since the low-$\ell$ and high-$\ell$ modes are independent, we simply add the corresponding Fisher matrices. 
\begin{table}
	\begin{center}
		\begin{tabular}{l S[table-format=4.1] S[table-format=4.1] S[table-format=4.1] S[table-format=4.1] S[table-format=4.1] S[table-format=4.1] S[table-format=4.1]}
				\toprule
			Frequency [\si{\giga\hertz}]		& 30	& 44	& 70	& 100	& 143	& 217	& 353	\\
				\midrule[0.065em]
			$\theta_b$ [\si{arcmin}]			& 33	& 23	& 14	& 10	& 7		& 5		& 5		\\
			$\Delta T$ [\si{\muKelvin.arcmin}]	& 145	& 149	& 137	& 65	& 43	& 66	& 200	\\
			$\Delta P$ [\si{\muKelvin.arcmin}]	& {--}	& {--}	& 450	& 103	& 81	& 134	& 406	\\
				\bottomrule 
		\end{tabular}
		\caption{Specifications for the Planck-like experiment used in~\cite{Allison:2015qca} and in the CMB-S4 Science Book~\cite{Abazajian:2016yjj}. The dashes in the first two columns for $\Delta P$ indicate that those frequency channels are not sensitive to polarization.}
		\label{tab:planckSpecs}
	\end{center}
\end{table}

\vskip4pt
We parameterize future CMB experiments in terms of a single effective frequency with noise level $\Delta T$, beam width $\theta_b$ and sky fraction $\fsky$. We will present constraints as a function of these three parameters. We take $\theta_b=\SI{3}{\arcmin}$, $\Delta T=\SI{5}{\muKelvin.arcmin}$ and $\fsky=0.3$ as the fiducial configuration of a CMB-S3-like experiment. For a representative CMB-S4 mission, we adopt the same configuration as in the CMB-S4 Science Book~\cite{Abazajian:2016yjj}: $\theta_b=\SI{2}{\arcmin}$, $\Delta T=\SI{1}{\muKelvin.arcmin}$ and $\fsky=0.4$. For both experiments, we use unlensed temperature and polarization spectra with $\lmin=30$, $\lmax^T = 3000$, $\lmax^P = 5000$. We add the low-$\ell$ Planck data as described above, include high-$\ell$ Planck data with $\fsky=0.3$ and $\fsky=0.2$ for CMB-S3 and CMB-S4, respectively, and impose the same Gaussian prior on the optical depth $\tau$ as for Planck.

\vskip4pt
Unlike the CMB-S4 Science Book, we do not include delensing of the T- and E-modes. For $\Neff$~forecasts, this was shown to have a negligible impact~\cite{Green:2016cjr}, while using unlensed spectra overestimates the constraining power of the CMB by roughly \SI{30}{\percent} for $\Neff$+$Y_p$. We are primarily interested in the improvement in parameters from adding LSS data, which should be robust to these relatively small differences. We also ignore the lensing convergence as it does not impact the constraints on these parameters.

\subsection{Future Constraints}
\label{app:futureCMB}

As a point of reference, we present constraints derived from CMB observations alone. In Table~\ref{tab:cmbForecast_LCDM_LCDM+Neff},%
\begin{table}[b]
	\centering
	\begin{tabular}{c S[table-format=2.4] S[table-format=2.4] S[table-format=2.4] S[table-format=2.4] S[table-format=2.4] S[table-format=2.4]}	
		\toprule
		Parameter				& {Planck}	& {CMB-S3}	& {CMB-S4}	& {Planck}	& {CMB-S3}	& {CMB-S4}	\\
		\midrule[0.065em]
		$\num{e5}\,\omega_b$	& 16		& 5.1		& 2.7		& 26		& 8.3		& 3.8		\\
		$\num{e4}\,\omega_c$	& 16		& 8.3		& 7.1		& 26		& 10		& 7.9		\\
		$\num{e7}\,\theta_s$	& 29		& 9.4		& 5.9		& 44		& 13		& 6.7		\\
		$\ln(\num{e10}\As)$		& 0.020		& 0.020		& 0.020		& 0.021		& 0.020		& 0.020		\\
		$\ns$					& 0.0040	& 0.0023	& 0.0020	& 0.0093	& 0.0040	& 0.0030	\\
		$\tau$					& 0.010		& 0.010		& 0.010		& 0.010		& 0.010		& 0.010		\\
		$\Neff$					& {--}		& {--}		& {--}		& 0.18		& 0.054		& 0.030		\\
		\bottomrule
	\end{tabular}
	\caption{Forecasted sensitivities of Planck, CMB-S3 and CMB-S4 for the parameters of $\Lambda\mathrm{CDM}$ and $\Lambda\mathrm{CDM}$+$\Neff$.}
	\label{tab:cmbForecast_LCDM_LCDM+Neff}
\end{table}
we show the $1\sigma$ constraints for Planck and the described representative configurations of CMB-S3 and CMB-S4. In Table~\ref{tab:cmbForecast_LCDM+Yp_LCDM+Neff+Yp}, we display how these constraints vary when we allow the primordial helium fraction~$Y_p$ to be an additional free parameter. The differences in the forecasted sensitivities for Planck compared to the constraints published in~\cite{Ade:2015xua} can be attributed entirely to the improvement in $\sigma(\tau)$ which arises from the imposed prior on the optical depth $\tau$. The forecast of $\Neff$ for CMB-S3 is a rough estimate and will be subject to the precise specifications of the respective experiment. While the precise design of CMB-S4 is also undetermined at this point, $\sigma(\Neff) = 0.03$ is a primary science target and is therefore more likely to be a reliable estimate of the expected performance.
\begin{table}[t]
	\centering
	\begin{tabular}{c S[table-format=2.4] S[table-format=2.4] S[table-format=2.4] S[table-format=2.4] S[table-format=2.4] S[table-format=2.4]}	
			\toprule
		Parameter				& {Planck}	& {CMB-S3}	& {CMB-S4}	& {Planck}	& {CMB-S3}	& {CMB-S4}	\\
			\midrule[0.065em]
		$\num{e5}\,\omega_b$	& 24		& 8.2		& 3.8		& 26		& 8.4		& 3.8		\\
		$\num{e4}\,\omega_c$	& 17		& 8.6		& 7.2		& 49		& 21		& 14		\\
		$\num{e7}\,\theta_s$	& 33		& 9.9		& 6.3		& 89		& 27		& 15		\\
		$\ln(\num{e10}\As)$		& 0.020		& 0.020		& 0.020		& 0.022		& 0.020		& 0.020		\\
		$\ns$					& 0.0082	& 0.0038	& 0.0029	& 0.0093	& 0.0040	& 0.0030	\\
		$\tau$					& 0.010		& 0.010		& 0.010		& 0.010		& 0.010		& 0.010		\\
		$\Neff$					& {--}		& {--}		& {--}		& 0.32		& 0.12		& 0.081		\\
		$Y_p$					& 0.012		& 0.0037	& 0.0021	& 0.018		& 0.0069	& 0.0047	\\
			\bottomrule
	\end{tabular}
	\caption{Forecasted sensitivities of Planck, CMB-S3 and CMB-S4 for the parameters of $\Lambda\mathrm{CDM}$+$Y_p$ and $\Lambda\mathrm{CDM}$+$\Neff$+$Y_p$.}
	\label{tab:cmbForecast_LCDM+Yp_LCDM+Neff+Yp}
\end{table}

\begin{figure}[b!]
	\centering
	\includegraphics{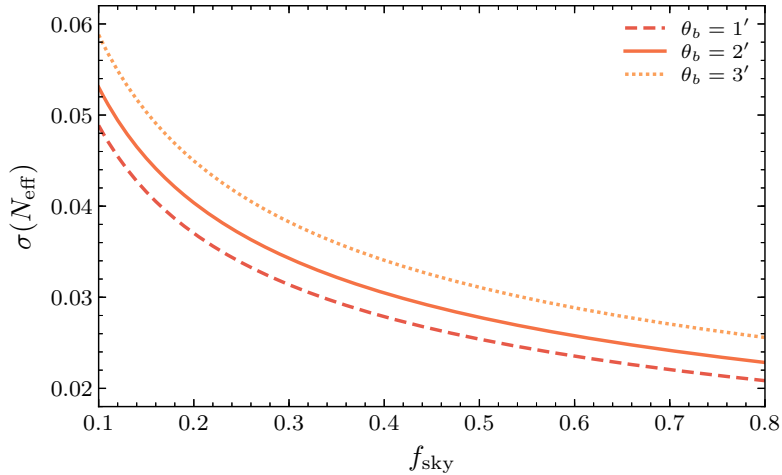}
	\caption{Marginalized constraints on $\Neff$ as a function of the sky fraction $\fsky$ for three values of the beam width $\theta_b$ and fixed noise level $\Delta T=\SI{1}{\muKelvin.arcmin}$.}
	\label{fig:cmbNeffFsky}
\end{figure}

\begin{figure}[t!]
	\begin{center}
		\includegraphics{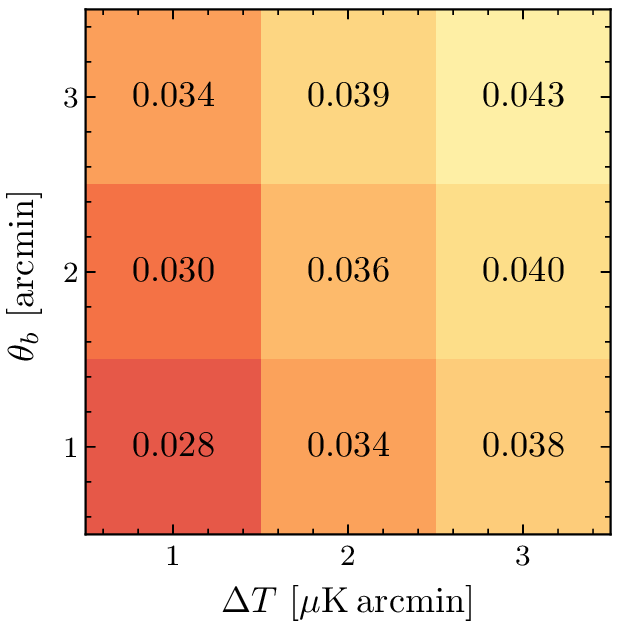}\hspace{1cm}
		\includegraphics{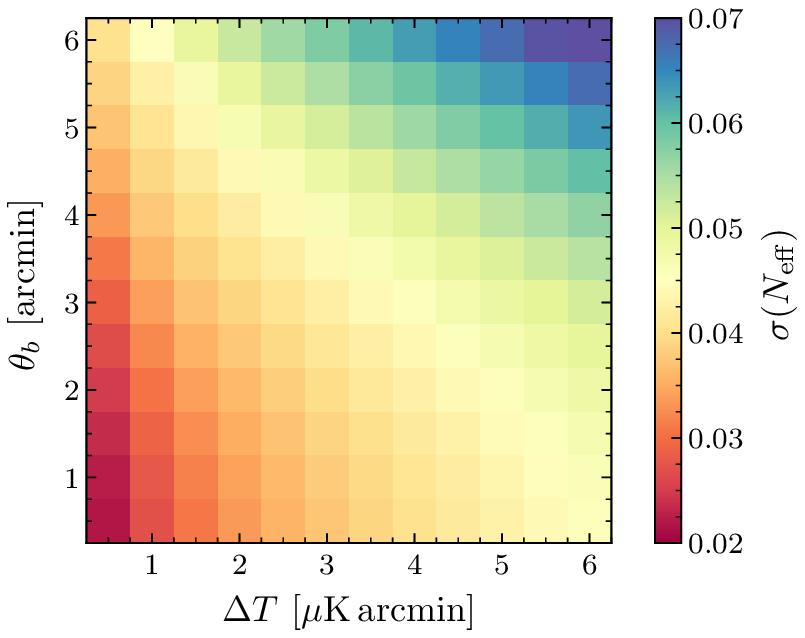}
		\caption{Marginalized constraints on $\Neff$ as a function of the beam size $\theta_b$ and the temperature noise level $\Delta T$, for fixed sky fraction $\fsky = 0.4$.}
		\label{fig:cmbNeffThetabDeltaT}
	\end{center}
\end{figure}

\vskip4pt
In Figure~\ref{fig:cmbNeffFsky},%
we demonstrate how the constraints on $\Neff$ depend on the sky fraction $\fsky$, for three different values of $\theta_b$ and fixed noise level $\Delta T = \SI{1}{\muKelvin.arcmin}$. When varying the total sky fraction, we also appropriately change the contribution of the included high-$\ell$ Planck data. In Figure~\ref{fig:cmbNeffThetabDeltaT}, we illustrate the constraint on $\Neff$ as a function of the beam size $\theta_b$ and the noise level~$\Delta T$, for fixed sky fraction $\fsky = 0.4$. Comparing Figure~\ref{fig:cmbNeffThetabDeltaT} to the equivalent figure in the CMB-S4 Science Book~\cite{Abazajian:2016yjj} (Fig.~22), we see that the difference between the two forecasts is $\Delta \sigma(\Neff) \approx 0.002$. This can be attributed to the effects of imperfect delensing and is completely negligible for our purposes.

\clearpage
\section{Forecasting LSS Constraints}
\label{app:specs}

In this appendix, we collect the specific information regarding the planned LSS surveys which we used in our Fisher and likelihood forecasts. We also provide the full set of constraints on all of the cosmological parameters and cosmologies that are studied in this paper.

\subsection{Survey Specifications}

Below, we provide the experimental specifications for the galaxy surveys used in our forecasts. We have slightly simplified the details compared to~\cite{Font-Ribera:2013rwa}, for example. In particular, we group different types of tracers (e.g.\ luminous red galaxies, emission line galaxies or quasars) into a single effective number density and bias. We find our results to be fairly insensitive to many of these details and well approximated by a fixed number of objects distributed with a constant comoving number density over the same redshift range.

\vskip4pt
The employed parametrization of the spectroscopic redshift surveys BOSS, eBOSS, DESI and Euclid are provided in Tables~\ref{tab:specsBOSS} to~\ref{tab:specsEuclid}. For eBOSS, we combine BOSS and the two eBOSS configurations of Table~\ref{tab:specseBOSS} into one survey neglecting the small overlap. We effectively treat each redshift bin with mean redshift $\bar{z}$ as an independent three-dimensional survey. Our Fisher matrix is the sum of the Fisher matrices associated with each bin, $F = \sum_{\bar z} F_{\bar{z}}$. We translated the survey specifications used in~\cite{Font-Ribera:2013rwa} into three numbers per redshift bin: the linear galaxy bias~$b$, the comoving number density of galaxies $\bar{n}_g$ and the bin volume $V$. This is sufficient to fully specify the Fisher matrix in each bin. The spherical bin volume is given by 
\beq
V = \frac{4\pi}{3}\, \fsky \left[d_c(\zmax)^3 - d_c(z_\mathrm{min})^3\right] , \qquad d_c(z) = \int_0^z \!\d z\, \frac{c}{H(z)}\, ,	\label{eq:binVolume}
\eeq
where $\fsky$ is the sky fraction, $d_c(z)$ is the comoving distance to redshift $z$, and $z_\mathrm{min} = \bar{z} - \Delta z/2$ and $\zmax = \bar{z} + \Delta z/2$ are the minimum and maximum redshift of the respective bin. Throughout this paper, we use redshift bins of width $\Delta z = 0.1$.

\vskip4pt
For the photometric surveys DES and LSST, we follow~\cite{Font-Ribera:2013rwa} and define the surveys by using $(\alpha, \beta, z_*, N_\mathrm{tot}, b_0) = (1.25, 2.29, 0.88, \SI{12}{arcmin^{-2}}, 0.95)$ and $(2.0, 1.0, 0.3, \SI{50}{arcmin^{-2}}, 0.95)$, respectively. These parameters are related to those used in our forecasts as follows:
\begin{align}
\bar{n}_g(\bar{z})	&= \frac{N_\mathrm{tot}}{V} \frac{\beta/z_*}{\Gamma\left[(\alpha+1)/\beta\right]} \int_{z_\mathrm{min}}^{\zmax} \!\d z\, \left(z/z_*\right)^\alpha \exp\left\{-\left(z/z_*\right)^\beta\right\} ,	\\[2pt]
b(\bar{z}) 			&= \frac{D_1(0)}{D_1(\bar{z}_i)}\, b_0\, ,
\end{align}
with gamma function $\Gamma$ and linear growth function $D_1(z)$.
\begin{table}
	\centering
	\begin{tabular}{l S[table-format=1.4] S[table-format=1.4] S[table-format=1.4] S[table-format=1.4] S[table-format=1.4] S[table-format=1.4] S[table-format=1.4] S[table-format=1.4]}
			\toprule
		$\bar{z}$													& 0.05		& 0.15	& 0.25	& 0.35	& 0.45	& 0.55	& 0.65	& 0.75		\\
			\midrule[0.065em]
		$b$ 														& 1.79		& 1.90	& 1.98	& 2.09	& 2.32	& 2.26	& 2.38	& 3.09		\\
		$\num{e3}\,\bar{n}_g\ [\si{\h\tothe{3}\per\Mpc\tothe{3}}]$ 	& 0.289		& 0.290	& 0.300	& 0.304	& 0.276	& 0.323	& 0.120	& 0.0100	\\
		$V\ [\si{\per\h\tothe{3}\Gpc\tothe{3}}]$ 					& 0.0255	& 0.164	& 0.402	& 0.704	& 1.04	& 1.38	& 1.72	& 2.04		\\
			\bottomrule
	\end{tabular}
	\caption{Basic specifications for BOSS derived from~\cite{Font-Ribera:2013rwa} with a sky area of $\Omega=\SI{10000}{deg^2}$ resulting in roughly \num{1.4e6} objects in a volume of about~\SI{7.5}{\per\h\tothe{3}\Gpc\tothe{3}}.}
	\label{tab:specsBOSS}
\end{table}
\begin{table}
	\centering
	\begin{tabular}{l S[table-format=2.3] S[table-format=2.3] S[table-format=2.3] S[table-format=2.3] S[table-format=2.4] S[table-format=2.3] S[table-format=2.3] S[table-format=2.3]}
			\toprule
		$\bar{z}$													& 0.55	& 0.65	& 0.75	& 0.85	& 0.95		& 1.05	& 1.15	& 1.25	\\
			\midrule[0.065em]
		$b$ 														& 3.07	& 2.07	& 1.57	& 1.57	& 1.61		& 3.51	& 1.98	& 2.35	\\
		$\num{e5}\,\bar{n}_g\ [\si{\h\tothe{3}\per\Mpc\tothe{3}}]$	& 0.463	& 21.3	& 35.5	& 23.6	& 5.40		& 0.563	& 1.53	& 1.48	\\
		$V\ [\si{\per\h\tothe{3}\Gpc\tothe{3}}]$ 					& 0.208	& 0.258	& 0.307	& 0.352	& 0.392		& 0.429	& 0.461	& 0.489	\\
			\midrule[0.065em]
		$b$ 														& 3.07	& 2.42	& 2.45	& 2.56	& 7.84		& 3.51	& 1.98	& 2.35	\\
		$\num{e5}\,\bar{n}_g\ [\si{\h\tothe{3}\per\Mpc\tothe{3}}]$	& 0.463	& 13.5	& 7.02	& 3.35	& 0.0412	& 0.563	& 1.53	& 1.48	\\
		$V\ [\si{\per\h\tothe{3}\Gpc\tothe{3}}]$ 					& 0.830	& 1.03	& 1.23	& 1.41	& 1.57		& 1.71	& 1.84	& 1.96	\\
			\bottomrule
	\end{tabular}\\[4pt]
	\begin{tabular}{l S[table-format=2.3] S[table-format=2.3] S[table-format=1.3] S[table-format=1.3] S[table-format=1.3] S[table-format=1.3] S[table-format=1.3] S[table-format=1.3] S[table-format=1.3]}
			\toprule
		$\bar{z}$																& 1.35	& 1.45	& 1.55	& 1.65	& 1.75	& 1.85	& 1.95	& 2.05	& 2.15	\\
			\midrule[0.065em]
		$b$ 																	& 3.65	& 2.40	& 2.42	& 2.08	& 2.10	& 3.33	& 3.35	& 1.72	& 1.73	\\
		$\num{e5}\,\bar{n}_g\ [\si{\h\tothe{3}\per\Mpc\tothe{3}}]\hskip1.5pt$	& 0.664	& 1.66	& 1.76	& 2.03	& 2.15	& 0.912	& 0.965	& 2.91	& 3.07	\\
		$V\ [\si{\per\h\tothe{3}\Gpc\tothe{3}}]$								& 0.513	& 0.533	& 0.551	& 0.565	& 0.577	& 0.587	& 0.594	& 0.600	& 0.604	\\
			\midrule[0.065em]
		$b$ 																	& 3.65	& 2.40	& 2.42	& 2.08	& 2.10	& 3.33	& 3.35	& 1.72	& 1.73	\\
		$\num{e5}\,\bar{n}_g\ [\si{\h\tothe{3}\per\Mpc\tothe{3}}]$				& 0.664	& 1.66	& 1.76	& 2.03	& 2.15	& 0.912	& 0.965	& 2.91	& 3.07	\\
		$V\ [\si{\per\h\tothe{3}\Gpc\tothe{3}}]$ 								& 2.05	& 2.13	& 2.20	& 2.26	& 2.31	& 2.35	& 2.38	& 2.40	& 2.42	\\
			\bottomrule
	\end{tabular}
	\caption{Basic specifications for eBOSS derived from~\cite{Font-Ribera:2013rwa}. The redshift range is covered twice, first showing the survey covering $\Omega=\SI{1500}{deg^2}$ that will include emission line galaxies (resulting in roughly \num{3.8e5} objects in a volume of about~\SI{8.0}{\si{\per\h\tothe{3}\Gpc\tothe{3}}}), and then the survey with $\Omega=\SI{6000}{deg^2}$ that will not (resulting in roughly \num{7.2e5} objects in a volume of about~\SI{32}{\per\h\tothe{3}\Gpc\tothe{3}}).}
	\label{tab:specseBOSS} 
\end{table}
\begin{table}
	\centering
	\begin{tabular}{l S[table-format=1.3] S[table-format=1.3] S[table-format=1.3] S[table-format=1.3] S[table-format=1.3] S[table-format=1.3] S[table-format=1.4] S[table-format=1.4] S[table-format=1.5, group-digits=false]}
			\toprule
		$\bar{z}$													& 0.15	& 0.25	& 0.35	& 0.45	& 0.55	& 0.65	& 0.75	& 0.85	& 0.95		\\
			\midrule[0.065em]
		$b$ 														& 1.13	& 1.39	& 1.64	& 1.81	& 1.87	& 1.89	& 1.90	& 1.82	& 1.53		\\
		$\num{e3}\,\bar{n}_g\ [\si{\h\tothe{3}\per\Mpc\tothe{3}}]$	& 2.38	& 1.07	& 0.684	& 0.568	& 0.600	& 0.696	& 0.810	& 0.719	& 0.558		\\
		$V\ [\si{\per\h\tothe{3}\Gpc\tothe{3}}]$					& 0.229	& 0.563	& 0.985	& 1.45	& 1.94	& 2.41	& 2.86	& 3.28	& 3.66		\\
			\bottomrule
	\end{tabular}\\[4pt]
	\begin{tabular}{l S[table-format=1.3] S[table-format=1.3] S[table-format=1.3] S[table-format=1.3] S[table-format=1.3] S[table-format=1.3] S[table-format=1.4] S[table-format=1.4] S[table-format=1.5, group-digits=false]}
			\toprule
		$\bar{z}$													& 1.05	& 1.15	& 1.25	& 1.35	& 1.45	& 1.55	& 1.65		& 1.75		& 1.85		\\
			\midrule[0.065em]
		$b$ 														& 1.47	& 1.49	& 1.58	& 1.62	& 1.73	& 2.01	& 1.98		& 2.56		& 4.17		\\
		$\num{e3}\,\bar{n}_g\ [\si{\h\tothe{3}\per\Mpc\tothe{3}}]$	& 0.522	& 0.506	& 0.454	& 0.356	& 0.242	& 0.127	& 0.0736	& 0.0289	& 0.00875	\\
		$V\ [\si{\per\h\tothe{3}\Gpc\tothe{3}}]$					& 4.00	& 4.30	& 4.56	& 4.79	& 4.98	& 5.14	& 5.28		& 5.39		& 5.48		\\
			\bottomrule
	\end{tabular}
	\caption{Basic specifications for DESI derived from~\cite{Font-Ribera:2013rwa}, covering a sky area $\Omega=\SI{14000}{deg^2}$ and resulting in roughly \num{2.3e7} objects in a volume of about~\SI{61}{\per\h\tothe{3}\Gpc\tothe{3}}.}
	\label{tab:specsDESI} 
\end{table}
\begin{table}
	\centering
	\begin{tabular}{l S[table-format=1.3] S[table-format=1.3] S[table-format=1.3] S[table-format=1.3] S[table-format=1.3] S[table-format=1.3] S[table-format=1.3] S[table-format=1.3]}
			\toprule
		$\bar{z}$														& 0.65	& 0.75	& 0.85	& 0.95	& 1.05	& 1.15	& 1.25	& 1.35	\\
			\midrule[0.065em]
		$b$ 															& 1.06	& 1.11	& 1.16	& 1.21	& 1.27	& 1.33	& 1.38	& 1.44	\\
		$\num{e3}\,\bar{n}_g\ [\si{\h\tothe{3}\per\Mpc\tothe{3}}]\,$	& 0.637	& 1.46	& 1.63	& 1.50	& 1.33	& 1.14	& 1.00	& 0.837	\\
		$V\ [\si{\per\h\tothe{3}\Gpc\tothe{3}}]$						& 2.58	& 3.07	& 3.52	& 3.92	& 4.29	& 4.61	& 4.89	& 5.13	\\
			\bottomrule
	\end{tabular}\\[4pt]
	\begin{tabular}{l S[table-format=1.4] S[table-format=1.4] S[table-format=1.4] S[table-format=1.4] S[table-format=1.4] S[table-format=1.4] S[table-format=1.4]}
			\toprule
		$\bar{z}$ 													& 1.45	& 1.55	& 1.65	& 1.75	& 1.85	& 1.95		& 2.05		\\
			\midrule[0.065em]
		$b$ 														& 1.51	& 1.54	& 1.63	& 1.70	& 1.85	& 1.90		& 1.26		\\
		$\num{e3}\,\bar{n}_g\ [\si{\h\tothe{3}\per\Mpc\tothe{3}}]$	& 0.652	& 0.512	& 0.357	& 0.246	& 0.149	& 0.0904	& 0.0721	\\
		$V\ [\si{\per\h\tothe{3}\Gpc\tothe{3}}]$					& 5.33	& 5.51	& 5.65	& 5.77	& 5.87	& 5.94		& 6.00		\\
			\bottomrule
	\end{tabular}
	\caption{Basic specifications for Euclid derived from~\cite{Font-Ribera:2013rwa}, covering a sky area $\Omega=\SI{15000}{deg^2}$ and resulting in roughly \num{5.0e7} objects in a volume of about~\SI{72}{\per\h\tothe{3}\Gpc\tothe{3}}.}
	\label{tab:specsEuclid}
\end{table}

\vskip4pt
For DES, we employ a survey area of $\Omega=\SI{5000}{deg^2}$ and a redshift coverage of $0.1 \leq z \leq 2.0$, while we take \SI{20000}{deg^2} and $0.1 \leq z \leq 3.5$ for LSST. This results in approximately \num{1.4e8} and \num{5.9e8} objects in a total survey volume of about \SI{24}{\per\h\tothe{3}\Gpc\tothe{3}} and \SI{215}{\per\h\tothe{3}\Gpc\tothe{3}} for the two surveys, respectively. We neglect the spectroscopic redshift error as it is expected to be comparable to (or smaller than) the longitudinal damping scale $\Sigma_\parallel$, but use a conservative root-mean-square photometric redshift error of $\sigma_{z0}=0.05$ for both DES and LSST. Finally, we reiterate that, by considering galaxy clustering alone, we only take a subset of the cosmological observables into account, in particular for photometric surveys, and we therefore expect to underestimate the full power of these experiments.

\subsection{Future Constraints}

Using these specifications, we generated forecasts for all of the cosmological parameters discussed in the main text in combination with the Fisher matrices for Planck, CMB-S3 and \mbox{CMB-S4}. We include both $P(k)$- and BAO-forecasts for $\Lambda\mathrm{CDM}$~(Table~\ref{tab:CMB+LSS_LCDM_full}), $\Lambda\mathrm{CDM}$+$\Neff$~(Table~\ref{tab:CMB+LSS_Neff_full}), $\Lambda\mathrm{CDM}$+$Y_p$~(Table~\ref{tab:CMB+LSS_Yp_full}), and $\Lambda\mathrm{CDM}$+$\Neff$+$Y_p$~(Table~\ref{tab:CMB+LSS_Neff+Yp_full}). As in~\S\ref{sec:constraints_planned}, the $P(k)$-forecasts use wavenumbers up to $\kmax=\SI{0.2}{\hPerMpc}$ and marginalize over the $b_{m\leq1}$-terms of~\eqref{eq:AB}. For the BAO-forecasts, we set $\kmax=\SI{0.5}{\hPerMpc}$ and marginalize over $a_{n\leq4}$ and $b_{m\leq3}$ in each redshift bin. Since we marginalize over galaxy bias, our forecasts show no improvements beyond the CMB for $\ln(\num{e10}\As)$ and $\tau$. We therefore do not include these two parameters in the following tables.

\vskip4pt
Apart from the improvements in the constraints on $\Neff$ and $Y_p$, which we already discussed in~\S\ref{sec:constraints_planned}, we see that mainly $\omega_b$ and $\omega_c$ benefit from combining the discussed LSS surveys with CMB experiments. The sensitivities may be enhanced by factors of three (two) and more compared to Planck (CMB-S3). We note that the DESI specifications of Table~\ref{tab:specsDESI} are slightly more optimistic overall than what was considered in~\cite{Aghamousa:2016zmz} resulting in roughly the same BAO-forecasts and up to about \SI{5}{\percent} better $P(k)$-forecasts.

\vskip4pt
Comparing our forecasts with the ones obtained from the BAO scale alone (combined with Planck), we see that the BOSS analysis for $\Lambda\mathrm{CDM}$ is nearly optimal, but improvements on the constraints of more than \SI{10}{\percent} can be achieved in extended cosmologies. For instance, the constraints on $\omega_b$, $\ns$ and $\Neff$ improve by \SI{3}{\percent} or more, and $\omega_c$ in $\Lambda\mathrm{CDM}$+$Y_p$ even by \SI{12}{\percent}. For DESI, the obtained sensitivities can generally be increased by a larger amount, e.g.\ up to \SI{15}{\percent} for $\omega_b$ and $\ns$ in $\Lambda\mathrm{CDM}$+$\Neff$, and for $\omega_c$ in $\Lambda\mathrm{CDM}$+$Y_p$.
\begin{table}
	\centering
	\scriptsize
	\subfloat[Planck + $P(k)$]{
		\begin{tabular}{c S[table-format=2.4] S[table-format=2.4] S[table-format=2.4] S[table-format=2.4] S[table-format=2.4] S[table-format=2.4] S[table-format=2.4]}	
				\toprule
			Parameter				& {Planck}	& {BOSS}	& {eBOSS}	& {DESI}	& {Euclid}	& {DES}		& {LSST}	\\
				\midrule[0.065em]
			$\num{e5}\,\omega_b$	& 16		& 13		& 13		& 12		& 11		& 14		& 12		\\
			$\num{e4}\,\omega_c$	& 16		& 8.9		& 7.7		& 4.6		& 4.3		& 13		& 8.2		\\
			$\num{e7}\,\theta_s$	& 29		& 28		& 27		& 27		& 27		& 29		& 28		\\
			$\ns$					& 0.0040	& 0.0033	& 0.0032	& 0.0028	& 0.0027	& 0.0037	& 0.0033	\\
				\bottomrule
		\end{tabular}
	}\\[2pt]
	\subfloat[CMB-S3 + $P(k)$]{
		\begin{tabular}{c S[table-format=2.4] S[table-format=2.4] S[table-format=2.4] S[table-format=2.4] S[table-format=2.4] S[table-format=2.4] S[table-format=2.4]}	
				\toprule
			Parameter				& {CMB-S3}	& {BOSS}	& {eBOSS}	& {DESI}	& {Euclid}	& {DES}		& {LSST}	\\
				\midrule[0.065em]
			$\num{e5}\,\omega_b$	& 5.1		& 4.9		& 4.9		& 4.7		& 4.6		& 5.0		& 4.8		\\
			$\num{e4}\,\omega_c$	& 8.3		& 6.7		& 6.1		& 4.0		& 3.7		& 7.8		& 6.3		\\
			$\num{e7}\,\theta_s$	& 9.4		& 9.1		& 9.0		& 8.7		& 8.6		& 9.3		& 9.1		\\
			$\ns$					& 0.0023	& 0.0021	& 0.0021	& 0.0019	& 0.0019	& 0.0022	& 0.0021	\\
				\bottomrule
		\end{tabular}
	}\\[2pt]
	\subfloat[S4 + $P(k)$]{
		\begin{tabular}{c S[table-format=2.4] S[table-format=2.4] S[table-format=2.4] S[table-format=2.4] S[table-format=2.4] S[table-format=2.4] S[table-format=2.4]}	
				\toprule
			Parameter				& {CMB-S4}	& {BOSS}	& {eBOSS}	& {DESI}	& {Euclid}	& {DES}		& {LSST}	\\
				\midrule[0.065em]
			$\num{e5}\,\omega_b$	& 2.7		& 2.7		& 2.7		& 2.6		& 2.6		& 2.7		& 2.6		\\
			$\num{e4}\,\omega_c$	& 7.1		& 6.0		& 5.6		& 3.9		& 3.6		& 6.8		& 5.8		\\
			$\num{e7}\,\theta_s$	& 5.9		& 5.7		& 5.6		& 5.3		& 5.2		& 5.9		& 5.7		\\
			$\ns$					& 0.0020	& 0.0018	& 0.0018	& 0.0016	& 0.0016	& 0.0019	& 0.0018	\\
				\bottomrule
		\end{tabular}
	}\\[2pt]
	\subfloat[Planck + BAO]{
		\begin{tabular}{c S[table-format=2.4] S[table-format=2.4] S[table-format=2.4] S[table-format=2.4] S[table-format=2.4] S[table-format=2.4] S[table-format=2.4]}	
				\toprule
			Parameter				& {Planck}	& {BOSS}	& {eBOSS}	& {DESI}	& {Euclid}	& {DES}		& {LSST}	\\
				\midrule[0.065em]
			$\num{e5}\,\omega_b$	& 16		& 13		& 13		& 13		& 13		& 15		& 14		\\
			$\num{e4}\,\omega_c$	& 16		& 8.7		& 8.0		& 5.1		& 5.5		& 13		& 9.4		\\
			$\num{e7}\,\theta_s$	& 29		& 27		& 27		& 27		& 26		& 29		& 27		\\
			$\ns$					& 0.0040	& 0.0031	& 0.0031	& 0.0028	& 0.0028	& 0.0037	& 0.0032	\\
				\bottomrule
		\end{tabular}
	}\\[2pt]
	\subfloat[CMB-S3 + BAO]{
		\begin{tabular}{c S[table-format=2.4] S[table-format=2.4] S[table-format=2.4] S[table-format=2.4] S[table-format=2.4] S[table-format=2.4] S[table-format=2.4]}	
				\toprule
			Parameter				& {CMB-S3}	& {BOSS}	& {eBOSS}	& {DESI}	& {Euclid}	& {DES}		& {LSST}	\\
				\midrule[0.065em]
			$\num{e5}\,\omega_b$	& 5.1		& 5.0		& 5.0		& 4.9		& 4.9		& 5.1		& 5.0		\\
			$\num{e4}\,\omega_c$	& 8.3		& 6.5		& 6.2		& 4.4		& 4.6		& 7.9		& 6.8		\\
			$\num{e7}\,\theta_s$	& 9.4		& 9.0		& 8.9		& 8.6		& 8.6		& 9.3		& 9.0		\\
			$\ns$					& 0.0023	& 0.0021	& 0.0020	& 0.0019	& 0.0019	& 0.0022	& 0.0021	\\
				\bottomrule
		\end{tabular}
	}\\[2pt]
	\subfloat[S4 + BAO]{
		\begin{tabular}{c S[table-format=2.4] S[table-format=2.4] S[table-format=2.4] S[table-format=2.4] S[table-format=2.4] S[table-format=2.4] S[table-format=2.4]}	
				\toprule
			Parameter				& {CMB-S4}	& {BOSS}	& {eBOSS}	& {DESI}	& {Euclid}	& {DES}		& {LSST}	\\
				\midrule[0.065em]
			$\num{e5}\,\omega_b$	& 2.7		& 2.7		& 2.7		& 2.7		& 2.7		& 2.7		& 2.7		\\
			$\num{e4}\,\omega_c$	& 7.1		& 5.9		& 5.7		& 4.2		& 4.3		& 6.8		& 6.1		\\
			$\num{e7}\,\theta_s$	& 5.9		& 5.6		& 5.6		& 5.2		& 5.2		& 5.9		& 5.7		\\
			$\ns$					& 0.0020	& 0.0018	& 0.0018	& 0.0016	& 0.0016	& 0.0019	& 0.0018	\\
				\bottomrule
		\end{tabular}
	}
	\caption{Full set of forecasted $1\sigma$ constraints in a $\Lambda\mathrm{CDM}$ cosmology for current and future LSS surveys in combination with the CMB experiments Planck, CMB-S3 and CMB-S4. We do not quote the sensitivities to $\ln(\num{e10}\As)$ and $\tau$ as they are the same as in Table~\ref{tab:cmbForecast_LCDM_LCDM+Neff} for all combinations.}
	\label{tab:CMB+LSS_LCDM_full}
\end{table}
\begin{table}
	\centering
	\scriptsize
	\subfloat[Planck + $P(k)$]{
		\begin{tabular}{c S[table-format=2.4] S[table-format=2.4] S[table-format=2.4] S[table-format=2.4] S[table-format=2.4] S[table-format=2.4] S[table-format=2.4]}	
				\toprule
			Parameter				& {Planck}	& {BOSS}	& {eBOSS}	& {DESI}	& {Euclid}	& {DES}		& {LSST}	\\
				\midrule[0.065em]
			$\num{e5}\,\omega_b$	& 26		& 19		& 18		& 15		& 15		& 24		& 20		\\
			$\num{e4}\,\omega_c$	& 26		& 23		& 21		& 15		& 13		& 25		& 19		\\
			$\num{e7}\,\theta_s$	& 44		& 41		& 40		& 35		& 34		& 43		& 39		\\
			$\ns$					& 0.0093	& 0.0068	& 0.0061	& 0.0039	& 0.0035	& 0.0085	& 0.0069	\\
			$\Neff$					& 0.18		& 0.14		& 0.13		& 0.087		& 0.079		& 0.17		& 0.14		\\
				\bottomrule
		\end{tabular}
	}\\[2pt]
	\subfloat[CMB-S3 + $P(k)$]{
		\begin{tabular}{c S[table-format=2.4] S[table-format=2.4] S[table-format=2.4] S[table-format=2.4] S[table-format=2.4] S[table-format=2.4] S[table-format=2.4]}	
				\toprule
			Parameter				& {CMB-S3}	& {BOSS}	& {eBOSS}	& {DESI}	& {Euclid}	& {DES}		& {LSST}	\\
				\midrule[0.065em]
			$\num{e5}\,\omega_b$	& 8.3		& 7.9		& 7.8		& 7.3		& 7.1		& 8.2		& 8.0		\\
			$\num{e4}\,\omega_c$	& 10		& 9.6		& 9.2		& 7.8		& 7.5		& 10		& 8.8		\\
			$\num{e7}\,\theta_s$	& 13		& 12		& 12		& 12		& 12		& 12		& 12		\\
			$\ns$					& 0.0040	& 0.0037	& 0.0036	& 0.0029	& 0.0028	& 0.0039	& 0.0037	\\
			$\Neff$					& 0.054		& 0.052		& 0.051		& 0.045		& 0.043		& 0.054		& 0.052		\\
				\bottomrule
		\end{tabular}
	}\\[2pt]
	\subfloat[S4 + $P(k)$]{
		\begin{tabular}{c S[table-format=2.4] S[table-format=2.4] S[table-format=2.4] S[table-format=2.4] S[table-format=2.4] S[table-format=2.4] S[table-format=2.4]}	
				\toprule
			Parameter				& {CMB-S4}	& {BOSS}	& {eBOSS}	& {DESI}	& {Euclid}	& {DES}		& {LSST}	\\
				\midrule[0.065em]
			$\num{e5}\,\omega_b$	& 3.8		& 3.7		& 3.7		& 3.6		& 3.6		& 3.8		& 3.7		\\
			$\num{e4}\,\omega_c$	& 7.9		& 7.1		& 6.8		& 5.5		& 5.3		& 7.6		& 6.7		\\
			$\num{e7}\,\theta_s$	& 6.7		& 6.6		& 6.5		& 6.2		& 6.2		& 6.7		& 6.5		\\
			$\ns$					& 0.0030	& 0.0029	& 0.0028	& 0.0025	& 0.0024	& 0.0030	& 0.0029	\\
			$\Neff$					& 0.030		& 0.030		& 0.030		& 0.028		& 0.027		& 0.030		& 0.030		\\
				\bottomrule
		\end{tabular}
	}\\[2pt]
	\subfloat[Planck + BAO]{
		\begin{tabular}{c S[table-format=2.4] S[table-format=2.4] S[table-format=2.4] S[table-format=2.4] S[table-format=2.4] S[table-format=2.4] S[table-format=2.4]}	
				\toprule
			Parameter				& {Planck}	& {BOSS}	& {eBOSS}	& {DESI}	& {Euclid}	& {DES}		& {LSST}	\\
				\midrule[0.065em]
			$\num{e5}\,\omega_b$	& 26		& 18		& 18		& 17		& 17		& 22		& 19		\\
			$\num{e4}\,\omega_c$	& 26		& 26		& 26		& 26		& 26		& 26		& 26		\\
			$\num{e7}\,\theta_s$	& 44		& 43		& 43		& 43		& 43		& 44		& 44		\\
			$\ns$					& 0.0093	& 0.0065	& 0.0063	& 0.0059	& 0.0059	& 0.0081	& 0.0067	\\
			$\Neff$					& 0.18		& 0.15		& 0.15		& 0.14		& 0.14		& 0.16		& 0.15		\\
				\bottomrule
		\end{tabular}
	}\\[2pt]
	\subfloat[CMB-S3 + BAO]{
		\begin{tabular}{c S[table-format=2.4] S[table-format=2.4] S[table-format=2.4] S[table-format=2.4] S[table-format=2.4] S[table-format=2.4] S[table-format=2.4]}	
				\toprule
			Parameter				& {CMB-S3}	& {BOSS}	& {eBOSS}	& {DESI}	& {Euclid}	& {DES}		& {LSST}	\\
				\midrule[0.065em]
			$\num{e5}\,\omega_b$	& 8.3		& 7.8		& 7.7		& 7.4		& 7.4		& 8.2		& 7.8		\\
			$\num{e4}\,\omega_c$	& 10		& 10		& 9.9		& 9.6		& 9.6		& 10		& 10		\\
			$\num{e7}\,\theta_s$	& 13		& 13		& 13		& 13		& 13		& 13		& 13		\\
			$\ns$					& 0.0040	& 0.0035	& 0.0035	& 0.0031	& 0.0032	& 0.0039	& 0.0036	\\
			$\Neff$					& 0.054		& 0.052		& 0.052		& 0.050		& 0.050		& 0.054		& 0.052		\\
				\bottomrule
		\end{tabular}
	}\\[2pt]
	\subfloat[S4 + BAO]{
		\begin{tabular}{c S[table-format=2.4] S[table-format=2.4] S[table-format=2.4] S[table-format=2.4] S[table-format=2.4] S[table-format=2.4] S[table-format=2.4]}	
				\toprule
			Parameter				& {CMB-S4}	& {BOSS}	& {eBOSS}	& {DESI}	& {Euclid}	& {DES}		& {LSST}	\\
				\midrule[0.065em]
			$\num{e5}\,\omega_b$	& 3.8		& 3.7		& 3.7		& 3.7		& 3.7		& 3.8		& 3.7		\\
			$\num{e4}\,\omega_c$	& 7.9		& 7.2		& 7.1		& 6.5		& 6.5		& 7.8		& 7.3		\\
			$\num{e7}\,\theta_s$	& 6.7		& 6.6		& 6.6		& 6.5		& 6.5		& 6.7		& 6.6		\\
			$\ns$					& 0.0030	& 0.0028	& 0.0028	& 0.0025	& 0.0025	& 0.0030	& 0.0028	\\
			$\Neff$					& 0.030		& 0.030		& 0.030		& 0.029		& 0.029		& 0.030		& 0.030		\\
				\bottomrule
		\end{tabular}
	}
	\caption{As in Table~\ref{tab:CMB+LSS_LCDM_full}, but for an extended $\Lambda\mathrm{CDM}$+$\Neff$ cosmology.}
	\label{tab:CMB+LSS_Neff_full}
\end{table}
\begin{table}
	\centering
	\scriptsize
	\subfloat[Planck + $P(k)$]{
		\begin{tabular}{c S[table-format=2.4] S[table-format=2.4] S[table-format=2.4] S[table-format=2.4] S[table-format=2.4] S[table-format=2.4] S[table-format=2.4]}	
				\toprule
			Parameter				& {Planck}	& {BOSS}	& {eBOSS}	& {DESI}	& {Euclid}	& {DES}		& {LSST}	\\
				\midrule[0.065em]
			$\num{e5}\,\omega_b$	& 24		& 19		& 19		& 17		& 16		& 22		& 20		\\
			$\num{e4}\,\omega_c$	& 17		& 8.9		& 7.8		& 4.7		& 4.3		& 13		& 8.6		\\
			$\num{e7}\,\theta_s$	& 33		& 30		& 29		& 27		& 27		& 32		& 30		\\
			$\ns$					& 0.0082	& 0.0066	& 0.0063	& 0.0048	& 0.0044	& 0.0077	& 0.0068	\\
			$Y_p$					& 0.012		& 0.011		& 0.0100	& 0.0087	& 0.0082	& 0.011		& 0.011		\\
				\bottomrule
		\end{tabular}
	}\\[2pt]
	\subfloat[CMB-S3 + $P(k)$]{
		\begin{tabular}{c S[table-format=2.4] S[table-format=2.4] S[table-format=2.4] S[table-format=2.4] S[table-format=2.4] S[table-format=2.4] S[table-format=2.4]}	
				\toprule
			Parameter				& {CMB-S3}	& {BOSS}	& {eBOSS}	& {DESI}	& {Euclid}	& {DES}		& {LSST}	\\
				\midrule[0.065em]
			$\num{e5}\,\omega_b$	& 8.2		& 7.9		& 7.8		& 7.5		& 7.4		& 8.1		& 7.9		\\
			$\num{e4}\,\omega_c$	& 8.6		& 6.8		& 6.3		& 4.0		& 3.7		& 8.1		& 6.6		\\
			$\num{e7}\,\theta_s$	& 9.9		& 9.5		& 9.4		& 8.9		& 8.8		& 9.8		& 9.5		\\
			$\ns$					& 0.0038	& 0.0035	& 0.0034	& 0.0030	& 0.0029	& 0.0037	& 0.0036	\\
			$Y_p$					& 0.0037	& 0.0036	& 0.0036	& 0.0034	& 0.0033	& 0.0037	& 0.0036	\\
				\bottomrule
		\end{tabular}
	}\\[2pt]
	\subfloat[S4 + $P(k)$]{
		\begin{tabular}{c S[table-format=2.4] S[table-format=2.4] S[table-format=2.4] S[table-format=2.4] S[table-format=2.4] S[table-format=2.4] S[table-format=2.4]}	
				\toprule
			Parameter				& {CMB-S4}	& {BOSS}	& {eBOSS}	& {DESI}	& {Euclid}	& {DES}		& {LSST}	\\
				\midrule[0.065em]
			$\num{e5}\,\omega_b$	& 3.8		& 3.8		& 3.8		& 3.7		& 3.7		& 3.8		& 3.8		\\
			$\num{e4}\,\omega_c$	& 7.2		& 6.1		& 5.7		& 3.9		& 3.6		& 6.9		& 5.9		\\
			$\num{e7}\,\theta_s$	& 6.3		& 6.0		& 5.9		& 5.5		& 5.4		& 6.2		& 6.0		\\
			$\ns$					& 0.0029	& 0.0028	& 0.0028	& 0.0025	& 0.0024	& 0.0029	& 0.0028	\\
			$Y_p$					& 0.0021	& 0.0021	& 0.0021	& 0.0020	& 0.0020	& 0.0021	& 0.0021	\\
				\bottomrule
		\end{tabular}
	}\\[2pt]
	\subfloat[Planck + BAO]{
		\begin{tabular}{c S[table-format=2.4] S[table-format=2.4] S[table-format=2.4] S[table-format=2.4] S[table-format=2.4] S[table-format=2.4] S[table-format=2.4]}	
				\toprule
			Parameter				& {Planck}	& {BOSS}	& {eBOSS}	& {DESI}	& {Euclid}	& {DES}		& {LSST}	\\
				\midrule[0.065em]
			$\num{e5}\,\omega_b$	& 24		& 19		& 19		& 18		& 18		& 22		& 19		\\
			$\num{e4}\,\omega_c$	& 17		& 8.7		& 8.0		& 5.5		& 5.7		& 14		& 9.4		\\
			$\num{e7}\,\theta_s$	& 33		& 29		& 29		& 28		& 28		& 31		& 29		\\
			$\ns$					& 0.0082	& 0.0063	& 0.0062	& 0.0059	& 0.0059	& 0.0075	& 0.0065	\\
			$Y_p$					& 0.012		& 0.011		& 0.011		& 0.0100	& 0.011		& 0.011		& 0.011		\\
				\bottomrule
		\end{tabular}
	}\\[2pt]
	\subfloat[CMB-S3 + BAO]{
		\begin{tabular}{c S[table-format=2.4] S[table-format=2.4] S[table-format=2.4] S[table-format=2.4] S[table-format=2.4] S[table-format=2.4] S[table-format=2.4]}	
				\toprule
			Parameter				& {CMB-S3}	& {BOSS}	& {eBOSS}	& {DESI}	& {Euclid}	& {DES}		& {LSST}	\\
				\midrule[0.065em]		
			$\num{e5}\,\omega_b$	& 8.2		& 7.8		& 7.8		& 7.6		& 7.6		& 8.1		& 7.9		\\
			$\num{e4}\,\omega_c$	& 8.6		& 6.6		& 6.3		& 4.4		& 4.6		& 8.2		& 6.9		\\
			$\num{e7}\,\theta_s$	& 9.9		& 9.3		& 9.3		& 8.8		& 8.9		& 9.8		& 9.4		\\
			$\ns$					& 0.0038	& 0.0034	& 0.0034	& 0.0031	& 0.0031	& 0.0037	& 0.0035	\\
			$Y_p$					& 0.0037	& 0.0036	& 0.0036	& 0.0035	& 0.0035	& 0.0037	& 0.0036	\\
				\bottomrule
		\end{tabular}
	}\\[2pt]
	\subfloat[S4 + BAO]{
		\begin{tabular}{c S[table-format=2.4] S[table-format=2.4] S[table-format=2.4] S[table-format=2.4] S[table-format=2.4] S[table-format=2.4] S[table-format=2.4]}	
				\toprule
			Parameter				& {CMB-S4}	& {BOSS}	& {eBOSS}	& {DESI}	& {Euclid}	& {DES}		& {LSST}	\\
				\midrule[0.065em]
			$\num{e5}\,\omega_b$	& 3.8		& 3.8		& 3.8		& 3.8		& 3.8		& 3.8		& 3.8		\\
			$\num{e4}\,\omega_c$	& 7.2		& 5.9		& 5.7		& 4.2		& 4.3		& 6.9		& 6.1		\\
			$\num{e7}\,\theta_s$	& 6.3		& 5.9		& 5.8		& 5.4		& 5.5		& 6.2		& 5.9		\\
			$\ns$					& 0.0029	& 0.0027	& 0.0027	& 0.0025	& 0.0025	& 0.0029	& 0.0028	\\
			$Y_p$					& 0.0021	& 0.0021	& 0.0021	& 0.0021	& 0.0021	& 0.0021	& 0.0021	\\
				\bottomrule
		\end{tabular}
	}
	\caption{As in Table~\ref{tab:CMB+LSS_LCDM_full}, but for an extended $\Lambda\mathrm{CDM}$+$Y_p$ cosmology. The constraints on $\ln(\num{e10}\As)$ and $\tau$ are the same as in Table~\ref{tab:cmbForecast_LCDM+Yp_LCDM+Neff+Yp} for all combinations.}
	\label{tab:CMB+LSS_Yp_full}
\end{table}
\begin{table}
	\centering
	\scriptsize
	\subfloat[Planck + $P(k)$]{
		\begin{tabular}{c S[table-format=2.4] S[table-format=2.4] S[table-format=2.4] S[table-format=2.4] S[table-format=2.4] S[table-format=2.4] S[table-format=2.4]}	
				\toprule
			Parameter				& {Planck}	& {BOSS}	& {eBOSS}	& {DESI}	& {Euclid}	& {DES}		& {LSST}	\\
				\midrule[0.065em]
			$\num{e5}\,\omega_b$	& 26		& 20		& 19		& 17		& 16		& 24		& 21		\\
			$\num{e4}\,\omega_c$	& 49		& 40		& 35		& 23		& 21		& 45		& 34		\\
			$\num{e7}\,\theta_s$	& 89		& 76		& 70		& 53		& 50		& 84		& 69		\\
			$\ns$					& 0.0093	& 0.0069	& 0.0065	& 0.0048	& 0.0045	& 0.0086	& 0.0071	\\
			$\Neff$					& 0.32		& 0.25		& 0.22		& 0.14		& 0.13		& 0.29		& 0.23		\\
			$Y_p$					& 0.018		& 0.016		& 0.016		& 0.013		& 0.012		& 0.017		& 0.015		\\
				\bottomrule
		\end{tabular}
	}\\[0.26pt]
	\subfloat[CMB-S3 + $P(k)$]{
		\begin{tabular}{c S[table-format=2.4] S[table-format=2.4] S[table-format=2.4] S[table-format=2.4] S[table-format=2.4] S[table-format=2.4] S[table-format=2.4]}	
				\toprule
			Parameter				& {CMB-S3}	& {BOSS}	& {eBOSS}	& {DESI}	& {Euclid}	& {DES}		& {LSST}	\\
				\midrule[0.065em]
			$\num{e5}\,\omega_b$	& 8.4		& 8.0		& 7.9		& 7.5		& 7.5		& 8.3		& 8.1		\\
			$\num{e4}\,\omega_c$	& 21		& 20		& 19		& 15		& 14		& 20		& 18		\\
			$\num{e7}\,\theta_s$	& 27		& 26		& 26		& 22		& 21		& 27		& 25		\\
			$\ns$					& 0.0040	& 0.0037	& 0.0036	& 0.0030	& 0.0029	& 0.0039	& 0.0037	\\
			$\Neff$					& 0.12		& 0.12		& 0.11		& 0.094		& 0.088		& 0.12		& 0.11		\\
			$Y_p$					& 0.0069	& 0.0068	& 0.0067	& 0.0060	& 0.0058	& 0.0069	& 0.0066	\\
				\bottomrule
		\end{tabular}
	}\\[0.26pt]
	\subfloat[S4 + $P(k)$]{
		\begin{tabular}{c S[table-format=2.4] S[table-format=2.4] S[table-format=2.4] S[table-format=2.4] S[table-format=2.4] S[table-format=2.4] S[table-format=2.4]}	
				\toprule
			Parameter				& {CMB-S4}	& {BOSS}	& {eBOSS}	& {DESI}	& {Euclid}	& {DES}		& {LSST}	\\
				\midrule[0.065em]
			$\num{e5}\,\omega_b$	& 3.8		& 3.8		& 3.8		& 3.7		& 3.7		& 3.8		& 3.8		\\
			$\num{e4}\,\omega_c$	& 14		& 14		& 13		& 12		& 11		& 14		& 13		\\
			$\num{e7}\,\theta_s$	& 15		& 15		& 14		& 13		& 13		& 15		& 14		\\
			$\ns$					& 0.0030	& 0.0029	& 0.0028	& 0.0025	& 0.0024	& 0.0030	& 0.0029	\\
			$\Neff$					& 0.081		& 0.079		& 0.078		& 0.070		& 0.067		& 0.081		& 0.078		\\
			$Y_p$					& 0.0047	& 0.0046	& 0.0046	& 0.0043	& 0.0042	& 0.0047	& 0.0046	\\
				\bottomrule
		\end{tabular}
	}\\[0.26pt]
	\subfloat[Planck + BAO]{
		\begin{tabular}{c S[table-format=2.4] S[table-format=2.4] S[table-format=2.4] S[table-format=2.4] S[table-format=2.4] S[table-format=2.4] S[table-format=2.4]}	
				\toprule
			Parameter				& {Planck}	& {BOSS}	& {eBOSS}	& {DESI}	& {Euclid}	& {DES}		& {LSST}	\\
				\midrule[0.065em]
			$\num{e5}\,\omega_b$	& 26		& 19		& 19		& 18		& 18		& 23		& 20		\\
			$\num{e4}\,\omega_c$	& 49		& 49		& 49		& 48		& 48		& 49		& 49		\\
			$\num{e7}\,\theta_s$	& 89		& 87		& 87		& 87		& 87		& 88		& 88		\\
			$\ns$					& 0.0093	& 0.0066	& 0.0065	& 0.0060	& 0.0061	& 0.0081	& 0.0068	\\
			$\Neff$					& 0.32		& 0.29		& 0.29		& 0.28		& 0.28		& 0.30		& 0.29		\\
			$Y_p$					& 0.018		& 0.018		& 0.018		& 0.018		& 0.018		& 0.018		& 0.018		\\
				\bottomrule
		\end{tabular}
	}\\[0.26pt]
	\subfloat[CMB-S3 + BAO]{
		\begin{tabular}{c S[table-format=2.4] S[table-format=2.4] S[table-format=2.4] S[table-format=2.4] S[table-format=2.4] S[table-format=2.4] S[table-format=2.4]}	
				\toprule
			Parameter				& {CMB-S3}	& {BOSS}	& {eBOSS}	& {DESI}	& {Euclid}	& {DES}		& {LSST}	\\
				\midrule[0.065em]
			$\num{e5}\,\omega_b$	& 8.4		& 7.9		& 7.9		& 7.6		& 7.7		& 8.3		& 8.0		\\
			$\num{e4}\,\omega_c$	& 21		& 21		& 21		& 21		& 21		& 21		& 21		\\
			$\num{e7}\,\theta_s$	& 27		& 27		& 27		& 27		& 27		& 27		& 27		\\
			$\ns$					& 0.0040	& 0.0035	& 0.0035	& 0.0032	& 0.0032	& 0.0039	& 0.0036	\\
			$\Neff$					& 0.12		& 0.12		& 0.12		& 0.12		& 0.12		& 0.12		& 0.12		\\
			$Y_p$					& 0.0069	& 0.0069	& 0.0069	& 0.0069	& 0.0069	& 0.0069	& 0.0069	\\
				\bottomrule
		\end{tabular}
	}\\[0.26pt]
	\subfloat[S4 + BAO]{
		\begin{tabular}{c S[table-format=2.4] S[table-format=2.4] S[table-format=2.4] S[table-format=2.4] S[table-format=2.4] S[table-format=2.4] S[table-format=2.4]}	
				\toprule
			Parameter				& {CMB-S4}	& {BOSS}	& {eBOSS}	& {DESI}	& {Euclid}	& {DES}		& {LSST}	\\
				\midrule[0.065em]
			$\num{e5}\,\omega_b$	& 3.8		& 3.8		& 3.8		& 3.8		& 3.8		& 3.8		& 3.8		\\
			$\num{e4}\,\omega_c$	& 14		& 14		& 14		& 14		& 14		& 14		& 14		\\
			$\num{e7}\,\theta_s$	& 15		& 15		& 15		& 15		& 15		& 15		& 15		\\
			$\ns$					& 0.0030	& 0.0028	& 0.0028	& 0.0025	& 0.0026	& 0.0030	& 0.0028	\\
			$\Neff$					& 0.081		& 0.080		& 0.080		& 0.079		& 0.079		& 0.081		& 0.080		\\
			$Y_p$					& 0.0047	& 0.0047	& 0.0047	& 0.0046	& 0.0046	& 0.0047	& 0.0047	\\
				\bottomrule
		\end{tabular}
	}
	\caption{As in Table~\ref{tab:CMB+LSS_Yp_full}, but for an extended $\Lambda\mathrm{CDM}$+$\Neff$+$Y_p$ cosmology.}
	\label{tab:CMB+LSS_Neff+Yp_full}
\end{table}

\clearpage
\section{Broadband and Phase Shift Extraction} 
\label{app:broadband+phaseShiftExtraction}

In this appendix, we describe our implementation of a robust broadband extraction method and the computation of the phase shift template.

\subsection{Broadband Extraction}
\label{app:broadbandExtraction}

The split of the matter power spectrum into a broadband (`no-wiggle') part and an oscillatory (`wiggle') part, $P(k) = \Pnw(k) + \Pw(k)$, is not uniquely defined, but depends on the method that is being used. In the following, we describe our method for extracting the broadband spectrum which is robust and stable over a very large parameter space.

\vskip4pt
Computationally it is easier to identify a bump over a smooth background than to find the zeros of oscillations on top of a smooth background. This suggests that it is convenient to sine transform the matter power spectrum to discrete real space where the oscillations map to a localized bump. We then remove this bump and inverse transform back to Fourier space. 

\vskip4pt
An algorithm for the discrete spectral method was outlined in \S A.1 of~\cite{Hamann:2010pw}. Concretely, the relevant steps of our implementation are:
\begin{enumerate}
	\item Provide $P(k)$: Compute the theoretical matter power spectrum $P(k)$ using \texttt{CLASS}~\cite{Blas:2011rf} for discrete wavenumbers $k$ up to a chosen $\kmax$ and log-log interpolate using cubic splines.
	
	\item Sample $\log[k P(k)]$: Sample $\log[k P(k)]$ in $2^n$ points for an integer number $n$. These points are chosen equidistant in $k$.
	
	\item Fast sine transform: Perform a fast sine transform of the $\log[k P(k)]$-array using the orthonormalized type-II sine transform. Denoting the index of the resulting array by $i$, split the even and odd entries, i.e.\ those entries with even $i$ and odd $i$, into separate arrays.
	
	\item Interpolate arrays: Linearly interpolate the two arrays separately using cubic splines.
	
	\item Identify baryonic bumps: Compute the second derivative separately for the interpolated even and odd arrays, and average over next-neighboring array entries to minimize noise. Choose $i_\mathrm{min} = i_* - 3$, where $i_*$ is the array index of the first minimum of the second derivative. Set $i_\mathrm{max} = i^* + \Delta i$, where $i^*$ is the array index of the second maximum of the second derivative, and $\Delta i = 10\text{ and }20$ for the even and odd array, respectively. These shifts were obtained empirically, but are found to give reliable and stable results for a large range of $n$ and $\kmax$.
	
	\item Cut baryonic bumps: Having found the location of the bumps within $[i_\mathrm{min}, i_\mathrm{max}]$ for the even and odd arrays, respectively, remove the elements within this range from the arrays. Then, fill the gap by interpolating the arrays rescaled by a factor of $(i+1)^2$ using cubic splines. This is analogous to interpolating $r^2\,\xi(r)$ instead of the correlation function $\xi(r)$ at separation $r$.
	
	\item Inverse fast sine transform: Merge the two arrays containing the respective elements without the bumps, and without the rescaling factor of $(i+1)^2$, and inversely fast sine transform. This leads to a discretized version of $\log[k \Pnw(k)]$.
	
	\item Provide $\Pnw(k)$ and $\Pw(k)$: In order to cut off numerical noise at low and high wavenumbers, perform two cuts at $k_1$ and $k_2$, where $k_1 = 3 \cdot 2^{-n}$ and the value of $k_2$ is found as the trough of $|P(k)-\Pnw(k)|/\Pnw(k)$ following the smallest maximum (before the oscillation amplitude increases again due to the numerical artefacts intrinsic to the procedure). The reliably extracted no-wiggle spectrum $\Pnw(k)$ is then valid for $k \in [k_1,k_2]$. In practice, choose $n$ and $\kmax$ large enough initially so that $k_{1,2}$ are outside the range of wavenumbers of interest, e.g.\ those covered by a survey. The wiggle spectrum in this range is then given by $\Pw(k) = P(k) - \Pnw(k)$.
\end{enumerate}

\begin{figure}[b!]
	\begin{center}
		\includegraphics{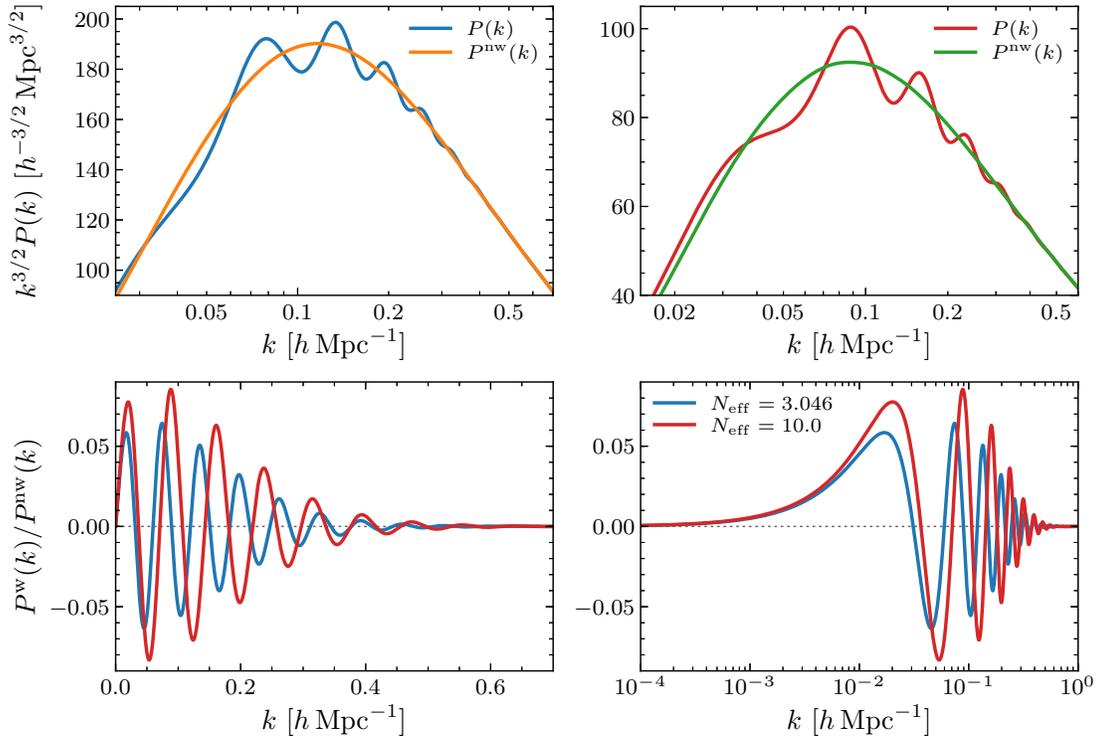}
		\caption{\textit{Top}: Extracted broadband spectrum $\Pnw(k)$ compared to the full power spectrum~$P(k)$ for $\Neff=3.046\text{ (\textit{left}) and }10$ (\textit{right}). The spectra are rescaled by $k^{3/2}$ to exaggerate any oscillations. \textit{Bottom}: Extracted BAO spectrum $\Pw(k)/\Pnw(k)$ for $\Neff=3.046\text{ and }10$ with linear (\textit{left}) and logarithmic (\textit{right}) $k$-axis.}
		\label{fig:broadbandExtraction}
	\end{center}
\end{figure}

\noindent
Examples of the broadband extraction using this procedure are shown in Fig.~\ref{fig:broadbandExtraction}. We see that the extraction method is unbiased, i.e.\ the resulting wiggle spectrum both oscillates around zero and asymptotes to zero for large wavenumbers. In addition, it is robust and stable over a large parameter space at small computation time (depending on $n$). Since the position of the first~BAO~peak is close to the peak of the matter power spectrum, it is sensitive to how exactly the baryonic bump is removed. However, we have checked that the computed constraints on cosmological parameters are insensitive to this uncertainty. The same holds for varying the parameters $n$ and~$\kmax$ with fixed shifts in step~5 as long as $k_{1,2}$ are outside the range of wavenumbers of interest.

\subsection{Phase Shift Measurement}
\label{app:phaseMeasurement}

In the following, we describe our method for computing the phase shift template used in the likelihood analysis of Section~\ref{sec:phaseShift}.

\vskip4pt
First, we compute the BAO spectrum using \texttt{CLASS} and the broadband extraction method detailed above for a given value of $\Neff$. In practice, we set the primordial helium fraction $Y_p$ to the fiducial value, but the final template is independent of this choice. As discussed in~\S\ref{sec:template}, we keep the time of matter-radiation equality fixed at its fiducial value by changing the dark matter density $\omega_c$ according to
\beq
\omega_c = \frac{a_\nu + \Neff}{a_\nu + N_\mathrm{eff}^\mathrm{fid}} \left(\omega_c^\mathrm{fid}+\omega_b^\mathrm{fid}\right) - \omega_b^\mathrm{fid}\, ,
\eeq
where $a_\nu$ is defined in~\eqref{eq:Nnu}. We then fit the following envelope function to the maxima of the absolute value of the BAO spectrum:
\beq
a(k) \equiv e(k) d(k)\, , \quad \text{where} \quad	\begin{aligned}	e(k) &\equiv 1 - A_e \exp\left\{-a_e\, (k/k_e)^{\kappa_e}\right\} , \\
																	d(k) &\equiv A_d \exp\left\{-a_d \, (k/k_d)^{\kappa_d}\right\} .
													\end{aligned}
\eeq
The parameters $A_i$, $a_i$, $\kappa_i$, with $i=d,e$, are fitting parameters, while $k_e$ is the location of the peak of $\Pnw(k)$ and $k_d$ is the wavenumber associated with the mean squared diffusion distance. These fitting functions are motivated by the modeling in~\cite{Follin:2015hya, Baldauf:2015xfa}. We define the `undamped spectrum' as 
\beq
\mathcal{O}(k) \equiv a(k)^{-1} \Pw(k)/\Pnw(k)\, .
\eeq
For the fiducial cosmology, for instance, we find the following parameters: $A_e\approx0.141$, $a_e\approx0.0035$, $\kappa_e\approx5.5$, $k_e\approx\SI{0.016}{\hPerMpc}$, and $A_d\approx0.072$, $a_d\approx0.32$, $\kappa_d\approx1.9$, $k_d\approx\SI{0.12}{\hPerMpc}$.

\vskip4pt
Before we can measure the phase shift, we have to match the sound horizon at the drag epoch,~$r_s$, to that in the fiducial cosmology to remove the change to the BAO frequency induced by~$\Neff$. By rescaling the wavenumbers as $k \to r_s^\mathrm{fid}/r_s\,k$, we fix $r_s k$ to the fiducial model for all wavenumbers $k$. For convenience, we also normalize the spectrum such that the amplitude of the fourth peak is the same as in the fiducial cosmology.

\vskip4pt
Finally, we can extract the phase shift as the shift of the peaks/troughs and zeros of $\mathcal{O}(k)$ relative to the fiducial cosmology, $\delta k_* = k_* - k_*^\mathrm{fid}$. To obtain the template $f(k)$, we sample 100 cosmologies with varying $\Neff \in [0,3.33]$,\footnote{We restrict to this range of values of $\Neff$ as we observed a small, but unexpected jump in the peak locations around $\Neff\sim3.33$. Below and above, the peak locations change coherently with $\Neff$. This range was then chosen as we are mostly interested in smaller $\Neff$. However, we expect the template to be also valid for larger $\Neff$ outside the sampling range.} and define
\beq
f(k) \equiv \left\langle \frac{1}{1-\beta(\Neff)}\, \frac{\delta k_*(k;\Neff)}{r_s^\mathrm{fid}} \right\rangle_{\!\Neff} \, ,
\eeq
where $\beta(\Neff)$ is the normalization introduced in~\eqref{eq:phi_norm}. The bars in Fig.~\ref{fig:phaseShiftTemplate} indicate the locations of the peaks/troughs/zeros of the fiducial spectrum $\mathcal{O}(k)$ and their length shows the standard deviation in these measurements which is generally small.

\clearpage
\section{Convergence and Stability Tests}
\label{app:convergence}

One of the motivations for including our full list of forecasts in Appendix~\ref{app:specs} is to make the results reproducible. It is therefore also important that we explain how the numerical derivatives were computed in the Fisher matrix, including the employed step sizes. In this appendix, we provide this information and demonstrate that the step sizes are appropriate for the convergence and stability of our calculations.

\vskip4pt
The numerical derivatives in~\eqref{eq:galaxyFisherMatrix} and~\eqref{eq:baoFisherMatrix} are computed using a symmetric difference quotient or two-point stencil, $f'(\theta) = \left[f(\theta+h)-f(\theta-h)\right]\!/\!\left(2h\right)$, with fiducial parameter value $\theta$ and absolute step size $h$. For each parameter, we choose the step sizes given in Table~\ref{tab:parameterSpacings} resulting in relative step sizes, $h_\mathrm{rel} = h/\theta$, that are generally of order $\mathcal{O}\!\left(\num{e-2}\right)$.\medskip
\begin{table}[h]
	\centering
	\begin{tabular}{c S[table-format=1.4] S[table-format=1.1e-1]}
			\toprule
		Parameter						& {$h$}		& {$h_\mathrm{rel}$}	\\
			\midrule[0.065em]
		$\omega_b$						& 0.0008	& 3.6e-2				\\
		$\omega_c$						& 0.002		& 1.7e-2				\\
		$100\,\theta_s$					& 0.002		& 1.9e-3				\\
		$\ln(\num{e10}\As)$				& 0.05		& 1.6e-2				\\
		$\ns$							& 0.01		& 1.0e-2				\\
		$\tau$							& 0.02		& 3.0e-1				\\
			\midrule[0.065em]
		$\Neff$							& 0.08		& 2.6e-2				\\
		$Y_p$							& 0.005		& 2.0e-2				\\
			\bottomrule 
	\end{tabular}
	\caption{Absolute and relative step sizes, $h$ and $h_\mathrm{rel}$, used when computing the derivatives in the Fisher matrices.\bigskip}
	\label{tab:parameterSpacings}
\end{table}

\begin{figure}[t!]
	\begin{center}
		\includegraphics{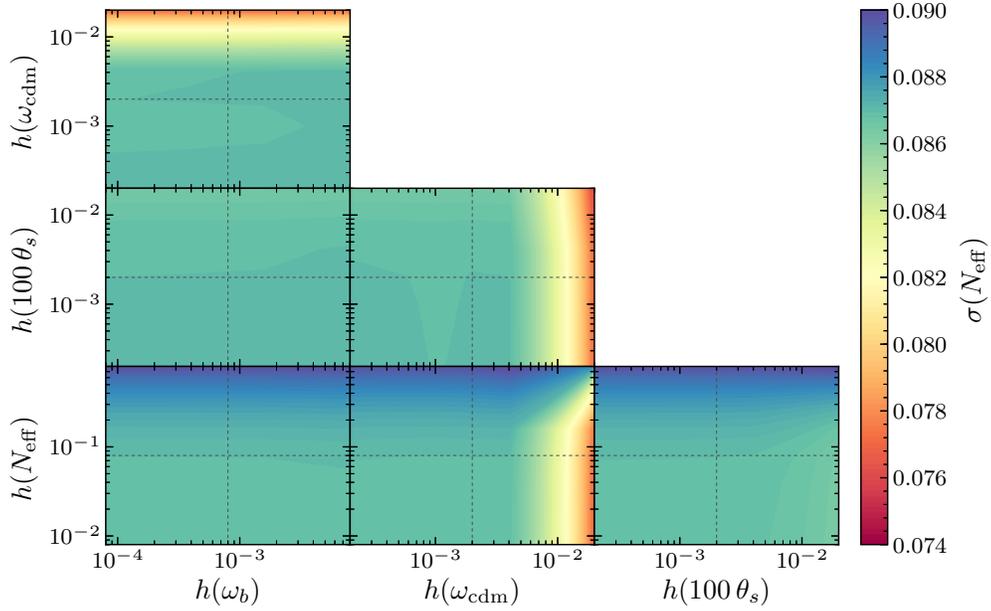}
		\caption{Results of the convergence test for the $P(k)$-forecasts of DESI in the fiducial $\Lambda\mathrm{CDM}$+$\Neff$ cosmology. The spectra for the numerical derivatives were computed using a high-accuracy setting of \texttt{CLASS}. The dashed lines indicate the step sizes employed in our forecasts.}
		\label{fig:convergencePk}
	\end{center}
\end{figure}

\begin{figure}[t!]
	\begin{center}
		\includegraphics{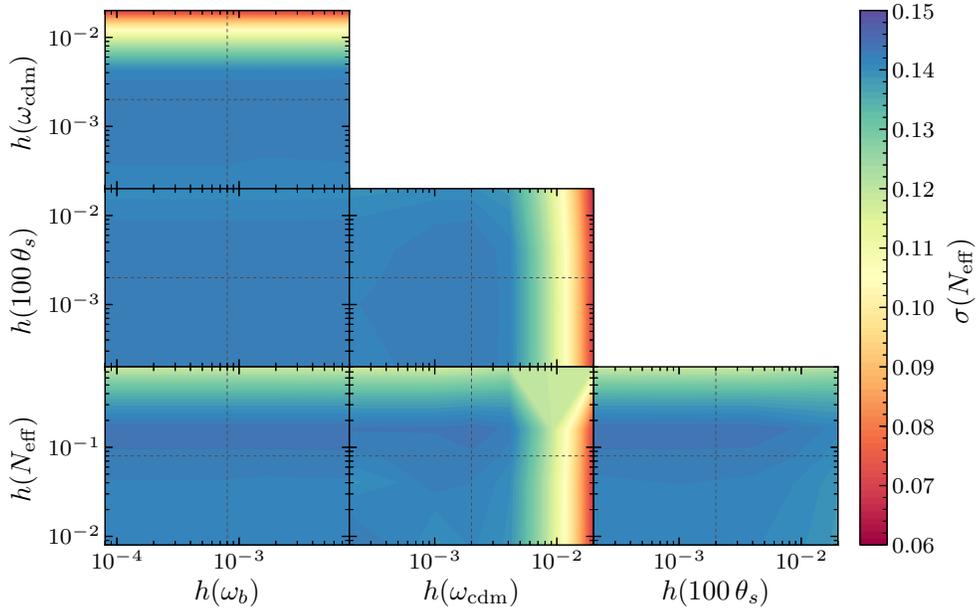}
		\caption{As in Figure~\ref{fig:convergencePk}, but for the BAO-forecasts of DESI.}
		\label{fig:convergenceBAO}
	\end{center}
\end{figure}

In Figures~\ref{fig:convergencePk} and~\ref{fig:convergenceBAO}, we show that our results are converged for both the $P(k)$- and BAO-forecasts. The results in these figures (as in the rest of this paper) use \texttt{CLASS} with a high-accuracy setting. We have also checked that the forecasted constraints are converged when employing the standard accuracy setting, but note that the results are slightly less stable to changes away from these values. For the $P(k)$-forecasts, we see that a sufficiently small step size is needed, but a further decrease in the step size still leads to converged results. The BAO-forecasts, by contrast, show islands of convergence where performance decreases both when the step size is increased and when it is decreased. This behavior is more noticeable using the standard accuracy setting of \texttt{CLASS}, but likely reflects the fact that the BAO feature is itself a small effect and small step sizes are therefore more likely to produce effects comparable to numeric or modeling errors.

\clearpage
\phantomsection
\addcontentsline{toc}{section}{References}
\bibliographystyle{utphys}
\bibliography{BAO-Forecast}

\providecommand{\href}[2]{#2}\begingroup\raggedright\begin{thebibliography}{10}

\bibitem{Bashinsky:2003tk}
S.~Bashinsky and U.~Seljak, ``{Neutrino Perturbations in CMB Anisotropy and
  Matter Clustering},''
  \href{http://dx.doi.org/10.1103/PhysRevD.69.083002}{{\em Phys. Rev. D}
  {\bfseries 69} (2004) 083002},
\href{http://arxiv.org/abs/astro-ph/0310198}{{\ttfamily arXiv:astro-ph/0310198
  [astro-ph]}}.

\bibitem{Hou:2011ec}
Z.~Hou, R.~Keisler, L.~Knox, M.~Millea, and C.~Reichardt, ``{How Massless
  Neutrinos Affect the Cosmic Microwave Background Damping Tail},''
  \href{http://dx.doi.org/10.1103/PhysRevD.87.083008}{{\em Phys. Rev. D}
  {\bfseries 87} (2013) 083008},
\href{http://arxiv.org/abs/1104.2333}{{\ttfamily arXiv:1104.2333
  [astro-ph.CO]}}.

\bibitem{Baumann:2015rya}
D.~Baumann, D.~Green, J.~Meyers, and B.~Wallisch, ``{Phases of New Physics in
  the CMB},'' \href{http://dx.doi.org/10.1088/1475-7516/2016/01/007}{{\em JCAP}
  {\bfseries 1601} (2016) 007},
\href{http://arxiv.org/abs/1508.06342}{{\ttfamily arXiv:1508.06342
  [astro-ph.CO]}}.

\bibitem{Pan:2016zla}
Z.~Pan, L.~Knox, B.~Mulroe, and A.~Narimani, ``{Cosmic Microwave Background
  Acoustic Peak Locations},''
  \href{http://dx.doi.org/10.1093/mnras/stw833}{{\em Mon. Not. Roy. Astron.
  Soc.} {\bfseries 459} (2016) 2513},
\href{http://arxiv.org/abs/1603.03091}{{\ttfamily arXiv:1603.03091
  [astro-ph.CO]}}.

\bibitem{Follin:2015hya}
B.~Follin, L.~Knox, M.~Millea, and Z.~Pan, ``{First Detection of the Acoustic
  Oscillation Phase Shift Expected from the Cosmic Neutrino Background},''
  \href{http://dx.doi.org/10.1103/PhysRevLett.115.091301}{{\em Phys. Rev.
  Lett.} {\bfseries 115} (2015) 091301},
\href{http://arxiv.org/abs/1503.07863}{{\ttfamily arXiv:1503.07863
  [astro-ph.CO]}}.

\bibitem{Essig:2013lka}
R.~Essig {\em et~al.}, ``{Working Group Report: New Light Weakly Coupled
  Particles},'' in {\em {Community Summer Study 2013: Snowmass on the
  Mississippi (CSS2013), Minneapolis, MN, USA, July 29-August 6, 2013}}.
\newblock 2013.
\newblock
\href{http://arxiv.org/abs/1311.0029}{{\ttfamily arXiv:1311.0029 [hep-ph]}}.
\newblock

\bibitem{Peccei:1977hh}
R.~Peccei and H.~Quinn, ``{CP Conservation in the Presence of
  Pseudoparticles},''
\href{http://dx.doi.org/10.1103/PhysRevLett.38.1440}{{\em Phys. Rev. Lett.}
  {\bfseries 38} (1977) 1440}.

\bibitem{Weinberg:1977ma}
S.~Weinberg, ``{A New Light Boson?},''
\href{http://dx.doi.org/10.1103/PhysRevLett.40.223}{{\em Phys. Rev. Lett.}
  {\bfseries 40} (1978) 223}.

\bibitem{Wilczek:1977pj}
F.~Wilczek, ``{Problem of Strong $P$ and $T$ Invariance in the Presence of
  Instantons},''
\href{http://dx.doi.org/10.1103/PhysRevLett.40.279}{{\em Phys. Rev. Lett.}
  {\bfseries 40} (1978) 279}.

\bibitem{Arvanitaki:2009fg}
A.~Arvanitaki, S.~Dimopoulos, S.~Dubovsky, N.~Kaloper, and J.~March-Russell,
  ``{String Axiverse},''
  \href{http://dx.doi.org/10.1103/PhysRevD.81.123530}{{\em Phys. Rev. D}
  {\bfseries 81} (2010) 123530},
\href{http://arxiv.org/abs/0905.4720}{{\ttfamily arXiv:0905.4720 [hep-th]}}.

\bibitem{Holdom:1985ag}
B.~Holdom, ``{Two $U(1)$'s and Epsilon Charge Shifts},''
\href{http://dx.doi.org/10.1016/0370-2693(86)91377-8}{{\em Phys. Lett. B}
  {\bfseries 166} (1986) 196}.

\bibitem{Galison:1983pa}
P.~Galison and A.~Manohar, ``{Two $Z$'s or Not Two $Z$'s?},''
\href{http://dx.doi.org/10.1016/0370-2693(84)91161-4}{{\em Phys. Lett. B}
  {\bfseries 136} (1984) 279}.

\bibitem{Abazajian:2012ys}
K.~Abazajian {\em et~al.}, ``{Light Sterile Neutrinos: A White Paper},''
\href{http://arxiv.org/abs/1204.5379}{{\ttfamily arXiv:1204.5379 [hep-ph]}}.

\bibitem{Raffelt:1996wa}
G.~Raffelt, {\em {Stars as Laboratories for Fundamental Physics}}.
\newblock University of Chicago Press, Chicago, IL, 1996.

\bibitem{Brust:2013xpv}
C.~Brust, D.~E. Kaplan, and M.~Walters, ``{New Light Species and the CMB},''
  \href{http://dx.doi.org/10.1007/JHEP12(2013)058}{{\em JHEP} {\bfseries 12}
  (2013) 058},
\href{http://arxiv.org/abs/1303.5379}{{\ttfamily arXiv:1303.5379 [hep-ph]}}.

\bibitem{Chacko:2015noa}
Z.~Chacko, Y.~Cui, S.~Hong, and T.~Okui, ``{Hidden Dark Matter Sector, Dark
  Radiation and the CMB},''
  \href{http://dx.doi.org/10.1103/PhysRevD.92.055033}{{\em Phys. Rev. D}
  {\bfseries 92} (2015) 055033},
\href{http://arxiv.org/abs/1505.04192}{{\ttfamily arXiv:1505.04192 [hep-ph]}}.

\bibitem{Baumann:2016wac}
D.~Baumann, D.~Green, and B.~Wallisch, ``{New Target for Cosmic Axion
  Searches},'' \href{http://dx.doi.org/10.1103/PhysRevLett.117.171301}{{\em
  Phys. Rev. Lett.} {\bfseries 117} (2016) 171301},
\href{http://arxiv.org/abs/1604.08614}{{\ttfamily arXiv:1604.08614
  [astro-ph.CO]}}.

\bibitem{Abazajian:2016yjj}
{K. Abazajian \textit{et al.} (CMB-S4 Collaboration)}, ``{CMB-S4 Science Book,
  First Edition},''
\href{http://arxiv.org/abs/1610.02743}{{\ttfamily arXiv:1610.02743
  [astro-ph.CO]}}.

\bibitem{Dawson:2012va}
{K. Dawson \textit{et al.} (BOSS Collaboration)}, ``{The Baryon Oscillation
  Spectroscopic Survey of SDSS-III},''
  \href{http://dx.doi.org/10.1088/0004-6256/145/1/10}{{\em Astron. J.}
  {\bfseries 145} (2013) 10},
\href{http://arxiv.org/abs/1208.0022}{{\ttfamily arXiv:1208.0022
  [astro-ph.CO]}}.

\bibitem{Dawson:2015wdb}
K.~Dawson {\em et~al.}, ``{The SDSS-IV extended Baryon Oscillation
  Spectroscopic Survey: Overview and Early Data},''
  \href{http://dx.doi.org/10.3847/0004-6256/151/2/44}{{\em Astron. J.}
  {\bfseries 151} (2016) 44},
\href{http://arxiv.org/abs/1508.04473}{{\ttfamily arXiv:1508.04473
  [astro-ph.CO]}}.

\bibitem{Abbott:2005bi}
{T. Abbott \textit{et al.} (DES Collaboration)}, ``{The Dark Energy Survey},''
\href{http://arxiv.org/abs/astro-ph/0510346}{{\ttfamily arXiv:astro-ph/0510346
  [astro-ph]}}.

\bibitem{Aghamousa:2016zmz}
{A. Aghamousa \textit{et al.} (DESI Collaboration)}, ``{The DESI Experiment
  Part I: Science, Targeting and Survey Design},''
\href{http://arxiv.org/abs/1611.00036}{{\ttfamily arXiv:1611.00036
  [astro-ph.IM]}}.

\bibitem{Ivezic:2008fe}
{\v{Z}. Ivezi{\'c} \textit{et al.} (LSST Collaboration)}, ``{LSST: From Science
  Drivers to Reference Design and Anticipated Data Products},''
\href{http://arxiv.org/abs/0805.2366}{{\ttfamily arXiv:0805.2366 [astro-ph]}}.

\bibitem{Laureijs:2011gra}
{R. Laureijs \textit{et al.} (Euclid Collaboration)}, ``{Euclid Definition
  Study Report},''
\href{http://arxiv.org/abs/1110.3193}{{\ttfamily arXiv:1110.3193
  [astro-ph.CO]}}.

\bibitem{Font-Ribera:2013rwa}
A.~Font-Ribera, P.~McDonald, N.~Mostek, B.~Reid, H.-J. Seo, and A.~Slosar,
  ``{DESI and Other Dark Energy Experiments in the Era of Neutrino Mass
  Measurements},'' \href{http://dx.doi.org/10.1088/1475-7516/2014/05/023}{{\em
  JCAP} {\bfseries 1405} (2014) 023},
\href{http://arxiv.org/abs/1308.4164}{{\ttfamily arXiv:1308.4164
  [astro-ph.CO]}}.

\bibitem{Dodelson:2016wal}
S.~Dodelson, K.~Heitmann, C.~Hirata, K.~Honscheid, A.~Roodman, U.~Seljak,
  A.~Slosar, and M.~Trodden, ``{Cosmic Visions Dark Energy: Science},''
\href{http://arxiv.org/abs/1604.07626}{{\ttfamily arXiv:1604.07626
  [astro-ph.CO]}}.

\bibitem{Obuljen:2017jiy}
A.~Obuljen, E.~Castorina, F.~Villaescusa-Navarro, and M.~Viel, ``{High-Redshift
  Post-Reionisation Cosmology with \SI{21}{cm} Intensity Mapping},''
  \href{http://dx.doi.org/10.1088/1475-7516/2018/05/004}{{\em JCAP} {\bfseries
  1805} (2018) 004},
\href{http://arxiv.org/abs/1709.07893}{{\ttfamily arXiv:1709.07893
  [astro-ph.CO]}}.

\bibitem{Baumann:2017lmt}
D.~Baumann, D.~Green, and M.~Zaldarriaga, ``{Phases of New Physics in the BAO
  Spectrum},'' \href{http://dx.doi.org/10.1088/1475-7516/2017/11/007}{{\em
  JCAP} {\bfseries 1711} (2017) 007},
\href{http://arxiv.org/abs/1703.00894}{{\ttfamily arXiv:1703.00894
  [astro-ph.CO]}}.

\bibitem{Eisenstein:2006nj}
D.~Eisenstein, H.-J. Seo, and M.~White, ``{On the Robustness of the Acoustic
  Scale in the Low-Redshift Clustering of Matter},''
  \href{http://dx.doi.org/10.1086/518755}{{\em Astrophys. J.} {\bfseries 664}
  (2007) 660},
\href{http://arxiv.org/abs/astro-ph/0604361}{{\ttfamily arXiv:astro-ph/0604361
  [astro-ph]}}.

\bibitem{Crocce:2007dt}
M.~Crocce and R.~Scoccimarro, ``{Nonlinear Evolution of Baryon Acoustic
  Oscillations},'' \href{http://dx.doi.org/10.1103/PhysRevD.77.023533}{{\em
  Phys. Rev. D} {\bfseries 77} (2008) 023533},
\href{http://arxiv.org/abs/0704.2783}{{\ttfamily arXiv:0704.2783 [astro-ph]}}.

\bibitem{Sugiyama:2013gza}
N.~Sugiyama and D.~Spergel, ``{How Does Nonlinear Dynamics Affect the Baryon
  Acoustic Oscillation?},''
  \href{http://dx.doi.org/10.1088/1475-7516/2014/02/042}{{\em JCAP} {\bfseries
  1402} (2014) 042},
\href{http://arxiv.org/abs/1306.6660}{{\ttfamily arXiv:1306.6660
  [astro-ph.CO]}}.

\bibitem{Hinshaw:2012aka}
{G. Hinshaw \textit{et al.} (WMAP Collaboration)}, ``{Nine-Year Wilkinson
  Microwave Anisotropy Probe (WMAP) Observations: Cosmological Parameter
  Results},'' \href{http://dx.doi.org/10.1088/0067-0049/208/2/19}{{\em
  Astrophys. J. Suppl.} {\bfseries 208} (2013) 19},
\href{http://arxiv.org/abs/1212.5226}{{\ttfamily arXiv:1212.5226
  [astro-ph.CO]}}.

\bibitem{Ade:2015xua}
{P. A. R. Ade \textit{et al.} (Planck Collaboration)}, ``{Planck 2015 Results.
  XIII. Cosmological Parameters},''
  \href{http://dx.doi.org/10.1051/0004-6361/201525830}{{\em Astron. Astrophys.}
  {\bfseries 594} (2016) A13},
\href{http://arxiv.org/abs/1502.01589}{{\ttfamily arXiv:1502.01589
  [astro-ph.CO]}}.

\bibitem{Alam:2016hwk}
{S. Alam \textit{et al.} (BOSS Collaboration)}, ``{The Clustering of Galaxies
  in the Completed SDSS-III Baryon Oscillation Spectroscopic Survey:
  Cosmological Analysis of the DR12 Galaxy Sample},''
  \href{http://dx.doi.org/10.1093/mnras/stx721}{{\em Mon. Not. Roy. Astron.
  Soc.} {\bfseries 470} (2017) 2617},
\href{http://arxiv.org/abs/1607.03155}{{\ttfamily arXiv:1607.03155
  [astro-ph.CO]}}.

\bibitem{Abbott:2017wau}
{T. Abbott \textit{et al.} (DES Collaboration)}, ``{Dark Energy Survey Year 1
  Results: Cosmological Constraints from Galaxy Clustering and Weak Lensing},''
\href{http://arxiv.org/abs/1708.01530}{{\ttfamily arXiv:1708.01530
  [astro-ph.CO]}}.

\bibitem{Cyburt:2015mya}
R.~Cyburt, B.~Fields, K.~Olive, and T.-H. Yeh, ``{Big Bang Nucleosynthesis:
  2015},'' \href{http://dx.doi.org/10.1103/RevModPhys.88.015004}{{\em Rev. Mod.
  Phys.} {\bfseries 88} (2016) 015004},
  \href{http://arxiv.org/abs/1505.01076}{{\ttfamily arXiv:1505.01076
  [astro-ph.CO]}}.

\bibitem{Borsanyi:2016ksw}
S.~Borsanyi {\em et~al.}, ``{Calculation of the Axion Mass Based on
  High-Temperature Lattice Quantum Chromodynamics},''
  \href{http://dx.doi.org/10.1038/nature20115}{{\em Nature} {\bfseries 539}
  (2016) 69},
\href{http://arxiv.org/abs/1606.07494}{{\ttfamily arXiv:1606.07494 [hep-lat]}}.

\bibitem{Keisler:2011aw}
R.~Keisler {\em et~al.}, ``{A Measurement of the Damping Tail of the Cosmic
  Microwave Background Power Spectrum with the South Pole Telescope},''
  \href{http://dx.doi.org/10.1088/0004-637X/743/1/28}{{\em Astrophys. J.}
  {\bfseries 743} (2011) 28},
\href{http://arxiv.org/abs/1105.3182}{{\ttfamily arXiv:1105.3182
  [astro-ph.CO]}}.

\bibitem{Bassett:2009mm}
B.~Bassett and R.~Hlozek, ``{Baryon Acoustic Oscillations},''
\href{http://arxiv.org/abs/0910.5224}{{\ttfamily arXiv:0910.5224
  [astro-ph.CO]}}.

\bibitem{Verde:2009tu}
L.~Verde, ``{Statistical Methods in Cosmology},''
  \href{http://dx.doi.org/10.1007/978-3-642-10598-2_4}{{\em Lect. Notes Phys.}
  {\bfseries 800} (2010) 147},
\href{http://arxiv.org/abs/0911.3105}{{\ttfamily arXiv:0911.3105
  [astro-ph.CO]}}.

\bibitem{Tegmark:1997rp}
M.~Tegmark, ``{Measuring Cosmological Parameters with Galaxy Surveys},''
  \href{http://dx.doi.org/10.1103/PhysRevLett.79.3806}{{\em Phys. Rev. Lett.}
  {\bfseries 79} (1997) 3806},
\href{http://arxiv.org/abs/astro-ph/9706198}{{\ttfamily arXiv:astro-ph/9706198
  [astro-ph]}}.

\bibitem{Blas:2016sfa}
D.~Blas, M.~Garny, M.~Ivanov, and S.~Sibiryakov, ``{Time-Sliced Perturbation
  Theory II: Baryon Acoustic Oscillations and Infrared Resummation},''
  \href{http://dx.doi.org/10.1088/1475-7516/2016/07/028}{{\em JCAP} {\bfseries
  1607} (2016) 028},
\href{http://arxiv.org/abs/1605.02149}{{\ttfamily arXiv:1605.02149
  [astro-ph.CO]}}.

\bibitem{Hand:2017ilm}
N.~Hand, U.~Seljak, F.~Beutler, and Z.~Vlah, ``{Extending the Modeling of the
  Anisotropic Galaxy Power Spectrum to $k = \SI{0.4}{\hPerMpc}$},''
  \href{http://dx.doi.org/10.1088/1475-7516/2017/10/009}{{\em JCAP} {\bfseries
  1710} (2017) 009},
\href{http://arxiv.org/abs/1706.02362}{{\ttfamily arXiv:1706.02362
  [astro-ph.CO]}}.

\bibitem{Ding:2017gad}
Z.~Ding, H.-J. Seo, Z.~Vlah, Y.~Feng, M.~Schmittfull, and F.~Beutler,
  ``{Theoretical Systematics of Future Baryon Acoustic Oscillation Surveys},''
  \href{http://dx.doi.org/10.1093/mnras/sty1413}{{\em Mon. Not. Roy. Astron.
  Soc.} {\bfseries 479} (2018) 1021},
\href{http://arxiv.org/abs/1708.01297}{{\ttfamily arXiv:1708.01297
  [astro-ph.CO]}}.

\bibitem{Ballinger:1996cd}
W.~Ballinger, J.~Peacock, and A.~Heavens, ``{Measuring the Cosmological
  Constant with Redshift Surveys},''
  \href{http://dx.doi.org/10.1093/mnras/282.3.877}{{\em Mon. Not. Roy. Astron.
  Soc.} {\bfseries 282} (1996) 877},
\href{http://arxiv.org/abs/astro-ph/9605017}{{\ttfamily arXiv:astro-ph/9605017
  [astro-ph]}}.

\bibitem{Kaiser:1987qv}
N.~Kaiser, ``{Clustering in Real Space and in Redshift Space},''
\href{http://dx.doi.org/10.1093/mnras/227.1.1}{{\em Mon. Not. Roy. Astron.
  Soc.} {\bfseries 227} (1987) 1}.

\bibitem{Seo:2007ns}
H.-J. Seo and D.~Eisenstein, ``{Improved Forecasts for the Baryon Acoustic
  Oscillations and Cosmological Distance Scale},''
  \href{http://dx.doi.org/10.1086/519549}{{\em Astrophys. J.} {\bfseries 665}
  (2007) 14},
\href{http://arxiv.org/abs/astro-ph/0701079}{{\ttfamily arXiv:astro-ph/0701079
  [astro-ph]}}.

\bibitem{White:2010qd}
M.~White, ``{Shot Noise and Reconstruction of the Acoustic Peak},''
\href{http://arxiv.org/abs/1004.0250}{{\ttfamily arXiv:1004.0250
  [astro-ph.CO]}}.

\bibitem{Seo:2003pu}
H.-J. Seo and D.~Eisenstein, ``{Probing Dark Energy with Baryonic Acoustic
  Oscillations from Future Large Galaxy Redshift Surveys},''
  \href{http://dx.doi.org/10.1086/379122}{{\em Astrophys. J.} {\bfseries 598}
  (2003) 720},
\href{http://arxiv.org/abs/astro-ph/0307460}{{\ttfamily arXiv:astro-ph/0307460
  [astro-ph]}}.

\bibitem{Zhan:2005ki}
H.~Zhan and L.~Knox, ``{Baryon Oscillations and Consistency Tests for
  Photometrically-Determined Redshifts of Very Faint Galaxies},''
  \href{http://dx.doi.org/10.1086/503622}{{\em Astrophys. J.} {\bfseries 644}
  (2006) 663},
\href{http://arxiv.org/abs/astro-ph/0509260}{{\ttfamily arXiv:astro-ph/0509260
  [astro-ph]}}.

\bibitem{Gleyzes:2016tdh}
J.~Gleyzes, R.~de~Putter, D.~Green, and O.~Dor\'{e}, ``{Biasing and the Search
  for Primordial Non-Gaussianity Beyond the Local Type},''
  \href{http://dx.doi.org/10.1088/1475-7516/2017/04/002}{{\em JCAP} {\bfseries
  1704} (2017) 002},
\href{http://arxiv.org/abs/1612.06366}{{\ttfamily arXiv:1612.06366
  [astro-ph.CO]}}.

\bibitem{Beutler:2016ixs}
{F. Beutler \textit{et al.} (BOSS Collaboration)}, ``{The Clustering of
  Galaxies in the Completed SDSS-III Baryon Oscillation Spectroscopic Survey:
  Baryon Acoustic Oscillations in Fourier Space},''
  \href{http://dx.doi.org/10.1093/mnras/stw2373}{{\em Mon. Not. Roy. Astron.
  Soc.} {\bfseries 464} (2017) 3409},
\href{http://arxiv.org/abs/1607.03149}{{\ttfamily arXiv:1607.03149
  [astro-ph.CO]}}.

\bibitem{Seo:2015eyw}
H.-J. Seo, F.~Beutler, A.~Ross, and S.~Saito, ``{Modeling the Reconstructed BAO
  in Fourier Space},'' \href{http://dx.doi.org/10.1093/mnras/stw1138}{{\em Mon.
  Not. Roy. Astron. Soc.} {\bfseries 460} (2016) 2453},
\href{http://arxiv.org/abs/1511.00663}{{\ttfamily arXiv:1511.00663
  [astro-ph.CO]}}.

\bibitem{Baumann:2018qnt}
D.~{Baumann}, F.~{Beutler}, R.~{Flauger}, D.~{Green}, A.~{Slosar},
  M.~{Vargas-Maga\~{n}a}, B.~{Wallisch}, and C.~{Y\`{e}che}, ``{First
  Measurement of Neutrinos in the BAO Spectrum},''
\href{http://arxiv.org/abs/1803.10741}{{\ttfamily arXiv:1803.10741
  [astro-ph.CO]}}.

\bibitem{Freedman:2017yms}
W.~Freedman, ``{Cosmology at a Crossroads: Tension with the Hubble Constant},''
  \href{http://dx.doi.org/10.1038/s41550-017-0121}{{\em Nat. Astron.}
  {\bfseries 1} (2017) 0121},
\href{http://arxiv.org/abs/1706.02739}{{\ttfamily arXiv:1706.02739
  [astro-ph.CO]}}.

\bibitem{Blas:2011rf}
D.~Blas, J.~Lesgourgues, and T.~Tram, ``{The Cosmic Linear Anisotropy Solving
  System (CLASS) II: Approximation Schemes},''
  \href{http://dx.doi.org/10.1088/1475-7516/2011/07/034}{{\em JCAP} {\bfseries
  1107} (2011) 034},
\href{http://arxiv.org/abs/1104.2933}{{\ttfamily arXiv:1104.2933
  [astro-ph.CO]}}.

\bibitem{Perez:2007ipy}
F.~P\'{e}rez and B.~Granger, ``{IPython: A System for Interactive Scientific
  Computing},'' \href{http://dx.doi.org/10.1109/MCSE.2007.53}{{\em Comput. Sci.
  Eng.} {\bfseries 9} (2007) 21}.

\bibitem{Hunter:2007mat}
J.~Hunter, ``{Matplotlib: A 2D Graphics Environment},''
  \href{http://dx.doi.org/10.1109/MCSE.2007.55}{{\em Comput. Sci. Eng.}
  {\bfseries 9} (2007) 90}.

\bibitem{Walt:2011num}
S.~van~der Walt, S.~Colbert, and G.~Varoquaux, ``{The NumPy Array: A Structure
  for Efficient Numerical Computation},''
  \href{http://dx.doi.org/10.1109/MCSE.2011.37}{{\em Comput. Sci. Eng.}
  {\bfseries 13} (2011) 22}, \href{http://arxiv.org/abs/1102.1523}{{\ttfamily
  arXiv:1102.1523 [cs.MS]}}.

\bibitem{Wu:2014hta}
W.~Wu, J.~Errard, C.~Dvorkin, C.~Kuo, A.~Lee, P.~McDonald, A.~Slosar, and
  O.~Zahn, ``{A Guide to Designing Future Ground-Based Cosmic Microwave
  Background Experiments},''
  \href{http://dx.doi.org/10.1088/0004-637X/788/2/138}{{\em Astrophys. J.}
  {\bfseries 788} (2014) 138},
\href{http://arxiv.org/abs/1402.4108}{{\ttfamily arXiv:1402.4108
  [astro-ph.CO]}}.

\bibitem{Galli:2014kla}
S.~Galli, K.~Benabed, F.~Bouchet, J.-F. Cardoso, F.~Elsner, E.~Hivon,
  A.~Mangilli, S.~Prunet, and B.~Wandelt, ``{CMB Polarization Can Constrain
  Cosmology Better Than CMB Temperature},''
  \href{http://dx.doi.org/10.1103/PhysRevD.90.063504}{{\em Phys. Rev. D}
  {\bfseries 90} (2014) 063504},
\href{http://arxiv.org/abs/1403.5271}{{\ttfamily arXiv:1403.5271
  [astro-ph.CO]}}.

\bibitem{Allison:2015qca}
R.~Allison, P.~Caucal, E.~Calabrese, J.~Dunkley, and T.~Louis, ``{Towards a
  Cosmological Neutrino Mass Detection},''
  \href{http://dx.doi.org/10.1103/PhysRevD.92.123535}{{\em Phys. Rev. D}
  {\bfseries 92} (2015) 123535},
\href{http://arxiv.org/abs/1509.07471}{{\ttfamily arXiv:1509.07471
  [astro-ph.CO]}}.

\bibitem{Green:2016cjr}
D.~Green, J.~Meyers, and A.~van Engelen, ``{CMB Delensing Beyond the
  B-Modes},'' \href{http://dx.doi.org/10.1088/1475-7516/2017/12/005}{{\em JCAP}
  {\bfseries 1712} (2017) 005},
\href{http://arxiv.org/abs/1609.08143}{{\ttfamily arXiv:1609.08143
  [astro-ph.CO]}}.

\bibitem{Hamann:2010pw}
J.~Hamann, S.~Hannestad, J.~Lesgourgues, C.~Rampf, and Y.~Y.~Y. Wong,
  ``{Cosmological Parameters from Large-Scale Structure -- Geometric Versus
  Shape Information},''
  \href{http://dx.doi.org/10.1088/1475-7516/2010/07/022}{{\em JCAP} {\bfseries
  1007} (2010) 022},
\href{http://arxiv.org/abs/1003.3999}{{\ttfamily arXiv:1003.3999
  [astro-ph.CO]}}.

\bibitem{Baldauf:2015xfa}
T.~Baldauf, M.~Mirbabayi, M.~Simonovi\'{c}, and M.~Zaldarriaga, ``{Equivalence
  Principle and the Baryon Acoustic Peak},''
  \href{http://dx.doi.org/10.1103/PhysRevD.92.043514}{{\em Phys. Rev. D}
  {\bfseries 92} (2015) 043514},
\href{http://arxiv.org/abs/1504.04366}{{\ttfamily arXiv:1504.04366
  [astro-ph.CO]}}.

\end{thebibliography}\endgroup

\end{document}